\documentclass[11pt,a4paper]{article}

\usepackage[
   textwidth=6.7in,
    top=0.9in,
    bottom=0.95in
]{geometry}

\usepackage{lineno}
\usepackage{enumerate}
\usepackage{amsthm,amssymb,amsmath,mathrsfs,mathtools}
\usepackage{bm,bbm,extarrows,cancel,manfnt,marvosym}
\usepackage{pst-node}
\usepackage{tikz-cd} 
\usepackage[titletoc, title]{appendix}
\usepackage{pstricks}
\usepackage{slashed}

\usepackage{booktabs}
\usepackage{tabularx}
\usepackage{array}
\usetikzlibrary{positioning}
\usetikzlibrary{calc}
\usetikzlibrary{decorations.markings, arrows.meta}
\usepackage{subcaption} 
\theoremstyle{definition}
\newtheorem{Def}{Definition}[section]

\newtheorem*{Prob_w/o_No.}{Problem}
\theoremstyle{remark}
\newtheorem{Rmk}[Def]{Remark}

\theoremstyle{plain}
\newtheorem{Prop}[Def]{Proposition}
\newtheorem{Lem}[Def]{Lemma}
\newtheorem{Thm}[Def]{Theorem}
\newtheorem{Cor}[Def]{Corollary}

\newtheorem{Ex}{Example}[section]
\newenvironment{Prf}{\vspace{4pt}\textbf{\\}{\color{black}\textit{Proof}}.\normalfont}{\color{black}\hfill$\blacksquare$\vspace{4pt}\par}	% 

\usepackage[
    colorlinks=true,
    linkcolor=blue,
    citecolor=blue,
    urlcolor=blue
]{hyperref}

\usepackage[numbers,sort&compress]{natbib}

\numberwithin{equation}{section}

\newcommand{\diag}{\operatorname{diag}}
\newcommand{\Hom}{\operatorname{Hom}}

\title{\boldmath Caterpillar Degenerations of Spectral Curves and Double Gelfand-Zeitlin Geometry}

\author{George Y. Liu}

\date{}

\begin{document}
\maketitle
\abstract{In this paper, we study the decomposition, in the caterpillar or weak-coupling limit, of a spectral curve with two irregular poles of order two, namely a spectral curve of type $(2,2)$, into two spectral curves each having one regular singularity(order-$1$ pole) and one irregular pole of order two, namely spectral curves of type $(1,2)$ or $(2,1)$. Such spectral curves can be identified with a double Gelfand--Zeitlin (DGZ) system. We show that the Duistermaat--Heckman measure on the DGZ side can be identified with the Weyl measure on the gauge-theory side arises from gluing two quivers. Furthermore, the Fourier expansion of the isomonodromic tau function can be interpreted as a Peter--Weyl expansion in the real polarization of the DGZ system. On the other hand, we introduce the nodal caterpillar spectral network. Applying the caterpillar spectral network, we then obtain an explicit expression for the Stokes matrix up to a normalization factor. This expression agrees with the existing analytic formulas. Moreover, we identify the variation of endpoint normalization arising in the Stokes computation with the equivariant Euler class of the fluctuation complex over the Coulomb moduli space. Finally, using the gluing technique for spectral curves and the DGZ system, we conjecturally predict the perturbative coefficients of the higher-rank $P_{\mathrm{III}}$ isomonodromic tau function from the viewpoint of spectral geometry.
 }
 
\section{Introduction}
We study the family of wild spectral curves
\begin{equation}
  \Gamma^{(n)}_{2,2}(u,A,B):\qquad
  \det\!\left[xI_n-\left(iu-\frac{B}{z}-\frac{A}{z^2}\right)\right]=0,
 \label{eq:22spc}
\end{equation}
where $u=\operatorname{diag}(u_1,\ldots,u_n)$ has pairwise distinct real
entries and $A,B$ are Hermitian or anti-Hermitian matrices. We further require that $A$ and all leading principal minors of $B$ are regular semisimple. The curve
$\Gamma^{(n)}_{2,2}$ is an $n$-sheeted branched cover of
$\mathbb{P}^1\setminus\{0,\infty\}$, with punctures above both irregular
ends.  It is the semiclassical spectral curve of the meromorphic system 
\begin{equation}
  \epsilon\frac{dF}{dz}
  =\left(iu-\frac{B}{z}-\frac{A}{z^2}\right)F,
  \label{eq:22ODE}
\end{equation}
and, equivalently, the characteristic curve of the meromorphic Higgs field
$\Phi(z)\,dz=(iu-B/z-A/z^2)\,dz$. The latter has poles of order two at
$z=0$ and $z=\infty$, and therefore defines a wild Hitchin system
\cite{Hitchin:1987mz,biquard2002wildnonabelianhodgetheory}. We also mention that the isomonodromic deformation equation of (\ref{eq:22ODE}) is the D6 Painlevé III, if $F$ is a $2\times 2$ solution matrix \cite{Jimbo:1981tov,Okamoto1986StudiesOT}. 

From the physical point of view, the relation between Seiberg-Witten/spectral curve and $4$d $N=2$ supersymmetric theories was first studied by Seiberg and Witten in \cite{Seiberg_1994}. In \cite{Witten_1997} Witten further derived the $4$d $N=2$ supersymmetric field theories from the configurations of branes. Namely, after lifting to $M$-theory, the spectral curves are wrapped by the $M$5 branes. Latter, Gaiotto started from $n$ coincidents $M5$ branes, the $A_{n-1}$ $6$d $(2,0)$ theory, and let them wrap $\mathbb{R}^{3,1} \times C$, here $C$ is a base Riemann surface. The compactification of the $6$d theory on $C$ then gives a $4$d $N=2$ supersymmetric field theory. By varying the punctures of the base curve $C$ and the $6$d theory, the field theories arising from the compactification vary as well, and those theories are known as the class-$\mathcal{S}$ theories\cite{Gaiotto_2012}. From the class-$\mathcal S$ perspective, the spectral curve \ref{eq:22spc} data describe the Seiberg--Witten geometry obtained from the six-dimensional $A_{n-1}$ type $(2,0)$ theory (up to the decoupled center-of-mass $U(1)$ sector) on a twice-punctured sphere with irregular defects \cite{nanopoulos2010hitchinequationirregularsingularity}.  In this interpretation, $\lambda=x\,dz$ is the Seiberg--Witten differential, and its periods give the special and dual special coordinates of the low-energy theory.

The problem addressed in this paper is to understand the structure when
both irregular ends are simultaneously taken to a caterpillar/weak-coupling regime. An explicit definition of such a limit is given in \ref{Def:dc}. The caterpillar limit is worth studying for several reasons. Mathematically, it produces hierarchical structure reflecting the chain $U(1)\subset U(2)\subset\cdots\subset U(n)$, and is closely related to the Gelfand--Zeitlin system, as well as the crystal structures emerging in the Poisson--Lie \cite{Boalch_2001}/quantum group \cite{TOLEDANOLAREDO2023109189,xu2024regularizedlimitsstokesmatrices,Xu2020RepresentationsOQ} description of Stokes data. From the physical viewpoint, the caterpillar limit is actually the weak-coupling limit of the corresponding quiver theory, as we will discuss in more detail in section~5. For the one-sided $(1,2)$ system, regularized Stokes limits and the corresponding caterpillar geometry have been studied in
\cite{xu2024regularizedlimitsstokesmatrices,alekseev2024wkbasymptoticsstokesmatrices}.  In particular, Alekseev--Neitzke--Xu--Zhou relate the leading WKB exponents to periods of a degenerate spectral curve and to Gelfand--Zeitlin patterns. The symplectic geometry of the two-sided system with two second-order poles was studied by Boalch \cite{Boalch_2007}, who related the moduli space of connections with two second-order poles, $T^*G$, to the Lu--Weinstein symplectic double groupoid. Recently, such a system was also studied analytically by Wang and Xu, who determine asymptotic parameters and explicit Stokes and connection matrices \cite{wang2025asymptoticmonodromyproblemshigherorder}. Related boundary-value problems for higher-dimensional isomonodromy equations were studied by Tang and Xu \cite{Tang:2024qyr}, who constructed solutions with prescribed boundary asymptotics and related them to generic isomonodromic solutions. Independently, the double Gelfand--Zeitlin system on $T^*U(n)$ was introduced in \cite{crooks2023doublegelfandcetlinsysteminvariance}. What remains unclear is a common geometric framework that connects the simultaneous two-sided degeneration, the
double Gelfand--Zeitlin integrable system, the limiting spectral network,
and the perturbative coefficients of the associated isomonodromic tau
functions.

\begin{figure}[htbp]
\centering

\begin{tikzpicture}[
box/.style={
    draw,
    rounded corners,
    thick,
    align=center,
    minimum width=4.2cm,
    minimum height=1cm,
    font=\small
},
widebox/.style={
    draw,
    rounded corners,
    thick,
    align=center,
    minimum width=7cm,
    minimum height=1.1cm,
    font=\small
},
arrow/.style={
    thick,
    ->,
    >=latex
}
]

\node[widebox] (curve) at (0,6)
{
$(2,2)$ spectral curve $\Gamma_{2,2}^{(n)}$
\\
and spectral network
};

\node[widebox] (deg) at (0,4.8)
{
Caterpillar degeneration
\\
$(2,2)\rightarrow (1,2)+(2,1)$
with nodal gluing
};

\draw[arrow] (curve.south)--(deg.north);

\node[box] (tail) at (-3.8,3)
{
Wild-tail calculation
\\
$(1,2)/(2,1)$ components
\\
Gelfand--Zeitlin structure
\\
Stokes data
};

\node[box] (glue) at (3.8,3)
{
Nodal gluing
\\
Double Gelfand--Zeitlin
\\
Duistermaat--Heckman measure
};

\draw[arrow] (deg.south)--(tail.north);
\draw[arrow] (deg.south)--(glue.north);

\node[box] (endpoint) at (-3.8,1.2)
{
Endpoint normalization $g_i^{(k)}$ obtained from  \\
$\frac{T_{k,i}g_i^{(k)}}{g_i^{(k)}}= c_{i,k}
|\ell^-_{i,k}|^2
|\ell^+_{i,k}|^2$

};

\node[box] (weyl) at (3.8,1.2)
{
Weyl measure
\\
$
\Delta(a)^2 da
$
\\
$Z_{\rm vec}^{\rm neck}$
};

\draw[arrow] (tail.south)--(endpoint.north);
\draw[arrow] (glue.south)--(weyl.north);

\node[widebox] (euler) at (-3.8,-0.8)
{
\\[1mm]
$
\frac{T_{k,i}g_i^{(k)}}{g_i^{(k)}}
\simeq
e_T(\mathcal G^{\rm vir}_{k,i})
$
\\
Coulomb-branch fluctuation complex
};

\draw[arrow] (endpoint.south)--(euler.north);

\node[box] (zloop) at (-3.8,-3)
{
Tail one-loop factor
\\
$Z_{\rm tail}^{\rm 1-loop}$
};

\draw[arrow] (euler.south)--(zloop.north);

\node[widebox] (final) at (0,-4.8)
{
Perturbative Nekrasov coefficient
\\[1mm]
$
Z_{\rm Nek}^{\rm pert}
=
Z_{\rm cl}Z_{\rm 1-loop}
=
Z_{\rm tail}^{\rm 1-loop}
Z_{\rm vec}^{\rm neck}
$
};

\draw[arrow] (zloop.south)--(final.north west);
\draw[arrow] (weyl.south)--(final.north east);

\end{tikzpicture}

\caption{
Main structure of the paper
}

\end{figure}
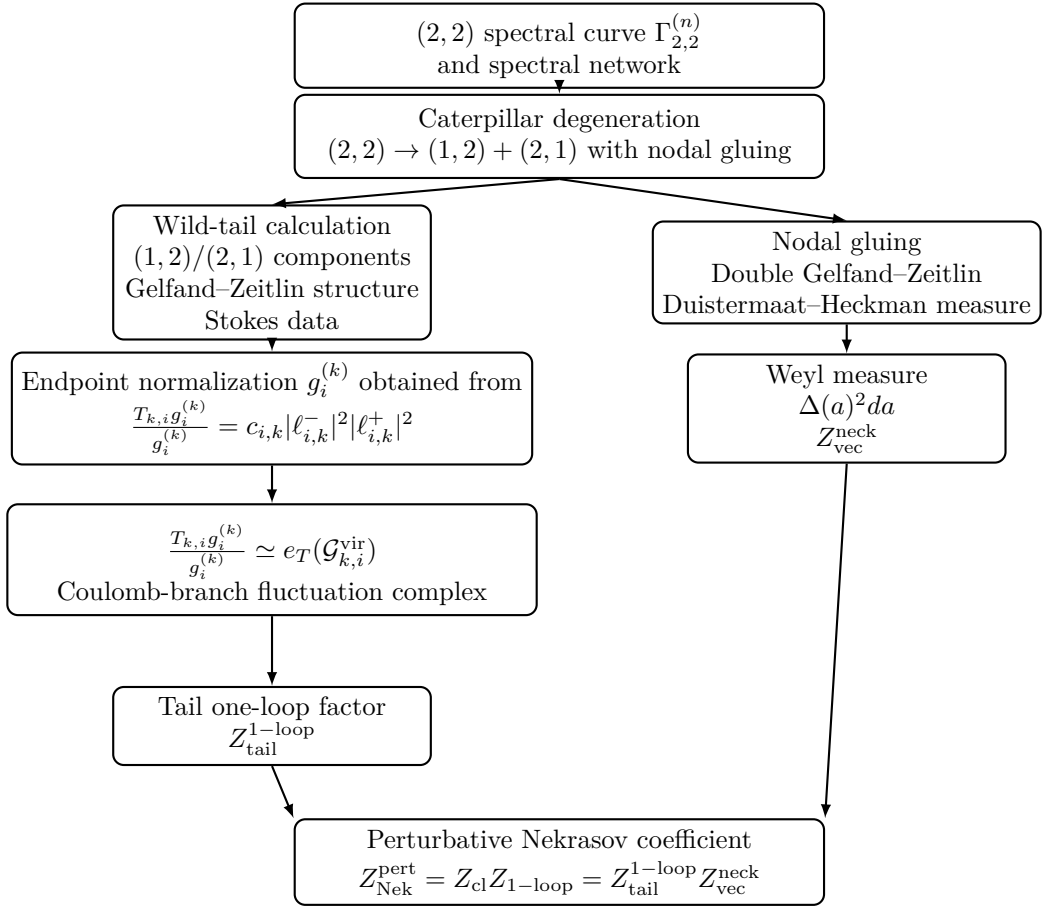

The main novelty of the present work is this geometric connection. It is
not a new closed formula for the Stokes matrices by itself: the relevant
analytic formulas are already available in
\cite{xu2024regularizedlimitsstokesmatrices,wang2025asymptoticmonodromyproblemshigherorder}.  Instead, we explain how their
caterpillar structure is encoded by a nodal spectral curve and by the
gluing of its spectral line bundle.  Relative to the one-sided analysis of
\cite{alekseev2024wkbasymptoticsstokesmatrices}, the simultaneous degeneration couples two Gelfand--Zeitlin tails through an internal neck and gives a geometric realization of the double Gelfand--Zeitlin system. Relative to \cite{wang2025asymptoticmonodromyproblemshigherorder}, our construction
provides a spectral-curve and spectral-network interpretation of the
Stokes data and identifies the neck contribution that later becomes the vector gluing factor in the Nekrasov partition function and, conjecturally, in the corresponding tau-function expansion..

Explicit formulas for Stokes matrices are useful beyond the computation of
monodromy data itself. In the one-sided/$(1,2)$ setting, such formulas and their
regularized caterpillar limits have already found applications in several
different directions. For instance, the Stokes description gives an explicit
realization of the Alekseev--Meinrenken diffeomorphism, and hence of a
Ginzburg--Weinstein linearization \cite{Xu2024TheAD}; the WKB asymptotics of the
regularized Stokes matrices provide a realization of crystal
structures and their wall-crossing behavior \cite{xu2024regularizedlimitsstokesmatrices}; and the
same WKB/Stokes data are closely related to cluster coordinates and their
tropical limits \cite{alekseev2024wkbasymptoticsstokesmatrices,Iwaki:2014vad}. Explicit
control of monodromy and Stokes data is also an important ingredient in the
study of special, and in particular algebraic, solutions of isomonodromy
equations. From this viewpoint, the geometric description developed here is
intended not only to reproduce the caterpillar Stokes formulas, but also to
explain the spectral geometry underlying these applications and to provide a
framework for their extension to the two-sided $(2,2)$ setting.

\subsection{Summary of results}

Our results are organized as follows. See figure $1$ for an overview, we explain how spectral curves give a geometric realization to the Double Gelfand-Zeitlin system and how the DGZ system explains the perturbative physical partition function. In Section~2 we analyze the double caterpillar degeneration of $\Gamma^{(n)}_{2,2}$.  We prove in theorem \ref{thm:22to21+12} that the curve degenerates into a
$(1,2)$ component and a $(2,1)$ component,
\begin{equation}
  \Gamma^{(n)}_{1,2}:\quad
  \det\!\left(xI_n-iu+\frac{B}{z}\right)=0,
  \qquad
  \Gamma^{(n)}_{2,1}:\quad
  \det\!\left(wI_n+\frac{B}{v}+\frac{A_0}{v^2}\right)=0,
  \label{eq:intro-two-components}
\end{equation}
with $A_0=A/t$ in the inner scaling.  The two components meet at $n$
nodes, and the Seiberg--Witten differential agrees across the corresponding
necks. Iterating the one-sided caterpillar decomposition on both
components yields a complete degeneration into genus-zero pieces. This
description determines a distinguished symplectic basis of cycles and
shows that the limiting action variables are the eigenvalues of the
leading principal minors of $-B$ and $GBG^{-1}$, where $G$ is the frame
diagonalizing $A$.

In Section~3 we explain the double caterpillar limit spectral curve as a geometric realization of the Double Gelfand-Zeitlin (DGZ) system \cite{crooks2023doublegelfandcetlinsysteminvariance}:  
\begin{Thm}[Theorem \ref{thm: dgz=spec}]
    The spectral curve degenerating system is a double Gelfand-Zeitlin system:
    \begin{equation}
    \begin{aligned}
        \phi_{\text{spec}}: \mathcal{M}_{\text{s-reg}}^{\text{fr}} &\to \mathfrak{G}^{\perp} \\
        (g,B) &\mapsto \bigl(\mu(GBG^{-1}),\mu(-B) \bigr).
    \end{aligned}
    \end{equation}
    Here we define $\mu:=(\mu^{(n)}, \dots, \mu^{(1)})$ and $\mu^{(k)}:=(\mu^{(k)}_1, \dots,\mu^{(k)}_k)$.  
\end{Thm}
The conjugate
angle variables have a direct spectral interpretation: they are the phases
of the gluing isomorphisms of the spectral line bundle at the normalized
nodes. More precisely, 
\begin{Prop}[Proposition \ref{Prop: angleCoorGluing}]
On the strongly regular locus, the phases of the framed spectral-line
gluing parameters are angle coordinates for the effective double
Gelfand--Zeitlin torus action.
\end{Prop}

This identifies both halves of the action--angle data, rather than
only the limiting eigenvalue pattern. Finally we give some physical insights. We identify the Duistermaat-Heckman measure with the Weyl measure arising from gluing two quivers, and interpret the Fourier lattice summation of isomonodromy tau as a Peter-Weyl expansion in the real polarization of the DGZ system. 

Section~4 introduces a nodal caterpillar spectral network. The object is a collection of local spectral networks on the rescaled components, together with sheet-channel identifications at the necks; it is not an ordinary spectral network on a smooth base. We prove that the regularized non-Abelian parallel transport decomposes into
componentwise transports connected by the gluing maps of the
spectral eigenlines at the nodes. After matching the canonical
endpoint frames, this composition is compatible with the regularized
Stokes data of the original differential equation, and should be viewed as a limit of the original smooth network. Assuming the exact WKB conjecture \cite{alekseev2024wkbasymptoticsstokesmatrices,Hollands_2020,gaiotto2011wallcrossinghitchinsystemswkb}, in our nodal framework, the path-lifting rule is componentwise and the neck contributes only a framed gluing operator on the shared eigenlines. As a result, the S-walls are restricted to the local components. See definition \ref{def: CatPathLifting} for a more comprehensive version. At the neck, the regularized non-Abelian transport on $E$ converges to the push-forward of the spectral-line gluing datum. Its elementary Schlesinger shift satisfies a finite-difference equation whose solution is expressed
in Gamma functions: 
\begin{Prop}[Proposition \ref{prop: variantEquation}] Let $ T_{k,i}: \frac{\xi_i^{(k)}}{\epsilon}\longmapsto \frac{\xi_i^{(k)}}{\epsilon}+1 $ be the elementary Schlesinger shift ($\epsilon \in \mathbb{R}^{\times}$) of the formal exponent associated with the intermediate Gelfand--Zeitlin eigenline $v_i^{(k)}$. Then 
\begin{equation} 
\frac{T_{k,i}g_i^{(k)}}{g_i^{(k)}}= \pm\frac{ \prod_{l=1}^{k-1}(\frac{\xi_i^{(k)}-\xi_l^{(k-1)}}{\epsilon}) \prod_{l=1}^{k+1}(\frac{\xi_i^{(k)}-\xi_l^{(k+1)}}{\epsilon})}{ \left[\prod_{j\neq i}(\frac{\xi_i^{(k)}-\xi_j^{(k)}}{\epsilon}) \right]^2}, \label{eq: 1.6}  
\end{equation} 
for anti-Hermitian, Hermitian $B/\epsilon$ respectively.  \end{Prop}
The resulting adjacent Stokes entries agree, up to
the stated normalization conventions, with the formulas of
\cite{xu2024regularizedlimitsstokesmatrices,alekseev2024wkbasymptoticsstokesmatrices,wang2025asymptoticmonodromyproblemshigherorder}. Indeed, we obtain the following theorem from the geometric/GZ aspect.
\begin{Thm} [Proposition \ref{prop:entryFormula}]
    The $(k,k+1)$ Stokes matrix entry of the ODE system (\ref{eq: symODE}) in the sector $S_+$ can be written as (up to an $\epsilon$-periodic factor)
    \begin{equation}
        (S_+)_{k,k+1}= N_k^{\rm reg} \sum_{i}^k  \langle e_k, P^{(k)}_i \frac{u_k}{\epsilon} \rangle   \frac{\prod_{l=1}^{k-1}\Gamma(i\mu_i^{(k)}/\epsilon-i\mu^{(k-1)}_l/\epsilon)\prod_{l=1}^{k+1}\Gamma(i\mu_i^{(k)}/\epsilon-i\mu^{(k+1)}_l/\epsilon)}{\left[\prod_{j\neq i}\Gamma(i\mu^{(k)}_i/\epsilon-i\mu^{(k)}_j/\epsilon) \right] ^2}. 
    \end{equation}
    Here $i\mu^{(k)}_i$ are the eigenvalues of principal minors of $iB$ appearing in \ref{eq: symODE}. 
\end{Thm}
 Assuming the exact-WKB conjecture, we also obtain the strict-caterpillar identification between the relevant
factorization coordinates and elementary spectral coordinates anticipated
in \cite{alekseev2024wkbasymptoticsstokesmatrices}.

The opening of Section~5 gives a perturbative gauge-theory
interpretation of the local gluing cocycles.  For the one-sided quiver tail
$U(n)-U(n-1)-\cdots-U(1)$, an elementary Schlesinger shift of a
Gelfand--Zeitlin eigenvalue acts as an elementary, or Hecke, modification
of the corresponding spectral eigenline.  The affected adjacent and
adjoint channels assemble into a virtual fluctuation complex
$\mathcal G^{\mathrm{vir}}_{k,i}$, whose equivariant Euler class reproduces,
up to the stated shift and normalization conventions, the finite-difference
cocycle (\ref{eq: 1.6}) obtained from the gluing factor in Section~4.  On the self-dual
$\Omega$-background, discrete integration of these local Euler classes
reconstructs the global one-loop Nekrasov factor
\cite{nekrasov2003seibergwittentheoryrandompartitions,Kimura_2016}. Thus, after identifying the local gluing cocycles with the equivariant Euler classes of the fluctuation complexes, the perturbative partition function is reconstructed from the spectral-network and factorization-coordinate data. Finally, as a conjecture, we apply the Gelfand--Zeitlin and spectral-network data to the perturbative coefficients of isomonodromic tau functions. For $n=2$, the resulting expression agrees sector by sector with the standard $P_{\mathrm{III}}(D_6)$ structure coefficient of \cite{Gamayun_2013,Iorgov_2014}; see Appendix \ref{app:PIII_Kiev_check}. For general rank, the classical factor and the perturbative Nekrasov factor are determined by the caterpillar and double Gelfand--Zeitlin data. We conjecture that, after the prescribed normalization, the latter gives the perturbative structure coefficient in the corresponding higher-rank isomonodromic tau-function expansion.

The appendices collect the exact-WKB input, the explicit rank-two spectral
network calculation, the Nekrasov factors, a direct computation of the
rank-two $(1,2)$ Jimbo--Miwa--Ueno tau function, the Abelian normalization,
and the comparison with the $P_{\mathrm{III}}(D_6)$ coefficient.

\section{Spectral curve}

In this section we consider the spectral curve
\begin{equation}
    P(x,z)=\text{det} \left[ x I_n - \left( iu -\frac{B}{z}- \frac{A}{z^2}  \right)   \right]=0, \label{eq: 2.1}
\end{equation}
We slightly extend our discussion to complex $u,A,B$. $u,A$ are required to be regular  semisimple and $u=\text{diag}(u_1, \dots ,u_n)$. Since $A$ is regular semisimple, we can choose $G\in GL_n(\mathbb{C})$ to be an ordered diagonalizing frame of $A$.  We further require that all the leading principal minors of $B$ and $GBG^{-1}$ are regular semisimple. The results in this section naturally apply to the real, Hermitian/anti-Hermitian slice. 

\subsection{General analysis and the caterpillar limit}

Since the spectral curve \ref{eq: 2.1} has order $2$ irregular singularities at $z=0$ and $z=\infty$, we call this curve a $(2,2)$ type wild spectral curve and denote it by $\Gamma^{(n)}_{2,2}$, here $n$ is the rank of the matrices. We also define $\Phi:= iu-\frac{B}{z}-\frac{A}{z^2}$ and $\Phi dz$ is then a Higgs field. 

\begin{Def}
    For a rank $n$ spectral curve with an order $p$ singularity at $z=0$ and an order $q$ singularity at $z=\infty$, we call it a $(p,q)$ type spectral curve and denote it by $\Gamma^{(n)}_{p,q}$.    
\end{Def}

Near $z= \infty$, we have $\text{det}\left( x-iu  \right) =0$. Therefore, there are $n$ distinct punctures at $(x,z)=(iu_i,\infty)$. At $z=0$, the spectral curve behaves as 
\begin{equation}
    \text{det} \left( xI_n +\frac{A}{z^2}+ \frac{B}{z}\right)=0, 
\end{equation}
and punctures behave as
\begin{equation}
    x_i = -\frac{\lambda_i}{z^2}-\frac{\beta_{ii}}{z} \dots.
\end{equation}
Here $\lambda_i$ is the $i$th eigenvalue of $A$ and $\beta_{ii}= w_i^{\dagger} Bv_i$, where $v_i, w_i$ is the $i$th right, left eigenvector of $A$ respectively. They are normalized by the condition $w_iv_j=\delta_{ij}$.  

We next calculate the genus. Recall that the spectral curve $\Gamma_{2,2,n}(u,A,B)$ is an $n$-to-1 cover of $\mathbb{P} - \{ 0,\infty \}$. By the Riemann-Hurwitz formula, we have
\begin{equation}
    2-2g(\bar{\Gamma}_{2,2,n})= n(2-2g_{\text{base}}) -R,
\end{equation}
here $R$ is the total ramification degree. Since we require the eigenvalues of $A$ and diagonals of $u$ to be distinct, there is no ramification at punctures. Then since the discriminant can be written as $\Delta =\prod_{i < j}(x_i-x_j)^2$. $x_i$ has order $2$ poles at $z=0$, hence $\Delta$ has an order $2n(n-1)$ pole at $z=0$ and no pole at $z=\infty$. Then the total number of zeros of $\Delta(z)$ on $\mathbb{P}^1\backslash\{ 0,\infty \}$ is $2n(n-1)$. For generic $B$, each zero corresponds to a simple branch point, and therefore $R= 2n(n-1)$. Hence
\begin{equation}
    2-2g(\bar{\Gamma}_{2,2,n})= 2n- 2n(n-1) \implies g(\bar{\Gamma}_{2,n})=(n-1)^2.
\end{equation}
For example, if $n=2$, then the spectral curve is a torus (with $2n$ punctures). 

Consider the Seiberg-Witten differential $\lambda =xdz$. Its residues are 
\begin{equation}
    \text{res}(\lambda)_{z=0}= -\beta_{ii}
\end{equation}
at $z=0$ and residues at $\infty$ can be calculated similarly. Since at $z=\infty$, the leading term is $u$ and $u$ is diagonal, the eigenvectors of $u$ are the standard basis vectors $e_i$, and the residues are 
\begin{equation}
    \text{res}(\lambda)_{z=\infty}= B_{ii}. 
\end{equation}

Now we introduce the following double caterpillar limit 
\begin{Def}[Double caterpillar limit]
    We define 
    \begin{equation}
    \begin{aligned}
        &z_0= u_1, \, z_1=u_2-u_1, \,z_2=\frac{u_3-u_1}{u_2-u_1}, \dots, z_{n-1}=\frac{u_n-u_1}{u_{n-1}-u_1}, \\
        & t=(u_n-u_1)(\lambda_n-\lambda_1), \\
        & w_{n-1}= \frac{\lambda_n-\lambda_1}{\lambda_{n-1}-\lambda_1}, \dots, w_{2}=\frac{\lambda_3-\lambda_1}{\lambda_2-\lambda_1}, w_0=\lambda_1, w_1=\lambda_2-\lambda_1. 
    \end{aligned} 
    \end{equation}  

    The double caterpillar limit is the ordered iterated limit
\begin{equation}
    t\longrightarrow0,
    \qquad
    z_{n-1}\longrightarrow\infty,\ldots,
    z_2\longrightarrow\infty,
    \qquad
    w_{n-1}\longrightarrow\infty,\ldots,
    w_2\longrightarrow\infty,
    \label{eq: double-caterpillar-limit}
\end{equation}
where the limits are taken successively in the displayed order.
All remaining parameters and framing data are assumed to converge.
    \label{Def:dc}
\end{Def}
The double caterpillar limit is a generalization of the caterpillar limit introduced in \cite{xu2024regularizedlimitsstokesmatrices}, and is named double because we apply the caterpillar limit on both $u$ and $A$. For a physical explanation, see section $5.1$. 

By the translation invariance, we fix $u_1=0, \lambda_1=0$. At the level of spectral curves, if we do the translation $u \mapsto u-u_1I_n, \,\,A \mapsto A-\lambda_1 I_n$, then 
\begin{equation}
    iu-\frac{B}{z}-\frac{A}{z^2} \mapsto iu-\frac{B}{z}-\frac{A}{z^2}-iu_1I_n+\frac{\lambda_1 }{z^2}I_n,
\end{equation}
which is equivalent to $x \mapsto x -iu_1 +\frac{\lambda_1}{z^2}$. Since the Seiberg-Witten form is $\lambda=xdz$, which differs from $\lambda^{\prime}=(x -iu_1 +\frac{\lambda_1}{z^2})dz$ by a (global) total differential, so the periods are unchanged. 

\subsection{Splitting}

We first consider the limit $t\to 0$ while still imposing the condition that $\lVert \lambda_1 \rvert<\lVert \lambda_2 \rvert< \dots <\lVert \lambda_n \rvert$ along the degeneration path. This limit also splits the spectral curve. If we fix $u$ and let $t\to 0$, then $\lambda_n=O(t)$. 

We consider a family of spectral curves parameterized by $\mathbb{C}_t$, and let $\mathcal{X}$ be this 1-parameter family of
spectral curves in $\mathbb{P}^1_{x} \times \mathbb{P}^1_{z} \times \mathbb{C}_t$. Then let $V=\mathbb{P}_x \times \mathbb{P}_z\times \mathbb{P}_{1/t}$. Now let $\tilde{V}$ be the blowup of $V$ at $z=0, x=\infty, t=0$, or $z=0, \frac{1}{x}=0, t=0$. Clearly $\tilde{V}$ is a closed subscheme of $\mathbb{P}^2\times V$. We also define $\tilde{\mathcal{X}}$ to be strict transform of $\bar{\mathcal{X}}$ in $\tilde{V}$. Let the coordinate of $\mathbb{P}^2$ be $(w^{\prime}, v, y_3)$, satisfying 
\begin{equation}
    \frac{y_3}{x}= tw^{\prime}, \,\,\,  tv= zy_3, 
\end{equation}
Letting $y_3=1$ and $w^{\prime}=\frac{1}{w}$, then $v=\frac{z}{t}, tx=w$. If we write $v^{\prime}=\frac{1}{v}$, then $zv^{\prime}=t$; this is exactly the plumbing equation where the nodes arise. Note that $\lambda=xdz=\frac{w}{t} d (vt)=wdv$. Substituting these expressions into $(\ref{eq: 2.1})$ then gives
\begin{equation}
    \text{det} \left[ \frac{1}{t}wI-(iu-\frac{B}{vt}-\frac{A}{v^2t^2}) \right]=0, 
\end{equation}
multiplying both sides by $t$,
\begin{equation}
    \text{det} \left[ wI-(itu-\frac{B}{v}-\frac{A}{v^2t}) \right]=0.
\end{equation}
As $u$ is fixed and $t\to 0$ , we are left with 
\begin{equation}
    \text{det} \left[ wI+\frac{B}{v}+\frac{A/t}{v^2} \right]=0.
\end{equation}
Also as $\lVert A \rVert=O(|t|)$, $A_0:=\frac{A}{t}$ is a well defined matrix. We emphasize that $t$ is only a bookkeeping parameter for the scaling of $A$, rather than an independent argument of $A$. Equivalently, one may simply set $A/t=A_0$. 

Away from the blowup region, namely where $\frac{1}{z}, x\in \mathbb{C}$, the last term $\frac{A}{z^2}$ is negligible comparing to $B$, therefore $(1)$ becomes 
\begin{equation}
    \text{det} \left[ xI-(iu-\frac{B}{z})  \right]=0. 
\end{equation}

The residues of $(1,2)$ type equation are $B_{ii}$ at punctures lying over $z=\infty$ and $-\mu_i$ at punctures lying over $z=0$, where $\mu_i$ is the $i$th eigenvalue of $B$. On the other hand, after a change of coordinates, the $(2,1)$ type equation becomes
\begin{equation}
\begin{aligned}
    \text{det} \left[ wI+Bv'+A_0(v')^2 \right]=0. 
\end{aligned}
\end{equation}
Hence at $v' = 0$ or $v= \infty$, $w_i =  -\mu_iv'+O((v')^2) \implies \lambda_i=w_idv=-\mu_i v' d\frac{1}{v'}= \frac{\mu_i}{v'}dv'$. Hence $\text{Res}_{v=\infty,i} \lambda=  \mu_i$. 

At $v=0$, multiplying both sides by $v^2$ gives 
\begin{equation}
    \text{det} \left[ wv^2+Bv+A_0\right]=0.
\end{equation}
Write $w=\frac{p}{v^2}+\frac{q}{v}+O(1)$, then as $v\to 0$ we have $\text{det} [ pI+A_0]=0$, which implies that $p_i=-\nu_i$, where $\nu_i=\frac{\lambda_i}{t}$ is the $i$th eigenvalue of $A_0$. Also $q_i= -f_i(A_0+Bv)_{v=0}=-\frac{d}{dv}(A_0+Bv)_{v=0}= -\beta_{ii}$. Here $f_i$ is the eigenvalue function. Hence
\begin{equation}
    w=-\frac{\nu_i}{v^2}-\frac{\beta_i}{v}+O(1) \implies \text{Res}_{v=0,i}=-\beta_{ii}. 
\end{equation}

Now when $z=v^{\prime}=0, t=0$, we are left with the determinant equation $\det(\eta I+B)=0$, which has $n$ distinct roots $\eta=-\mu_i$(recall $B$ has simple spectrum). The node coordinates are then $(\eta,z,v^{\prime})=(-\mu_i,0,0)$. By the implicit function theorem, $\eta=\eta_i(z,v^{\prime})$ is the unique analytic function near the nodes. Eliminating $\eta$, we are left with the local equation $zv^{\prime}=t$ precisely. This also gives us the flatness: outside the blow-up center, the blow-up map is an isomorphism, so $\tilde{\mathcal{X}}$ is identified with the original flat family, while near the blow-up center flatness follows from the local equation $zv^{\prime}=t$. Then the match of singularities implies that there are $n$ shared singularities: the $n$ punctures lying over $v=\infty$ of the $(2,1)$ curve should be glued with the $n$ punctures lying over $z=0$ of the $(1,2)$ curve respectively. The geometric picture is that there are $n$ tubes of length $O(- \text{log} \,|t| )$ that finally pinch off as $t\to 0$. This leaves us with $n$ additional punctures at both sides (actually this is the normalization of nodes). We can define vanishing cycles as well: namely they are the radial cycles surrounding the tube. 

Since we have $n$ distinct shared punctures, there are $n-1$ independent vanishing cycles. This corresponds to the difference of genus: $(n-1)^2 -2 \frac{(n-1)(n-2)}{2}=n-1$. This proves the following theorem:

\begin{tikzpicture}[x=0.9cm,y=0.9cm,>=Latex]

  \def\N{5} % number of punctures / tubes
  
  \def\LX{-2.55}
  \def\RX{ 2.55}
  \def\PX{-6.82}
  \def\QX{ 6.82}
  \tikzset{
    surf/.style={draw=black!70, fill=blue!7, line width=0.9pt},
    handle/.style={draw=black!70, line width=0.9pt},
    tubeouter/.style={draw=black!45, line width=6.2pt, line cap=round, line join=round},
    tubeinner/.style={draw=blue!10, line width=3.0pt, line cap=round, line join=round},
    puncture/.style={draw=black, fill=black, line width=0.3pt}
  }

  \path[surf]
    (-7.0,-3.0)
    .. controls (-7.8,-1.7) and (-7.8,1.7) .. (-7.0,3.0)
    .. controls (-6.2,4.3) and (-4.7,4.8) .. (-3.4,4.2)
    .. controls (-2.8,3.8) and (-2.4,3.0) .. (-2.3,2.1)
    -- (-2.3,-2.1)
    .. controls (-2.4,-3.0) and (-2.8,-3.8) .. (-3.4,-4.2)
    .. controls (-4.7,-4.8) and (-6.2,-4.3) .. (-7.0,-3.0)
    -- cycle;

  \path[surf]
    (7.0,-3.0)
    .. controls (7.8,-1.7) and (7.8,1.7) .. (7.0,3.0)
    .. controls (6.2,4.3) and (4.7,4.8) .. (3.4,4.2)
    .. controls (2.8,3.8) and (2.4,3.0) .. (2.3,2.1)
    -- (2.3,-2.1)
    .. controls (2.4,-3.0) and (2.8,-3.8) .. (3.4,-4.2)
    .. controls (4.7,-4.8) and (6.2,-4.3) .. (7.0,-3.0)
    -- cycle;

  \foreach \x/\y in {-5.9/2.25,-4.75/3.15,-3.6/2.25} {
    \draw[handle]
      (\x-0.42,\y)
      .. controls (\x-0.42,\y-0.55) and (\x+0.42,\y-0.55) .. (\x+0.42,\y)
      .. controls (\x+0.42,\y+0.28) and (\x-0.42,\y+0.28) .. (\x-0.42,\y)
      -- cycle;
  }

  \foreach \x/\y in {5.9/2.25,4.75/3.15,3.6/2.25} {
    \draw[handle]
      (\x-0.42,\y)
      .. controls (\x-0.42,\y-0.55) and (\x+0.42,\y-0.55) .. (\x+0.42,\y)
      .. controls (\x+0.42,\y+0.28) and (\x-0.42,\y+0.28) .. (\x-0.42,\y)
      -- cycle;
  }

  \node[align=center] at (-4.85,-0.2)
    {$\displaystyle \frac{(n-1)(n-2)}{2}$\\ \\[-1mm]genus};
  \node[align=center] at (4.85,-0.2)
    {$\displaystyle \frac{(n-1)(n-2)}{2}$\\ \\[-1mm]genus};

  \foreach \k in {1,...,\N} {
    \pgfmathsetmacro{\yy}{-2.15 + 4.30*(\k-1)/(\N-1)}
    \pgfmathsetmacro{\amp}{ifthenelse(mod(\k,2)==0,0.18,-0.18)}

    \draw[tubeouter]
      (\LX,\yy) .. controls (-1.0,\yy+\amp) and (1.0,\yy+\amp) .. (\RX,\yy);

    \draw[tubeinner]
      (\LX,\yy) .. controls (-1.0,\yy+\amp) and (1.0,\yy+\amp) .. (\RX,\yy);
  }

  \foreach \k in {1,...,\N} {
    \pgfmathsetmacro{\yy}{-2.15 + 4.30*(\k-1)/(\N-1)}
    \filldraw[puncture] (\PX,\yy) circle (1.9pt);
    \filldraw[puncture] (\QX,\yy) circle (1.9pt);
  }

  \node at (0,4.05) {$n$ tubes};

\end{tikzpicture}

\begin{Thm}[Degeneration of $(2,2)$ spectral curves via $t\to0$]
\label{thm:22to21+12}
Consider the family of spectral curves $\Gamma_{2,n}(u,A,B)$ defined by
\[
\det\!\Big( xI_n - iu + \frac{B}{z} + \frac{A}{z^2} \Big) = 0,
\]
with $u=\operatorname{diag}(u_1,\dots,u_n)$ fixed and regular semisimple, and $A,B$ regular semisimple. Let $t = u_n\lambda_n$, where $\lambda_n$ is the eigenvalue of $A$ with the largest norm, and take the limit $t\to 0$ while keeping $u$ fixed (hence $\|A\|\to 0$). Furthermore, $A_0=\frac{A}{t}$ is well defined. Then the family approaches a degenerate curve consisting of two generically irreducible components:
\begin{itemize}
\item a $(1,2)$-type curve $\Gamma_{(1,2)}$ given by
  \[
  \det\!\Big( xI_n - iu + \frac{B}{z} \Big) = 0,
  \]
  with $n$ first‑order punctures at $z=0$ (residues $-\mu_i$, where $\mu_i$ are the eigenvalues of $B$) and $n$ second‑order irregular punctures at $z=\infty$ (residues $B_{ii}$);
\item a $(2,1)$-type curve $\Gamma_{(2,1)}$ given, in suitable rescaled coordinates $v=z/t$, $w=tx$, by
  \[
  \det\!\Big( wI_n + \frac{B}{v} + \frac{A_0}{v^2} \Big) = 0,\qquad A_0 = \frac{A}{t},
  \]
  with $n$ first‑order punctures at $v=\infty$ (residues $\mu_i$) and $n$ second‑order irregular punctures at $v=0$ (residues $-\beta_{ii}$, where $\beta_{ii}=w_i B v_i$ with $v_i, w_i$ the right/left eigenvectors of $A_0$ respectively).
\end{itemize}
The two components intersect transversally in $n$ nodes, obtained by pairwise identifying the $n$ punctures of $\Gamma_{(1,2)}$ above $z=0$ with the $n$ punctures of $\Gamma_{(2,1)}$ above $v=\infty$. 
These nodes are the limits of tubes of length $O(-\log |t|)$ that pinch off in the limit. 
Consequently, the degenerate curve acquires $n$ additional marked points (punctures) on each component along the nodal locus, and there are exactly $n-1$ independent vanishing cycles, which matches the drop in genus:
\begin{equation}
    g(\Gamma_{2,n}) = (n-1)^2,\qquad 
g(\Gamma_{(1,2)}) = g(\Gamma_{(2,1)}) = \frac{(n-1)(n-2)}{2},
\end{equation}
so that $g(\Gamma_{2,n}) - (g(\Gamma_{(1,2)})+g(\Gamma_{(2,1)})) = n-1$. \label{Thm: mth1}
\end{Thm}

\begin{Rmk}
    We can compare those residues with those of the $(2,2)$ spectral curve. Recall that the for the $(2,2)$ curve $\text{Res}_{z=0,i}=-\beta_{ii}$ and $\text{Res}_{z=\infty,i}=B_{ii}$, this matches perfectly with the residues of the $(2,1)$ spectral curve lying over $v=0$ and those of the $(1,2)$ spectral curve lying over $z=\infty$ respectively.    \label{rmk:2.5}
\end{Rmk}

Also, the cycle count can be made precise using the
Mayer--Vietoris sequence. Let $\Gamma_t$ be a nearby smooth fiber and
choose an open covering $\Gamma_t=U_{1,2}\cup U_{2,1}$, where $U_{1,2}$ and $U_{2,1}$ deformation retract onto the two punctured limiting components, and $ U_{1,2}\cap U_{2,1}= \bigsqcup_{a=1}^{n}\mathcal A_a$ is the disjoint union of the $n$ neck annuli. We take homology with coefficients in $\mathbb C$. Since the surfaces are punctured,
their second homology groups vanish, and the Mayer--Vietoris sequence
gives
\begin{equation}
\begin{aligned}
0\longrightarrow& \frac{H_1(U_{1,2})\oplus H_1(U_{2,1})}{
\operatorname{im}H_1(U_{1,2}\cap U_{2,1})} \longrightarrow H_1(\Gamma_t)\\
&\longrightarrow \ker\left[ H_0(U_{1,2}\cap U_{2,1}) \longrightarrow
H_0(U_{1,2})\oplus H_0(U_{2,1}) \right] \longrightarrow0.
\end{aligned}
\end{equation}
For a connected Riemann surface of genus $g$ with $r$ punctures, $\dim H_1=2g+r-1$.
Therefore
\begin{equation}
\begin{aligned}
    \dim H_1(\Gamma_{2,2}^{(n)}) &= 2(n-1)^2+2n-1 = 2n^2-2n+1,\\
    \dim H_1(\Gamma_{1,2}^{(n)}) &= \dim H_1(\Gamma_{2,1}^{(n)})
    =(n-1)(n-2)+2n-1 = n^2-n+1.
\end{aligned}
\end{equation}
The map from the first homology of the $n$ annuli has rank $n$ and
imposes the $n$ boundary-matching relations. Hence the part inherited
from the two limiting components has dimension $ 2(n^2-n+1)-n =2n^2-3n+2$. 
Since the intersection has $n$ connected components while
$U_{1,2}$ and $U_{2,1}$ are connected,
\begin{equation}
\dim\ker\left[ H_0(U_{1,2}\cap U_{2,1}) \longrightarrow H_0(U_{1,2})\oplus H_0(U_{2,1})
\right] = n-1.
\end{equation}
These $n-1$ additional classes are represented by the longitudinal
cycles which pass through one neck and return through a fixed reference
neck. They are precisely the broken $B$-cycles.

\subsection{Further decomposition}

Proposition 3.7 of \cite{alekseev2024wkbasymptoticsstokesmatrices} states that $\Gamma_{1,2}$ can be further decomposed by taking the caterpillar limit $ z_{n-1},\dots,z_2 \to \infty$. We show this holds for $\Gamma_{2,1}$ as well by adapting the proof of proposition given in \cite{alekseev2024wkbasymptoticsstokesmatrices}. 

First under a coordinate change $v'=\frac{1}{v}$ and $\omega=wv^2$ (this change of coordinate is legitimate because it is biholomorphic away from the indicated divisor), $\Gamma_{2,1}$ can be written as
\begin{equation}
    \text{det} \left[ A_0 +\frac{B}{v'}+\omega I_n \right]=0,
\end{equation}
and after diagonalizing $A_0$, we obtain
\begin{equation}
    \text{det} \left[ \Lambda + \frac{GBG^{-1}}{v'}+\omega I_n \right]=0.
\end{equation}

For convenience, in this section we use $\lambda_i$ to denote the eigenvalues of $A_0$. Now we fix $w_2, w_3,\dots, w_{n-2}$ while letting $\lambda_n$ go to $\infty$. Let $V'=\mathbb{P}_\omega \times \mathbb{P}_{v'} \times \mathbb{P}_{\lambda_n}$ and $\tilde{V}'$ be the blow-up of $V'$ at $v'=0, \frac{1}{\omega}=0, \frac{1}{\lambda_n}=0$. Then $\tilde{V^{\prime}}$ is a subscheme of $V' \times \mathbb{P}^2$ and we may choose a coordinate $[p_1 :p_2:1]$ satisfying
\begin{equation}
    \frac{1}{\omega}=\frac{p_1}{\lambda_n},\,\,\, v'=\frac{p_2}{\lambda_n}.
\end{equation}
Therefore let $a=\frac{\omega}{\lambda_n}$ and $b=\lambda_n v'$ 
\begin{equation}
     \text{det} \left[ \Lambda + \frac{\lambda_nGBG^{-1}}{b}+a\lambda_n I_n \right]=0.
\end{equation}
Dividing both sides by $\lambda_n$ and taking the limit $\lambda_n\to \infty$ gives
\begin{equation}
\begin{aligned}
    \text{det} \left[ E_n + \frac{GBG^{-1}}{b}+aI_n \right]=0.
\end{aligned}
\end{equation}
We denote this curve by $\Gamma_{2,1}^{(n)}(E_n)$. On the other hand, away from $\omega=wv^2=\infty$ and $\frac{1}{v}=v'=0$, the Laplace expansion of the determinant respect to the last row yields
\begin{equation}
    (\lambda_n+\frac{(GBG^{-1})_{nn}}{v'}+\omega) \text{det}\left[ \Lambda^{(n-1)}+\frac{(GBG^{-1})^{(n-1)}}{v'}+\omega I_{n-1} \right] + f(\omega,\frac{1}{v'})=0.
\end{equation}
Here the first term is lower-right corner times the $(n-1)$-th leading principal minor, and $f$ denotes the rest part of the Laplace expansion. Note that $f$ is a polynomial in $\frac{1}{v^{\prime}}$ and $\omega$ and the coefficients of them are finite and contain no $\lambda_n$. Dividing $\lambda_n$ on both sides yields
\begin{equation}
\begin{aligned}
    &\text{det}\left[ \Lambda^{(n-1)}+\frac{(GBG^{-1})^{(n-1)}}{v'}+\omega I_{n-1} \right]=0 \\
    \implies &\text{det}\left[ \frac{\Lambda^{(n-1)}}{v^2}+\frac{(GBG^{-1})^{(n-1)}}{v}+w I_{n-1} \right]=0,
\end{aligned} 
\end{equation}
which is exactly $\Gamma^{(n-1)}_{2,1}$. Then the residues of $\Gamma_{2,1}^{(n-1)}$ at $v'=\infty$(or $v=0$) are $-\alpha_{ii}^{(n-1)}:= w_i^{(n-1)}(GBG^{-1})^{(n-1)}v_i^{(n-1)}$ with $v_i^{(n-1)}, w_i^{(n-1)}$ the right/left eigenvectors of $A_0^{(n-1)}$ respectively, and the residues at $v'=0$(or $v=\infty$) are $\mu^{(n-1)}_i(GBG^{-1})$, $i=1,2, \dots, n-1$. The residues of $\Gamma^{(n)}_{2,1}(E_n)$ at $b=0$ are $\mu^{(n)}_i$ for $i=1,2, \dots, n$ and these at $b=\infty$ are $-\mu^{(n-1)}_i(GBG^{-1})$ for $i=1,2, \dots n-1$, $-\alpha_{nn}$ for $i=n$. Therefore we see that there are $n-1$ shared punctures at $b=\infty$ and $v'=0$. Moreover, this decomposition can be repeated. 

In each step $\Gamma^{(k+1)}_{2,1}\to \Gamma^{(k)}_{2,1}+\Gamma^{(k+1)}_{2,1}(E_{k+1})$, there are $k$ tubes stretched to infinity finally leave $k$ punctures on both sides. The whole picture is 
\begin{equation}
    \Gamma_{2,2}=\left(\bigsqcup_{i=2, \rm nodes}^n \Gamma_{2,1}^{(i)}(E_i)  \right) \sqcup_{\rm nodes} \left(\bigsqcup_{j=2, \rm nodes}^n \Gamma_{1,2}^{(j)}(E_j)  \right).
\end{equation}
 At each side, we have $\sum_{i=1}^{n-1}i=\frac{n(n-1)}{2}$ shared punctures and $n$ punctures at the middle. And the total genus loss is $2\sum_{g=1}^{n-2}g+(n-1)=(n-1)^2$, which implies that all the handles are pinched off.

 We can construct similar $B$ cycles as in section $2.2$. Therefore we obtain a whole set of $A$-cycles and $B$-cycles of $\Gamma_{2,2}$. The periods of those cycles naturally give rise to an integrable system. Moreover, this is actually related to a double Gelfand-Zeitlin system, as we will show in the next section. Physically, these periods serve as electric charges and magnetic charges of the quiver $4d$ supersymmetric theory corresponding to $\Gamma_{2,2}$.

\section{The double Gelfand-Zeitlin system}
In this section, we go back to the real, Hermitian/anti-Hermitian slice, while keeping the semisimple condition for $u$, $A$ and all leading principal minors of $B$.

To identify the double Gelfand-Zeitlin system, we identify $(u,A,B)$ with $T^{*}U(n) \cong U(n) \times \mathfrak{u}(n)^{*}$. In fact the eigenvalues of $u$ and $A$
are \textbf{external} parameters and they can be fixed. What really involves is $B$ and the frame of $A$. Recall that for $\Gamma_{2,2}^{(n)}$, the residues are $\text{res}(\lambda)_{z=0}=-\beta_{ii}=-w_iBv_i$ and $\text{res}(\lambda)_{z=\infty}=B_{ii}$, and from remark \ref{rmk:2.5} these properties are inherited by $\Gamma_{1,2}^{(n)}$ and $\Gamma_{2,1}^{(n)}$. The point is that $A$ may have a different eigenbasis, and this reminds us that we should focus on the frame of $A$ instead of $A$ itself. Also, for convenience in this section we will use $\lambda_i$ to denote the eigenvalue of $A$ and use $\lambda_i(\cdot)$ as a function of extracting the $i$-th eigenvalue. 

\begin{Def}
For a fixed set of distinct, real eigenvalues, say $\Lambda_0=\{\lambda_1, \dots,\lambda_n  \}$, and let $D=\text{diag}\,\Lambda_0$. Define the framed space $\mathcal{M}_{\Lambda_0}^{\text{fr}}(A,F,B)$. Here $B\in \mathcal{H}(n)$, $A\in \mathcal{O}_{\Lambda_0}$(here $\mathcal{O}$ denotes the conjugate orbit of $D$), and $F=(v_1, v_2, \dots, v_n )$ is a tuple of ordered unitary eigenbasis satisfying $Av_i=\lambda_iv_i$. \label{Def: frmdspace}
\end{Def}

\begin{Prop}
    $\mathcal{M}_{\Lambda_0}^{\text{fr}}(A,F,B)$ has dimension $2n^2$. 
\end{Prop}
\begin{Prf}
    Define $g_F^{\dagger}=(v_1, \dots, v_n)\in U(n)$, then $g_F^{\dagger} \in U(n)$ and $A=g_F^{\dagger}Dg_F$. On the other hand, for each $g\in U(n)$, there is an $A$ satisfying $A=g^{\dagger}Dg$, and $F=\{g^{\dagger}e_1,g^{\dagger}e_2, \dots, g^{\dagger}e_n\}$. Therefore $(A,F)$ is in one-to-one correspondence with $U(n)$. Now $\mathcal{H}(n)$ and $U(n)$ both have dimension $n^2$, so $\text{dim}\,\mathcal{M}_{\Lambda_0}^{\text{fr}}(A,F,B)=2n^2 $. 
\end{Prf}

\begin{Prop}
     There exists a symplectomorphism between $\mathcal{M}_{\Lambda_0}^{\text{fr}}(A,F,B)$ and $T^{*}U(n)$. 
\end{Prop}

\begin{Prf}
    Define the symplectomorphism
    \begin{equation}
        \begin{aligned}
            \Psi:  \,\,  & \mathcal{M}_{\Lambda_0}^{\text{fr}}(A,F,B) \to  T^{*}U(n)    \\
             &(A,F,B) \mapsto (g_F,B).
        \end{aligned}
    \end{equation}
    There exists an inverse map
    \begin{equation}
        \Psi^{-1}: (g,B) \mapsto (g^{\dagger}Dg, \left( g^{\dagger}e_1,\dots, g^{\dagger}e_n \right),B ).
    \end{equation}
    Under the left trivialization, $T^*U(n) \cong U(n) \times \mathfrak{u}(n)^*$ , consider the pairing $\langle B,X \rangle:=i\text{Tr}(BX),\, X\in \mathfrak{u}(n)$, $\mathcal{H}(n)$ can be naturally identified as $\mathfrak{u}(n)^*$. Then the canonical $1$-form is 
    \begin{equation}
        \theta=\langle B, g^{-1}dg\rangle=i\text{Tr} (Bg^{-1}dg). 
    \end{equation}
    The symplectic form is then $d\theta$. On $\mathcal{M}_{\Lambda_0}^{\text{fr}}(A,F,B)$ we may define the canonical one form
    \begin{equation}
        \theta_{\mathcal{M}}= \langle B, g_F^{-1}dg_F  \rangle =i \text{Tr} \left(B g_F^{-1}dg_F\right).
    \end{equation}
    The symplectic form is then $d\theta_{\mathcal{M}}$. We can check that $\Psi^*(\theta)=\Psi^*(i\text{Tr}(Bg^{-1}dg)) = i\text{Tr}(Bg_F^{-1}dg_F)=\theta_{\mathcal{M}}$. Hence $\Psi^*(d\theta)= d\theta_{\mathcal{M}} \implies \Psi^*\omega=\omega_{\mathcal{M}}$. $\Psi$ is indeed a symplectomorphism.  
    
\end{Prf}

\textbullet \textbf{(Alternative definition of the framed space)} \,The preceding concrete description of the framed space $\mathcal{M}^{\text{fr}}_{\Lambda_0}$ admits an equivalent intrinsic formulation as a cotangent bundle. Therefore there is naturally a tautological $1$-form on $\mathcal{M}^{\text{fr}}_{\Lambda_0}$. We introduce the following definitions describing the same $\mathcal{M}^{\text{fr}}_{\Lambda_0}$.

\begin{Def}[Diagonalizing frame bundle]
Take $D$ in definition \ref{Def: frmdspace}. Define
\begin{equation}
    \mathcal F_D := \left\{ (A,G)\in\mathcal O_D\times U(n) \;:\;
    GAG^{-1}=D \right\}.
\end{equation}
No quotient by the diagonal torus is taken: the phases of the ordered
eigenvectors are retained as part of the framing data.
\end{Def}

\begin{Def}[Framed spectral phase space]
The framed phase space associated with the fixed irregular eigenvalues
$D=\text{diag} \Lambda_0$ is
\begin{equation}
    \mathcal M_{\Lambda_0}^{\mathrm{fr}}:= T^*\mathcal F_D.
\end{equation}
Using the pairing
$\langle B,X\rangle:=i\operatorname{Tr}(BX), \, B\in\mathcal H(n), \,
X\in\mathfrak u(n)$, and the left trivialization of $T^*\mathcal F_D$, a point of
$\mathcal M_{\Lambda_0}^{\mathrm{fr}}$ is represented by a triple $(A,G,B)$.
\end{Def}

Then we show the identification of the moment map. Before doing so we emphasize that the eigenvalue functions are smooth only on the open subset where all leading principal minors on both sides have simple spectra. Further, We define the strongly regular locus by the maximal-rank condition for the double Gelfand–Zeitlin map, as in \cite{crooks2023doublegelfandcetlinsysteminvariance}. On the Hermitian $U(n)$ locus, this is equivalently described by the corresponding strict interlacing conditions for consecutive leading principal minors; in particular, all relevant principal minors have simple spectra. Specifically, when $G=U(n)$, Cauchy interlacing theorem for eigenvalues of nested Hermitian principal submatrices enables the strongly regular locus to possess a Gelfand-Zeitlin pattern. Since we have the symplectomorphism, we can also define the strongly regular locus through pullback of $\Psi$. We will denote it by $\mathcal{M}_{\text{s-reg}}$. The following discussion will be restricted to the strongly regular locus. 

It is known that by the theorem of Guillemin-Sternberg\cite{GUILLEMIN1983106}, the eigenvalues of all leading
principal submatrices of a Hermitian matrix Poisson commute with respect to the Kirillov-Kostant-Souriau (KKS) Poisson bracket. Therefore elements of $\mu_i^{(k)}(-B)$ naturally commute, and so do elements of $\mu_i^{(k)}(GBG^{-1})$. It remains to prove that elements from two sides commute. This follows from the fact that the left and right moment maps Poisson-commute. Therefore the pullback of Hamiltonians by $\phi_1$ commute with those of $\phi_2$, which implies
\begin{equation}
    \{\mu_i^{(k)}(GBG^{-1}), \mu_j^{(l)}(-B) \}=0. 
\end{equation}

In \cite{crooks2023doublegelfandcetlinsysteminvariance}, the authors have stated that under the left trivialization, the moment map of the action $(h_L,h_R)\cdot(G,B)=(h_LGh_R^{-1}, h_RBh_R^{-1})$ can be written as
\begin{equation}
\begin{aligned}
    \phi:=(\phi_1,\phi_2): T^*U(n) &\to \mathfrak{u}(n)^* \times \mathfrak{u}(n)^* \\
    (g,\xi) &\mapsto  (\text{Ad}_g^*(\xi),-\xi),
\end{aligned}
\end{equation}
where $(g,\xi)\in U(n)\times \mathfrak{u}(n)^*$. In our case $G=U(n)$, $\mathfrak{u}(n)$ is the anti-Hermitian matrices and $\mathfrak{u}(n)^*$ is identified with Hermitian matrices, we have
\begin{equation}
    \langle \text{Ad}_g^*(\xi_B), X\rangle =\langle\xi_B, \text{Ad}_{g^{-1}}(X)\rangle=  \langle \xi_B,g^{-1}Xg  \rangle=i\text{Tr}(Bg^{-1}Xg)= i\text{Tr}(gBg^{-1}X).
\end{equation}
Therefore $\langle\text{Ad}_g^*(B), X \rangle=\langle gBg^{-1},X \rangle$ which yields
\begin{equation}
    \text{Ad}_g^*(B)=gBg^{-1}. 
\end{equation}

The degeneration cascade of \(\Gamma_{1,2}\) gives the eigenvalues of the
leading principal submatrices of \(B\), namely
\begin{equation}
    \mu_i^{(k)}(-B),\qquad 1\leq i\leq k,\quad 1\leq k\leq n.
\end{equation}
Similarly, the degeneration cascade of \(\Gamma_{2,1}\) gives the eigenvalues
of the leading principal submatrices of $GBG^{-1}$, namely
\begin{equation}
\mu_i^{(k)}(GBG^{-1}),\qquad 1\leq i\leq k,\quad 1\leq k\leq n.
\end{equation}
Now we identify $G$ and $g_F$ corresponding to the eigenvectors of $A$(i.e. the frame of $A$) and hence we identify $G=g$ via the symplectomorphism. Then these are precisely
$\mu_i^{(k)}(\phi_1),\,\mu_i^{(k)}(\phi_2)$. Therefore the spectral degeneration map agrees with the double Gelfand--Zeitlin moment map:
\begin{equation}
    \phi_{\text{spec}}=(\lambda,\lambda)\circ \phi \circ \Psi=(\lambda\circ\phi_1,\lambda\circ \phi_2) \circ \Psi. 
\end{equation}
We end up with a proposition:
\begin{Prop}
    Under the symplectomorphism $\Psi$, the spectral degeneration map 
    \begin{equation}
        \phi_{\text{spec}}(A,F,B)= \left(\mu_i^{(k)}(GBG^{-1}), \mu_i^{(k)}(-B)\right)
    \end{equation}
    coincides with the double Gelfand-Zeitlin map $\Phi= (\lambda,\lambda)\circ \phi$, i.e. $\phi_{\text{spec}}=\Phi\circ \Psi$ on the strongly regular locus $\mathcal{M}_{\text{s-reg}}$. As a result, on the strongly regular locus $\phi_{\text{sepc}}$ is the Hamiltonian $\mathbb{T}\times \mathbb{T}$-moment map, here $\mathbb{T}$ is the $\frac{n(n+1)}{2}$-dimensional torus. 
\end{Prop}

We have shown in section $2.2$, or more explicitly in Theorem \ref{thm:22to21+12}, that the left and right components share a group of eigenvalues, or $\text{Spec}(\phi_{\text{spec,L}})=-\text{Spec}(\phi_{\text{spec,R}})$. Since the Gelfand-Zeitlin pattern arranges the eigenvalues in decreasing order, the negative eigenvalues should be arranged in reverse order. This leads to
\begin{equation}
    \mu^{(n),L}_i+\mu_{n+1-i}^{(n),R}=0. \label{eq: 3.15}
\end{equation}

On the double Gelfand-Zeitlin side, the subtorus $\mathbb{S}$ is defined to be 
\begin{equation}
    \mathbb{S} :\left\{ ((t_1,\dots, t_n,1,\dots,1),(t_n,\dots,t_1,1,\dots,1)): (t_1,\dots,t_n)\in \mathbb{T}_0 \right\} \subset \mathbb{T} \times \mathbb{T},
\end{equation}
where $\mathbb{T}_0$ is the $n$-dimensional torus and $\mathbb{T}$ is the $\frac{n(n+1)}{2}$-dimensional torus. Then the lie algebra is
\begin{equation}
    \mathfrak{G} :\left\{ ((x_1,\dots, x_n,0,\dots,0),(x_n,\dots,x_1,0,\dots,0)): (x_1,\dots,x_n)\in \mathbb{R}^n \right\} \subset \mathbb{R}^{\frac{n(n+1)}{2}} \times \mathbb{R}^{\frac{n(n+1)}{2}},
\end{equation}
The moment map corresponding to $\mathbb{S}$ is
\begin{equation}
    \nu_{\mathbb{S}}(x)=\sum_{i=1}^n x_i\lambda_i^{(n),L} +\sum_{j=1}^n x_{n+1-j}\lambda^{(n),R}_j. 
\end{equation}
Now, by definition \cite{crooks2023doublegelfandcetlinsysteminvariance} $\Phi:T^*U(n)\to\mathfrak{G}^{\perp}$, and this implies $\nu_{\mathbb{S}}(x)=0$ for arbitrary $x_i\in \mathbb{R}$. This requires that $ \lambda^{(n),L}_i+\lambda_{n+1-i}^{(n),R}=0$. This is exactly the relation given in $(\ref{eq: 3.15})$. 

Finally we count the independent integrals of motion. We have $\sum_{k=1}^{n-1}k=\frac{n(n-1)}{2}$ eigenvalues on each side, so in total $n(n-1)$. Further, there are $n$ shared eigenvalues.  On the other hand, the effective torus action is given by $(\mathbb{T}\times \mathbb{T})/\mathbb{S}$ with dimension $\text{dim} (\mathbb{T}\times \mathbb{T})/\mathbb{S} = n(n+1)-n=n^2$. Therefore we have $n^2$ independent Hamiltonians as desired.
We've finally proved the equivalence between our spectral-curve system and the double Gelfand-Zeitlin system. 

\begin{Thm} \label{thm: dgz=spec}
    The spectral curve degenerating system is a double Gelfand-Zeitlin system:
    \begin{equation}
    \begin{aligned}
        \phi_{\text{spec}}: \mathcal{M}_{\text{s-reg}}^{\text{fr}} &\to \mathfrak{G}^{\perp} \\
        (g,B) &\mapsto \bigl(\mu(GBG^{-1}),\mu(-B) \bigr).
    \end{aligned}
    \end{equation}
    Here we define $\mu:=(\mu^{(n)}, \dots, \mu^{(1)})$ and $\mu^{(k)}:=(\mu^{(k)}_1, \dots,\mu^{(k)}_k)$.  
\end{Thm}

\subsection{Angle coordinates}

On the double Gelfand-Zeitlin side, we may examine the action of the two Hamiltonian flows. After fixing the eigenvalues, they actually rotate the eigenvectors. There is one more question remaining: what is the angle coordinate of the integrable system on the spectral curve side? There are two natural candidates. First we may consider the plumbing phases of the tubes. However, at each level of the decomposition, we use a single global parameter in the plumbing process. Namely we are considering the $xy=q_i$ in the splitting process, but in our case all the $q_i$s are the same. This implies that the gluing angles are frozen at each level of separation. 

The angle coordinates on the spectral curve side are actually the gluing phase of the line bundle on the spectral curve. 

\begin{Def}[The spectral line bundle \cite{Ramanan1989}]
    Let $(x,z)$ be the coordinate of a spectral curve $\Gamma: \text{det}\left(xI-\Phi\right)=0$, here $\Phi$ is a Higgs field. Then define 
    \begin{equation}
        \mathcal{L}_{(x,z)}:= \text{ker} (\Phi_z-xI).
    \end{equation}
    If the eigenvalue is simple then $\text{dim}\, \mathcal{L}_{(x,z)}=1$. Those $1$-dimensional vector spaces can be glued to  a line bundle $\mathcal{L} \to \Gamma$, i.e. $\mathcal{L}= \text{ker} (\pi^*\Phi-\lambda)$, where $\lambda$ is the eigenvalue function, and $\pi$ is the covering map $\Gamma\to C$.  \label{Def: speclinebundle}
\end{Def}

We now identify the angle coordinates on the spectral-curve side.
Consider a node $q$ of the degenerate spectral curve, with preimages
$q^+$ and $q^-$ on the normalization. Let $\mathcal L$ be the spectral
line bundle. After choosing unitary endpoint frames
$ s_q^+\in\mathcal L_{q^+}, \, s_q^-\in\mathcal L_{q^-}$, the gluing isomorphism
$\rho_q:\mathcal L_{q^+}\longrightarrow\mathcal L_{q^-}$ is represented by a nonzero scalar $c_q$:
\begin{equation}
    \rho_q(s_q^+)=c_qs_q^-.
\end{equation}
We further define
\begin{equation}
    \theta_q:=\arg c_q.
\end{equation}

\begin{Prop} \label{Prop: angleCoorGluing}
On the strongly regular locus, the phases of the framed spectral-line
gluing parameters are angle coordinates for the effective double
Gelfand--Zeitlin torus action.
\end{Prop}

\begin{Prf}
We first consider the Gelfand-Zeitlin side. Let $\mu_i^{(k)}(X)$ be a Gelfand--Zeitlin Hamiltonian and let $\widehat{P}_i^{(k)}(X)$ denote the extension of the spectral projector $P_i^{(k)}(X)$ corresponding to $\mu_i^{(k)}(X)$ by zero to $\mathbb C^n$. The corresponding circle flow is generated, up to the fixed moment-map normalization, by
\begin{equation}
    U_i^{(k)}(s) = \exp\left( \mathrm{i}s\widehat P_i^{(k)}(X) \right).
\end{equation}
Since $P_i^{(k)}(X)$ is the projector onto the eigenline
$L_i^{(k)}$, this flow acts on that eigenline by $v_i^{(k)} \longmapsto e^{\mathrm{i}s}v_i^{(k)}$, and leaves the other eigenlines unchanged. To be clear, $L^{(k)}_i$ is preserved, and the action of $U^{(k)}_i$ is restricted to a scalar multiplication and acts trivially on other spectral eigenspaces. 
 
On the spectral curve side, the essential point is that the limiting fibers of the
spectral line bundle at the normalized nodes are precisely the
eigenlines entering the two Gelfand--Zeitlin systems. On the
$(1,2)$ component, near $z=0$ one has $\Phi_{1,2}(z)= \frac{B}{z}-iu$,  and hence
\begin{equation}
    z\Phi_{1,2}(z)= B-iuz \longrightarrow B.
\end{equation}
Since $B$ has simple spectrum, analytic perturbation theory implies
that the corresponding rank-one spectral projectors of
$B-iuz$ depend analytically on $z$ near $z=0$ and converge to the
spectral projectors of $B$ as $z\to0$
\cite[Chapter II, Sections 1.3--1.4]{KatoPerturbation}.
Consequently, the spectral eigenlines of $\Phi_{1,2}(z)$ converge to
the eigenlines of $B$.

Similarly, on the $(2,1)$ component, in the coordinate $v'=1/v$,
$\Phi_{2,1}(v') = D_0(v')^2 + GBG^{-1}v'$,  so that
\begin{equation}
    \frac{1}{v'}\Phi_{2,1}(v') = GBG^{-1}+D_0v' \longrightarrow GBG^{-1}
\end{equation}
as $v'\to0$. Therefore the corresponding spectral eigenlines converge
to the eigenlines of $GBG^{-1}$.

Therefore the double Gelfand--Zeitlin Hamiltonian flows preserve those eigenlines of $B$ and $GBG^{-1}$ and act on their framed fibers with the appropriate $U(1)$-weights. Consequently, under the degeneration they change the endpoint frames
of the spectral line bundle. If the gluing isomorphism is written as $\rho_i(s_i^+)= c_i s_i^-$, then an eigenline $U(1)$ scalar multiplication at one endpoint gives
$c_i \longmapsto e^{\pm\mathrm i s}c_i$, where the sign depends on the orientation of the two branches. Hence
\begin{equation}
    \theta_i:= \arg c_i
\end{equation}
is translated by the corresponding Hamiltonian flow and is therefore
the angle coordinate conjugate to the associated Gelfand--Zeitlin
action variable.

The left and right Gelfand--Zeitlin flows act on the eigenline
framings inherited from the two limiting components separately. Thus,
for the corresponding effective torus action, the framing at one
endpoint is rotated while the framing at the opposite endpoint is
kept fixed. On the top level, the simultaneous rephasing of the two endpoint frames is
precisely part of the kernel subtorus and acts trivially on the relative
gluing parameter .

At the subsequent stages of the two caterpillar cascades, the same
argument applies with $B$ and $GBG^{-1}$ replaced by the corresponding
leading principal submatrices. Hence the limiting eigenlines at every
normalized node are precisely the eigenlines entering the associated
Gelfand--Zeitlin Hamiltonian. Since the Hamiltonian flow generated by $\mu_i^{(k)}$ acts on the corresponding gluing parameter by
$c_i^{(k)}\mapsto e^{\pm is}c_i^{(k)}$, it translates
$\theta_i^{(k)}=\arg c_i^{(k)}$ by $\theta_i^{(k)}\mapsto\theta_i^{(k)}\pm s$; hence
$X_{\mu_i^{(k)}}\theta_i^{(k)}=\pm1$, or equivalently
$\{\theta_i^{(k)},\mu_i^{(k)}\}=\pm1$, with the sign fixed by the orientation convention.
\end{Prf}

\begin{Ex} \label{Ex: explicit}
    We illustrate explicitly the action generated by the Hamiltonian
$\mu_i^{(k)}$. Write the $(k+1)\times(k+1)$ principal block in the
bordered form
\begin{equation}
B^{(k+1)} = \begin{pmatrix} B^{(k)} & u_k \\ u_k^\ast & b
\end{pmatrix},
\end{equation}
and let $v_i^{(k)}$ be a normalized eigenvector of $B^{(k)}$ with
eigenvalue $\mu_i^{(k)}$. Denote the corresponding spectral projector by
$P_i^{(k)} = v_i^{(k)}\bigl(v_i^{(k)}\bigr)^\ast$. The circle generated by $\mu_i^{(k)}$ is represented by the stabilizer element $h_i^{(k)}(s) =
\exp\left(\mathrm{i}sP_i^{(k)}\right)$.

Since $P_i^{(k)}$ is a spectral projector of $B^{(k)}$, one has $h_i^{(k)}(s)B^{(k)}h_i^{(k)}(s)^{-1} = B^{(k)}$. On the other hand, the Hamiltonian action on the bordered matrix is
\begin{equation}
B^{(k+1)}
\longmapsto
\begin{pmatrix}
h_i^{(k)}(s) & 0 \\
0 & 1
\end{pmatrix}
B^{(k+1)}
\begin{pmatrix}
h_i^{(k)}(s)^{-1} & 0 \\
0 & 1
\end{pmatrix},
\end{equation}
and therefore $B^{(k)} \longmapsto B^{(k)}, \,\, u_k \longmapsto h_i^{(k)}(s)u_k$. Writing $u_k = \sum_{j=1}^{k} c_jv_j^{(k)}, \,c_j=
\langle v_j^{(k)},u_k\rangle$, we obtain
\begin{equation}
u_k \longmapsto \sum_{j\neq i}c_jv_j^{(k)} + e^{\mathrm{i}s}c_iv_i^{(k)}.
\end{equation}
Thus, in a fixed eigenframe of $B^{(k)}$, the Gelfand--Zeitlin
Hamiltonian flow rotates precisely the $i$-th component of the
extension vector, $\langle v_i^{(k)},u_k\rangle \longmapsto e^{\mathrm{i}s}
\langle v_i^{(k)},u_k\rangle$.
Equivalently,
\begin{equation}
\langle e_k,P_i^{(k)}u_k\rangle
\longmapsto
e^{\mathrm{i}s}
\langle e_k,P_i^{(k)}u_k\rangle, \label{eq: rotation}
\end{equation}
while the analogous quantities with $j\neq i$ remain unchanged.

There is another, equivalent way to describe the same stabilizer
element. Since
\begin{equation}
h_i^{(k)}(s)v_i^{(k)}
=
e^{\mathrm{i}s}v_i^{(k)},
\end{equation}
one may say that $h_i^{(k)}(s)$ rotates the $i$-th eigenline. This
statement, however, should not be interpreted as an independent
Hamiltonian motion of the eigenvector. Indeed, $B^{(k)}$ itself is
fixed along the flow, and hence its eigenline $L_i^{(k)}$ is unchanged.
If one chooses a fixed eigenframe, the nontrivial Hamiltonian motion is
seen entirely in $u_k$. If instead one transports the eigenframe by
the same stabilizer element,
\begin{equation}
v_i^{(k)}(s) = h_i^{(k)}(s)v_i^{(k)} = e^{\mathrm{i}s}v_i^{(k)},
\end{equation}
then both the frame and $u_k$ are transformed simultaneously, and
\begin{equation}
\langle v_i^{(k)}(s),u_k(s)\rangle = \langle v_i^{(k)},u_k\rangle.
\end{equation}
Thus the two descriptions correspond to the same Gelfand--Zeitlin
circle action written in a fixed frame or in a co-moving frame,
respectively. In particular, a bare rephasing
$v_i^{(k)}\mapsto e^{\mathrm{i}\alpha}v_i^{(k)}$ with the matrix data
held fixed is only a change of eigenvector representative and should
not be confused with the Hamiltonian flow.
\end{Ex}

\begin{Rmk}
    One might notice that there is a dimension mismatch between the DGZ system and the SW type integrable system. Namely $\text{dim}_{\rm DGZ}/2=n^2$ and $\text{dim}_{\rm SW}/2=(n-1)^2$. Indeed, the difference between dimensions comes from the removal of the central $U(1)$ at each level. The action variables of SW system come from the independent $A$-periods, while the DGZ system includes the $U(1)$ directions. 
\end{Rmk}

\subsection{Physical interpretations}

\textbullet \textbf{Weyl measure from the double Gelfand-Zeitlin system} \,We briefly explain the physical implications of this identification. In section $5.1$ we discuss the physical descriptions of the caterpillar spectral curves: the weak coupling limit of a quiver gauge theory with two wide tails. The quiver gauge theory corresponding to a $(2,2)$ type spectral curve can be viewed as gluing the two one-sided quiver tails by gauging their diagonal $U(n)$ flavor symmetry, thereby producing the middle $U(n)$ gauge node. This corresponds naturally to the Duistermaat–Heckman measure on the Gelfand-Zeitlin side \cite{Sun_2016}. 

\begin{Prop}
\label{Prop: WeylMeasureDH}
Let $\Phi_{\rm DGZ}= \left( a,\mu_L^{<n},\mu_R^{<n} \right) $ be the double Gelfand--Zeitlin moment map on the strongly regular locus of $T^*U(n)$, where $a$ is the top row of the right moment map with $a_1>\cdots>a_n$,
\, and $a^*:=-w_0a=(-a_n,\ldots,-a_1)$. Let $p_M$ retain only the common top row $a$. Then
\begin{equation}
    (p_M\circ\Phi_{\rm DGZ})_*\left( \frac{\omega^{n^2}}{n^2!} \right)
    = C_n^{\rm DGZ}\Delta(a)^2\,da,
\end{equation}
where $C_n^{\rm DGZ}$ is independent of $a$.
\end{Prop}

\begin{Prf}
On the strongly regular locus, the double Gelfand--Zeitlin Hamiltonians
generate a completely integrable $U(1)^{n^2}$-action (note that this is the effective torus action of $\mathbb{T}\times \mathbb{T}/\mathbb{S}$). Hence, in action-angle coordinates,
\begin{equation}
    \frac{\omega^{n^2}}{n^2!} = da\,d\mu_L^{<n}\,d\mu_R^{<n}\,d\theta
\end{equation}
up to orientation.

As $\xi$ varies with decreasingly ordered spectrum
$\operatorname{Spec}(-\xi)=a$, the lower rows of the right
Gelfand--Zeitlin map range over $\operatorname{GT}_a$. For each such
$\xi$, varying $g\in U(n)$ makes
\begin{equation}
    \mu_L=\operatorname{Ad}_g^*\xi
\end{equation}
range over the full orbit $\mathcal O_{a^*}$. Conversely, for every
pair of points in $\operatorname{GT}_a$ and
$\operatorname{GT}_{a^*}$, one may first choose $\xi$ realizing the
right point and then choose $g$ realizing the left point. Hence the
fixed-$a$ action image is
\begin{equation}
    \operatorname{GT}_a\times\operatorname{GT}_{a^*}.
\end{equation}

Integration over the angles contributes only an $a$-independent
constant. Therefore
\begin{equation}
    (p_M\circ\Phi_{\rm DGZ})_* \left( \frac{\omega^{n^2}}{n^2!} \right) =
    C_{\rm ang}\, \operatorname{Vol}(\operatorname{GT}_a)
    \operatorname{Vol}(\operatorname{GT}_{a^*})\,da.
\end{equation}
Using the Gelfand-Zeitlin volume formula \cite{olshanski2013projectionsorbitalmeasuresgelfandtsetlin}
\begin{equation}
    \operatorname{Vol}(\operatorname{GT}_a) =\frac{\Delta(a)}{\prod_{r=1}^{n-1}r!},
    \qquad \Delta(a^*)=\Delta(a),
\end{equation}
gives the result. Since the complement of the strongly regular locus
has measure zero, it does not affect the pushforward measure.
\end{Prf}

\begin{Rmk}
\label{Rmk: DHmeasure}
The same density follows immediately from the Weyl integration formula,
since diagonalizing $\xi\in\mathfrak u(n)^*$ gives
\begin{equation}
    d\xi \propto \Delta(a)^2\,da\,d\mu_{U(n)/T^n}.
\end{equation}
The double Gelfand--Zeitlin description contains the additional
information that
\begin{equation}
    \Delta(a)^2 \propto \operatorname{Vol}(\operatorname{GT}_a)
    \operatorname{Vol}(\operatorname{GT}_{a^*}),
\end{equation}
so that the two Vandermonde factors arise from the two sides of the
double caterpillar degeneration.
\end{Rmk}

\textbullet \textbf{Fourier lattice from Peter-Weyl}\, The lattice $\Lambda_{\rm cat}$ (\ref{eq: lattice}) is the character lattice of the
$SU$-reduced caterpillar angle torus
\begin{equation}
    \mathbb T_{\rm cat}^{SU} = \prod_{k=2}^{n-1} \left(T_k/U(1)_{\rm diag}\right)_L
    \times \left(T_n/U(1)_{\rm diag}\right)_M
    \times \prod_{k=2}^{n-1} \left(T_k/U(1)_{\rm diag}\right)_R.
\end{equation}
Indeed,
\begin{equation}
    X^*\left(T_k/U(1)_{\rm diag}\right) = Q(A_{k-1}),
\end{equation}
and hence
\begin{equation}
    X^*(\mathbb T_{\rm cat}^{SU}) = \Lambda_{\rm cat}.
\end{equation}
The Fourier completion in equation (\ref{eq: FourierCompletion}) may therefore be viewed as the
Abelianized Peter--Weyl expansion associated with the double
Gelfand--Zeitlin real polarization. The full Bohr--Sommerfeld set is
more restrictive, since its integral Gelfand--Zeitlin patterns obey
the interlacing inequalities; $\Lambda_{\rm cat}$ records instead the
affine lattice of Fourier and Schlesinger translations.

\section{The nodal caterpillar spectral networks}

Although the spectral curve already encodes the Seiberg--Witten periods and the algebraic integrable system, it does not by itself specify how these periods appear in the asymptotic expansion of flat sections or in the Stokes matrices of the associated meromorphic connection. For this purpose one needs an additional WKB decomposition, depending on a phase. This is precisely the role of the spectral network \cite{gaiotto2011wallcrossinghitchinsystemswkb,Gaiotto_2013}. It organizes the trajectories of the Seiberg--Witten differential on the base curve and records how paths lift to the spectral cover. Therefore, the spectral network should be viewed as the bridge between the algebraic geometry of the spectral curve and the analytic/Stokes data of the differential equation. In the following section we use this bridge to study how the caterpillar degeneration of the spectral curve manifests itself at the level of path lifting, Stokes factors, and spectral coordinates.

Strictly speaking, the spectral network is not an additional curve but a phase-dependent graph on the base curve $C$, whose edges are projections of WKB trajectories associated with the sheets of the spectral cover $\pi:\Gamma\to C$. Therefore, once the caterpillar limit decomposes the spectral cover over different regions of the base, it also gives a natural decomposition problem for the spectral network. Away from the neck region, the WKB equations converge to those of the limiting outer and inner spectral curves respectively. Thus the outer and inner parts of the network are governed by the corresponding limiting components of the spectral curve, while the neck region contains the residual matching information. This is the sense in which the spectral curve decomposition induces, at the level of WKB geometry, a decomposition of the spectral network.

Let $\hat{\Gamma}(u,A,B)$ be the real oriented blow-up of $\Gamma(u,A,B)$ at the $2n$ preimages of $0, \infty$. We consider the relative homology 
\begin{equation}
    H(u,A,B,v)=H_1(\hat{\Gamma}(u,A,B); L(v)),
\end{equation}
here $L(v)$ is the set of $4n$ marked points, with $2$ marked points on the real blow-up of $\infty$ lying over the two anti-Stokes directions($v,v+\pi$) and $2$ marked points on the real blow-up of $0$ lying over the two anti-Stokes directions ($\pi -v, 2\pi-v$) on each of the $n$ sheets. Namely $L(v)=(\infty_1^+,\infty_1^-, \dots , \infty_n^+,\infty_n^-, 0_1^+,0_1^-, \dots, 0_n^+,0_n^-)$. 

The arguments of Sections 2 and 3 apply equally to
\begin{equation}
\Phi_\kappa(z)= iu-\kappa\frac{B}{z}-\kappa\frac{A}{z^2},\qquad \kappa\in\{1,i\}.
 \end{equation}
The reason is that the decomposition argument actually applies to all generic $A,B$. Furthermore, the left, right eigenvectors of anti-Hermitian $A$ are still pairwise conjugate. For the case $\kappa=i$, the Hermitian representatives $A$ and $B$ still define the double Gelfand--Zeitlin variables, while the spectral residues acquire the
common factor $i$.

For simplicity, in this section we consider the system 
\begin{equation}
    \epsilon\frac{dF}{dz} = \left( iu-(i\frac{B}{z}+i\frac{A}{z^2})   \right)F
    \label{eq: symODE}
\end{equation}
or the spectral curve
\begin{equation}
    \det \left( xI_n -\left( iu- i\frac{B}{z}-i\frac{A}{z^2}\right)  \right)=0,
    \label{eq: symSpec}
\end{equation}
so that there is an involution symmetry $(x,z) \to (-\bar{x},\bar{z})$ of the spectral curve. But we also mention almost all the results in this section can be extended to the Hermitian case, in particular the stable caterpillar spectral networks, the neck gluing structure, and the endpoint-normalization calculations. 

\subsection{n=2}

\begin{Prop}
    In the double caterpillar limit, $0$ is a double root of $P(z)$, so that the genus is pinched off. Moreover, $Z(\alpha)$ gives the residue contributions around $z=0$ and $z=\infty$. 
\end{Prop}
\begin{Prf}
    In the $n=2$ double caterpillar limit $t\to 0$, if we fix $\Delta u=u_2-u_1$, then $w_1=\lambda_2-\lambda_1 \to 0$. Since $A=G^{-1} \text{Diag} (\lambda_1, \dots, \lambda_n)G$ (we will then denote the diagonal matrix by $D$). We can write 
    $D= w_1E_2+\lambda_1 I_2$, and then 
    \begin{equation}
        A=G^{-1} (w_1E_2+\lambda_1 I_2)G= \lambda_1I+w_1G^{-1}E_2G. 
    \end{equation}
    This implies the off-diagonal terms are all of order $w_1$, namely $a_{12}=\bar{a}_{21}=O(w_1)$. Then by letting $w_1\to 0$
    \begin{equation}
    \begin{aligned}
        P(z)&= (\Delta u z^2-\Delta Bz-w_1\Delta (G^{-1}E_2G) )^2+4(b_{12}z+O(w_1))(\bar{b}_{12}z+O(\bar{w}_1)) \\
        &= (\Delta u z^2-\Delta Bz)^2+4\lvert b_{12} \rvert^2z^2 \\
        &= z^2P'(z). 
    \end{aligned}
    \end{equation}
    Therefore $z=0$ becomes a double root. Geometrically, $P'(z)$ now defines a genus $0$ curve. Because of the involution symmetry, the two colliding branch point must be exchanged by reflection across the real axis. If these colliding branch points lie to the left of the other two branch points, then $\alpha$ becomes $\partial_0$. Otherwise, we can view $\alpha$ as encircling the puncture at infinity, except for a singular point $z=0$. Therefore $Z(\alpha)$ is a combination of the residues at $0$ and $\infty$. 
\end{Prf}

First we need to study how the spectral network behaves under the caterpillar limit. 

\begin{Prop}
\label{prop: 4.2}
    Consider the rank two $(2,2)$ spectral curve $\Gamma_{2,2}^{(2)}$. Let
    \begin{equation}
        \omega_t:=\lambda_{+,t}-\lambda_{-,t} = (x_{+,t}-x_{-,t})dz
    \end{equation}
    be the difference of the two Seiberg--Witten differentials. Fix a
    non-critical WKB phase $\vartheta$, namely assume that no finite WKB trajectory of phase $\vartheta$ appears for sufficiently small $t$. Then, in the caterpillar limit $t\to0$, the spectral network of $\Gamma_t$ locally converges to the componentwise broken spectral network obtained by gluing the $(1,2)$-network of the rank-$2$ components given in \ref{eq:intro-two-components}, away from the branch points and nodes. 
    
\end{Prop}
\begin{Prf}
    We first consider the outer scale \(z\sim 1\). Let $K\subset \mathbb C_z^{\times}$be a compact subset avoiding $z=0$ and the branch points of $\Gamma_{\text{out}}$. On $K$, the matrix $iu-i\frac{B}{z}-i\frac{tA}{z^2}$ converges uniformly, together with all derivatives, to $iu-i\frac{B}{z}$. Since $K$ avoids the branch points, the two eigenvalues can be chosen holomorphically on $K$. Hence $x_{\pm,t}(z)\longrightarrow x_{\pm,\mathrm{out}}(z)$ in the \(C^{\infty}\)-topology on \(K\). Therefore
    \begin{equation}
        \omega_t = (x_{+,t}-x_{-,t})dz \longrightarrow \omega_{\mathrm{out}}
        := (x_{+,\mathrm{out}}-x_{-,\mathrm{out}})dz .
    \end{equation}

    A spectral-network wall of phase $\vartheta$ is locally a trajectory of
    the line field determined by $\text{Im} \,( e^{-iv}\int \omega_t)=\text{const}$.
    Equivalently, away from zeroes and poles of $\omega_t$, it is an
    integral curve of the corresponding WKB line field. Since
    $\omega_t\to \omega_{\mathrm{out}}$ in $C^{\infty}$ on $K$, these
    line fields converge in $C^{\infty}$. By the continuous dependence of
    solutions of ordinary differential equations on parameters, every finite
    segment of a spectral-network wall contained in $K$ converges to a wall
    of the $(1,2)$-network on $\Gamma_{\mathrm{out}}$.

    Next we consider the inner scale. Put $z=tv, x=\frac{X}{t}$, then $xdz=Xdv$. 
    Substituting $z=tv$ and $x=X/t$ into the equation of $\Gamma_t$, and
    multiplying the matrix inside the determinant by $t$, we obtain
    \begin{equation}
        \det\left(XI_2-\left(itu-i\frac{B}{v}-i\frac{A}{v^2}\right)\right)=0.
    \end{equation}
    Hence, as $t\to0$, the inner spectral curve converges to
    \begin{equation}
        \Gamma_{\mathrm{in}}: \det\left(XI_2+i\frac{B}{v}+i\frac{A}{v^2}\right)=0.
    \end{equation}
    Let $K'\subset \mathbb C_v^{\times}$ be a compact subset avoiding $v=0$, $v=\infty$, and the branch points of $\Gamma_{\mathrm{in}}$. On $K'$, the two eigenvalues
    $X_{\pm,t}(v)$ converge in $C^{\infty}$ to the eigenvalues
    $X_{\pm,\mathrm{in}}(v)$ of the limiting inner curve. Since $xdz=Xdv$, we have
    $\omega_t= (X_{+,t}-X_{-,t})dv$. 
    Therefore
    \begin{equation}
        \omega_t \longrightarrow \omega_{\mathrm{in}}:= (X_{+,\mathrm{in}}-X_{-,\mathrm{in}})dv
    \end{equation}
    in the $C^{\infty}$-topology on $K'$. The same ODE-continuity
    argument then implies that, in the rescaled coordinate $v=\frac{z}{t}$, the
    microscopic part of the spectral network converges to the spectral network of $\Gamma_{\mathrm{in}}$.

   As shown in Section $2.2$, the Seiberg-Witten differentials agree on the two sides of the neck. Therefore the limiting object is not the disjoint union of the two networks, but the broken spectral network
    \begin{equation}
        \mathcal W_{\mathrm{lim}}(v)=\mathcal W_{\mathrm{out}}(v)\cup_{0=\infty} \mathcal W_{\mathrm{in}}(v),
    \end{equation}
    where the gluing is performed by identifying $z=0_i$ on the outer
    component with $v=\infty_i$ on the inner component.
\end{Prf}

\subsection{General $n$}

In fact, this holds true for general $n$. Let the diagonalization of $A$ be $D_n$, then 
 \begin{equation}
     D=\lambda_1I_n+w_1E_2+w_2w_1E_3+\dots+ \prod^{n-1}_{j=1}w_jE_n.
 \end{equation}
Now, if we fix $u$, then under the double caterpillar limit $\lambda_n-\lambda_1\sim t=\prod^{n-1}_{j=1}w_j \to 0$. Therefore if $GDG^{-1}=A$, then $A=\lambda_1I_n+w_1GE_2G^{-1}+\dots+\prod^{n-1}_{j=1}w_jGE_nG^{-1}=\lambda_1I_n+\widetilde{A}(t)$ where $\widetilde{A}(t)\to 0$. 
\begin{Prop}
    Consider the rank $n$ spectral curve  where $u,B,A(t)\in \mathcal H(n)$, and assume that $u$ and $B$ are regular semisimple. Suppose furthermore that, in the caterpillar limit, $A(t)$ tends to a scalar matrix, namely
    \begin{equation}
        A(t)=\lambda I_n+\widetilde A(t), \qquad \widetilde A(t)\to 0.
    \end{equation}
    Then $n(n-1)$ branch points of $\Gamma^{(n)}_{2,2}(t)$ collapse to
    $z=0$. If the curve is invariant under the involution $z\mapsto \bar z$,
    and the branch points are generic, then these $n(n-1)$ branch points
    form $\frac{n(n-1)}{2}=\binom n2$ conjugate pairs.
\end{Prop}

\begin{Prf}
    The branch points of $\Gamma^{(n)}_{2,2}(t)\to \mathbb C_z$ occur
    when two eigenvalues of the Higgs field collide. Since multiplying the
    Higgs field by a scalar meromorphic function does not change the locus of
    eigenvalue collisions, the branch points are equivalently given by the
    discriminant of $M_t(z):=z^2u-zB-A(t)$. More precisely, define
    \begin{equation}
        \Delta_t(z):= \operatorname{Disc}_y \det\left(yI_n-M_t(z)\right).
    \end{equation}
    Then, away from the singular points of the projection, the zeroes of
    $\Delta_t(z)$ are precisely the branch points of the spectral cover.

    Since adding a scalar matrix does not change the difference of
    eigenvalues, the scalar part $\lambda(t)I_n$ of $A(t)$ does not
    affect the discriminant. Therefore, for the purpose of calculating
    $\Delta_t(z)$, we may replace $A(t)$ by $\widetilde A(t)$. In the
    caterpillar limit $t\to 0$, we have $\widetilde A(t)\to 0$. Hence the limiting matrix is $M_0(z)=z^2u-zB=z(zu-B)$. 

    Let $\nu_1(z),\dots,\nu_n(z)$ be the eigenvalues of $zu-B$. Then the eigenvalues of $M_0(z)$ are $z\nu_1(z),\dots,z\nu_n(z)$. Therefore the discriminant of \(M_0(z)\) is
    \begin{equation}
    \begin{aligned}
        \Delta_0(z) &= \prod_{1\leq a<b\leq n} \left(
        z\nu_a(z)-z\nu_b(z) \right)^2                                     
        = z^{2\binom n2} \prod_{1\leq a<b\leq n} \left( \nu_a(z)-\nu_b(z)\right)^2     
        =z^{n(n-1)} \operatorname{Disc}(zu-B).
    \end{aligned}
    \end{equation}
    Since $B$ is regular semisimple, $-B$ has distinct eigenvalues.
    Therefore
    \begin{equation}
        \operatorname{Disc}(zu-B)\big|_{z=0} = \operatorname{Disc}(-B) \neq 0.
    \end{equation}
    Hence $z=0$ is a zero of $\Delta_0(z)$ of order exactly $n(n-1)$.
   
    By the Weierstrass preparation theorem, for sufficiently small $t$,
    there exists a neighborhood $U$ of $z=0$ such that $\Delta_t(z)$ has exactly $n(n-1)$ zeroes in $U$, counted with multiplicity. Equivalently, $n(n-1)$ branch points of the spectral cover collapse to $z=0$ as $t\to 0$. For generic $t$, these zeroes are simple, since the condition that two zeroes of $\Delta_t(z)$ collide is a proper discriminant condition on the coefficients.

    Finally, suppose that the coefficients are chosen so that the spectral
    curve is invariant under complex conjugation. Then $\Delta_t(\bar z)=\overline{\Delta_t(z)}$. Hence if $z_\ast$ is a branch point, then $\bar z_\ast$ is also a branch point. For generic parameters, no collapsing branch point lies on the real axis. Therefore the $n(n-1)$ collapsing branch points are
    paired by the involution $z\mapsto \bar z$, and they form $\frac{n(n-1)}{2}=\binom n2$ conjugate pairs.
\end{Prf}

For general $n$, under the caterpillar limit the spectral networks can be further split. The proof will be basically the same, as the Seiberg-Witten differential agrees on the neck. This can be seen from the distribution of the branch points. On the degenerate spectral curve
\begin{equation}
    \Gamma^{(k)}:\det \left( xI_k-(iE_k-i\frac{A^{(k)}}{z})  \right)=0, 
\end{equation}
there are $k-1$ pairs of branch points. The picture is, as we take $k$-th layer of the caterpillar limit respectively, $k-1$ pairs of conjugate branch points coalesce; however, under a microscope rescaling the $2(k-1)$ branch points can be observed and remain separated. Therefore in a full process of single side caterpillar degeneration, there are in total $2\sum_{k=1}^{n-1}k=n(n-1)$, which is exactly the total number of branch points. 

Since $\Gamma^{(k)}$ is a blow-up using $u_k$, while in the caterpillar limit we have $u_2\ll u_3\ll \dots \ll u_{n-1} \ll u_n $, therefore $\Gamma^{(n)}$ should be viewed as the most microscopic component. 

\begin{tikzpicture}[
    >=latex,
    thick,
    node distance=2.6cm,
    comp/.style={
        draw,
        rounded corners,
        minimum width=1.9cm,
        minimum height=0.9cm,
        align=center
    },
    lab/.style={
        font=\small,
        fill=white,
        inner sep=1pt
    }
]
\node[comp] (G2) {$\Gamma^{(2)}$};
\node[comp, right=of G2] (G3) {$\Gamma^{(3)}$};
\node[right=1.6cm of G3] (dots) {$\cdots$};
\node[comp, right=1.6cm of dots] (Gn) {$\Gamma^{(n)}$};
\draw (G2.east) -- node[lab, above] {$2$ nodes} (G3.west);
\draw (G3.east) -- node[lab, above] {$3$ nodes} (dots.west);
\draw (dots.east) -- node[lab, above] {$(n-1)$ nodes} (Gn.west);
\end{tikzpicture}

Specifically, on the $k$-th component, the branch cuts are exactly of the type $(1,k)$, $(2,k)$, $\dots$, $(k-1,k)$.

The spectral network has a similar behavior. For each component $\Gamma^{(k)}$ there are $2(k-1)$ branch points on it. Similarly we may write 
\begin{equation}
    \mathcal{W}^{(2)}  \longleftrightarrow\mathcal{W}^{(3)} \longleftrightarrow\dots\longleftrightarrow\mathcal{W}^{(n-1)} \longleftrightarrow\mathcal{W}^{(n)}. 
\end{equation}

\begin{Rmk}
    Caterpillar limit of the spectral network requires clarification, because the limit is not an ordinary spectral network on a smooth base. We will discuss this object more rigorously in next section. 
\end{Rmk}

\subsection{The nodal caterpillar limit of spectral networks}

By Section~4.2, the rescaled spectral covers converge to the local covers $\pi^{(k)}:\Gamma^{(k)}\to C^{(k)}$, whose elementary branch cuts are $(k,1),\dots,(k,k-1)$. We now specify the network on this nodal curve and its relation to the original differential equation. In this subsection we use the convention of (\ref{eq: symODE}), with $B=B^{\dagger}$ and fixed $\epsilon>0$. The local scale ratios will be denoted by $t$, independently of the scale notation in Section~2.

\begin{Def} \label{def: nodalSp}
Fix a phase $\vartheta$ which is generic for the local spectral covers. The stable caterpillar spectral network is the collection
\begin{equation}
    \mathcal W_{\rm cat}(\vartheta):=\left(\mathcal W^{(2)}(\vartheta),\dots,\mathcal W^{(n)}(\vartheta);\mathcal N_{\rm neck}\right), \label{eq: structure}
\end{equation}
where $\mathcal W^{(k)}(\vartheta)$ is the GMN spectral network of $\pi^{(k)}:\Gamma^{(k)}\to C^{(k)}$. The datum $\mathcal N_{\rm neck}$ identifies the two branches of each spectral node and their spectral-line fibers. At the neck between $\Gamma^{(k+1)}$ and $\Gamma^{(k)}$, it matches the two copies of each eigenline $L_i^{(k)}$ of $B^{(k)}$. No continuation of wall germs through a node is prescribed.
\end{Def}

Thus the local networks are constructed directly from the limiting spectral covers. The convergence in Section~4.2 implies convergence of their WKB equations on compact subsets away from branch points and punctures. Near a simple branch point, $\lambda_i-\lambda_j=c(z-b)^{1/2}dz+\cdots$, with $c\neq0$, so the three outgoing wall germs also depend continuously on the rescaled data. Consequently, finite wall segments issuing from the corresponding branch points converge as long as they remain in the component and encounter no critical event. This statement does not identify the complete smooth wall set with (\ref{eq: structure}). The relation needed for the Stokes calculation is instead the convergence of regularized transport, which we establish below.

We first consider one step of the right caterpillar tail. After removing the scalar irregular term and rescaling, the original one-sided equation takes the form
\begin{equation}
    \epsilon\frac{dF}{dv}=\left(iM_t-\frac{iB^{(k+1)}}{v}\right)F,\qquad M_t=\diag(tD,1),\qquad B^{(k+1)}=\begin{pmatrix}B^{(k)}&u_k\\ u_k^*&b\end{pmatrix}, \label{eq: cat-local-system}
\end{equation}
where $D$ is a bounded real diagonal matrix and $t\to0^+$. For example, after setting $u_1=0$, one may take $t=u_k/u_{k+1}$, $D=\diag(u_1/u_k,\dots,u_k/u_k)$ and $v=u_{k+1}\zeta$, where $\zeta$ is the original coordinate. The outer coordinate is $z=tv$. Hence $s=v^{-1}$ gives $zs=t$ and $V=z\partial_z-s\partial_s=v\partial_v$.

\begin{Lem} \label{lem: cat-analytic-neck}
Let $V_k=\mathbb C^k\oplus0$ be the common block in (\ref{eq: cat-local-system}). For $0<t<\delta^2$ and sufficiently small $\delta$, let $T_{t,\delta}$ be transport along the positive real collar from $z=\delta$ to $z=t/\delta$. Then
\begin{equation}
    T_{t,\delta}=\diag\left((t/\delta^2)^{-iB^{(k)}/\epsilon},\chi_{t,\delta}\right)+O_\epsilon(\delta),\qquad |\chi_{t,\delta}|=1. \label{eq: cat-neck-estimate}
\end{equation}
In particular, mixing between $V_k$ and its complement tends to zero, and the regularized transport on $V_k$ converges to the identity in the common standard frame.
\end{Lem}

\begin{Prf}
The logarithmic equation is $\epsilon VF=(iM_t/s-iB^{(k+1)})F$. Set
\begin{equation}
    K_t=\begin{pmatrix}0&-(I-tD)^{-1}u_k\\ u_k^*(I-tD)^{-1}&0\end{pmatrix},\qquad H_t(s)=e^{sK_t}.
\end{equation}
Then $K_t^{\dagger}=-K_t$ and $[M_t,K_t]=B^{(k+1)}-\diag(B^{(k)},b)$. The actual analytic gauge $F=H_tY$ therefore gives
\begin{equation}
    \epsilon VY=\left[\diag(izD-iB^{(k)},i/s-ib)+R_t(s)\right]Y,\qquad \|R_t(s)\|\le C_\epsilon s. \label{eq: cat-analytic-block}
\end{equation}
The bound is uniform for small $t$ and bounded $D$. On the positive real collar both $H_t$ and the propagators are unitary. Comparing (\ref{eq: cat-analytic-block}) with $\epsilon VY=\diag(-iB^{(k)},i/s-ib)Y$, the Duhamel formula gives an error bounded by a constant times
\begin{equation}
    \int_{t/\delta}^{\delta}(z+t/z)\frac{dz}{z}=2(\delta-t/\delta)\le2\delta.
\end{equation}
At both endpoints $H_t=I+O(\delta)$, so returning to the original frame proves (\ref{eq: cat-neck-estimate}). The scalar $\chi_{t,\delta}$ is the transport of the complementary scalar equation. This argument controls the actual common block and does not require convergence of a formal block-diagonalizing series.
\end{Prf}

The power in (\ref{eq: cat-neck-estimate}) is expressed between finite collar endpoints. In the tangentially normalized node frames, the common-block solutions have asymptotics $z^{-iB^{(k)}/\epsilon}$ and $s^{iB^{(k)}/\epsilon}$ respectively. Since $zs=t$, the corresponding plumbing factor is
\begin{equation}
    P_k(t)=t^{-iB^{(k)}/\epsilon}=\sum_{i=1}^k t^{-i\mu_i^{(k)}/\epsilon}P_i^{(k)},\qquad z^{-iB^{(k)}/\epsilon}=s^{iB^{(k)}/\epsilon}P_k(t). \label{eq: cat-plumbing-power}
\end{equation}
Here $P_i^{(k)}$ is the spectral projector of $B^{(k)}$, and the logarithms are real on the positive axis; for a reverse transport we just use $P_k(t)^{-1}$. If the chosen spectral-line frames are obtained from the common standard frame by $C_-$ and $C_+$, the raw matching matrix is $M_k(t)=C_+^{-1}P_k(t)C_-+o(1)$. Thus
\begin{equation}
    M_k^{\rm reg}(t):=M_k(t)C_-^{-1}P_k(t)^{-1}C_-\longrightarrow C_+^{-1}C_-=: \mathcal G_k. \label{eq: cat-framed-gluing}
\end{equation}
The frame changes respect the corresponding eigenlines, so $\mathcal G_k$ is the push-forward of their individual gluing isomorphisms. This yields the following corollary:

\begin{Cor} \label{lem: nAconnection}
On the regular locus, the regularized non-Abelian transport through the neck, restricted to the common block $V_k$, converges to the push-forward of the spectral-line gluing datum.
\end{Cor}

Next, we match these node frames with the canonical Stokes frames. For each one-sided system, let $F_0$ be normalized by $F_0(v)v^{iB/\epsilon}\to I$ as $v\to0$, and let $F_+$ be the canonical sectorial solution normalized at infinity, with the logarithm real on the positive axis. We use the connection matrix convention $F_0=F_+C$. The canonical solutions exist also for a repeated leading eigenvalue, with the full residue block retained in the formal exponent; see Section~2.1 of \cite{xu2024regularizedlimitsstokesmatrices}.

\begin{Prop} \label{prop: cat-connection-matching}
Let $C(t)$ be the connection matrix of (\ref{eq: cat-local-system}), $C_E$ that of its inner limit with $M_0=E_{k+1}$, and $C_D$ that of the outer system $\epsilon F'=(iD-iB^{(k)}/z)F$. Write $d_k=\diag((B^{(k)})_{11},\dots,(B^{(k)})_{kk})$. Then
\begin{equation}
    C(t)=\diag\left(t^{-id_k/\epsilon}C_Dt^{iB^{(k)}/\epsilon},1\right)C_E+O_\epsilon(\sqrt t). \label{eq: cat-connection-factorization}
\end{equation}
The error is uniform for bounded $D$ as the lower scale ratios vary within the ordered chamber.
\end{Prop}

\begin{Prf}
All normalized fundamental solutions are unitary on the positive real axis. The difference between the full and inner limiting equations is $it\diag(D,0)$, and their normalizations at zero agree. Hence we have $F_{0,t}(v)=F_{0,E}(v)+O_\epsilon(tv)$ on this axis. At infinity, (\ref{eq: cat-analytic-block}) has remainder $O_\epsilon(v^{-2})$ in the ordinary differential equation. Integrating from the canonical normalization at infinity then gives
\begin{equation}
    F_{+,t}(v)=\diag\left(F_{+,D}(tv)t^{id_k/\epsilon},e^{iv/\epsilon}v^{-ib/\epsilon}\right)+O_\epsilon(v^{-1}).
\end{equation}
Indeed, the comparison matrix has exactly the same formal normalization as $F_{+,t}$ at infinity, and the integral of the remainder is $O_\epsilon(v^{-1})$. Similarly,
\begin{equation}
    F_{+,E}(v)=\diag\left(v^{-iB^{(k)}/\epsilon},e^{iv/\epsilon}v^{-ib/\epsilon}\right)+O_\epsilon(v^{-1}).
\end{equation}
Choose $v=t^{-1/2}$, so that $z=tv=\sqrt t$. Using $F_{+,D}(z)=F_{0,D}(z)C_D^{-1}$ and $F_{0,D}(z)=(I+O(z))z^{-iB^{(k)}/\epsilon}$, we obtain
\begin{equation}
    F_{+,t}(v)=F_{+,E}(v)\diag\left(t^{-iB^{(k)}/\epsilon}C_D^{-1}t^{id_k/\epsilon},1\right)+O_\epsilon(\sqrt t).
\end{equation}
Substituting this and $F_{0,t}=F_{0,E}+O_\epsilon(\sqrt t)$ into $C(t)=F_{+,t}^{-1}F_{0,t}$ proves (\ref{eq: cat-connection-factorization}). The powers and comparison solutions are unitary, so the errors are not amplified by the length of the neck. The estimate retains the entire lower-rank system and therefore applies recursively at subsequent caterpillar levels.
\end{Prf}

Formula (\ref{eq: cat-connection-factorization}) distinguishes the shared-block power $t^{iB^{(k)}/\epsilon}$ from the endpoint power $t^{-id_k/\epsilon}$ caused by the change of scale. Removing both factors leaves the component product $\diag(C_D,1)C_E$. Any remaining overall coordinate scale must also be retained: under $\zeta=av$, $a>0$, the connection matrices satisfy
\begin{equation}
    C_{\zeta}=a^{id_B/\epsilon}C_v a^{-iB/\epsilon},\qquad d_B=\diag(B_{11},\dots,B_{nn}). \label{eq: cat-endpoint-scale}
\end{equation}

\begin{Prop} \label{prop: cat-regularized-transport}
Let $\rho_{\rm cat}=\rho_m\circ\cdots\circ\rho_0$ be a broken path with specified node crossings $e_1,\dots,e_m$, tangential endpoints and componentwise relative homotopy classes. Smooth its node crossings using the fixed plumbing branches above. After removing the plumbing powers and retaining the canonical endpoint frame identifications, its transport has the limit
\begin{equation}
    \mathop{\rm catlim}\mathcal T_\rho^{\rm reg}=\mathcal T_{\rho_m}\mathcal G_{e_m}\mathcal T_{\rho_{m-1}}\cdots\mathcal G_{e_1}\mathcal T_{\rho_0}. \label{eq: cat-transport-limit}
\end{equation}
Each crossing matches the corresponding eigenline channels, and reverse crossings use $\mathcal G_e^{-1}$. In canonical Stokes frames, the one-sided composition gives the caterpillar regularized Stokes data.
\end{Prop}

\begin{Prf}
Cut the path at collar boundaries. On fixed component paths the differential equations converge, so their transports converge. Corollary \ref{lem: nAconnection} identifies the regularized collar transports with $\mathcal G_e$. Taking the scale limits and then shrinking the collars proves (\ref{eq: cat-transport-limit}). All factors are expressed in the same chosen node frames, so changes of an intermediate frame cancel in the composition. Proposition~\ref{prop: cat-connection-matching} gives the matching to canonical solutions when an endpoint is irregular.

For comparison with \cite{xu2024regularizedlimitsstokesmatrices}, divide the original equation by $\epsilon$ and set $\widetilde u=u/\epsilon$, $\widetilde A=-2\pi B/\epsilon$ (we mention that the ODE systems appeared in \cite{xu2024regularizedlimitsstokesmatrices} and \cite{alekseev2024wkbasymptoticsstokesmatrices} differ by an $\epsilon$). With the same scale and logarithm conventions, iteration of (\ref{eq: cat-connection-factorization}) gives the ordered product of elementary connection matrices defining the caterpillar connection matrix in Definition~3.18 of that reference. Its identification with the regularized smooth Stokes data follows from Theorem~1.5 and Definition~3.19 there. In particular, in the ordered chamber the monodromy relation is
\begin{equation}
    S_-^{\rm reg}S_+^{\rm reg}=C_{\rm cat}e^{-2\pi B/\epsilon}C_{\rm cat}^{-1}, \label{eq: cat-Stokes-matching}
\end{equation}
with the remaining scale factors treated as in (\ref{eq: cat-endpoint-scale}). The Stokes factors are recovered by triangular factorization with the prescribed diagonal normalization.
\end{Prf}

\begin{Rmk}
The same construction applies to the left caterpillar tail after inversion of the base coordinate and passage to the unitary eigenbasis of $A$. The residue sign and the direction of each crossing are transformed accordingly. At the middle neck of the $(2,2)$ degeneration, the logarithmic coefficient already has the form $-iB+O(z)+O(s)$, so the collar estimate applies directly with the full common block. Thus nodes carry spectral-line matching and regularized transport data on both tails and at the middle neck. No assertion about continuation of smooth walls through these necks is needed.
\end{Rmk}

\begin{Rmk} \label{Rmk: jointrule}
The GMN joint rule is applied separately on each component. In the $S_+$ convention, oriented primary walls with labels $(k,j)$ have a common source, while their reversed labels have a common target. Within either such family, two walls are not composable under the joint rule \cite{Gaiotto_2013}. This statement is understood in a common local sheet labeling; labels must be transported when crossing a branch cut. In particular, matching eigenline channels at a node does not create a new GMN joint or prescribe a wall on the next component.
\end{Rmk}

Now we assume the exact WKB conjecture \cite{alekseev2024wkbasymptoticsstokesmatrices,Hollands_2020,gaiotto2011wallcrossinghitchinsystemswkb}, or more precisely the exact nonabelianization statement in Conjecture~C.15(1) of \cite{alekseev2024wkbasymptoticsstokesmatrices}. Since we have (\ref{eq: cat-transport-limit}), we apply the exact nonabelianization statement to each component independently. Namely, for each limiting component, we assume
\begin{equation}
\begin{aligned}
     \exists\,\mathcal L_{\rm ab}^{(r)}(\epsilon) \in \operatorname{Loc}_1^{-}(\Gamma^{(r)}),\,\,\, \operatorname{Nab}_{\mathcal W^{(r)}(\vartheta)}
      \bigl(\mathcal L_{\rm ab}^{(r)}(\epsilon)\bigr)
      \simeq \mathcal L_{\nabla^{(r)}(\epsilon)},
\end{aligned}
\end{equation}
with the prescribed boundary flags and endpoint identifications. Here $\mathcal L_{\nabla^{(r)}(\epsilon)}$ denotes the rank-$r$ local system of horizontal sections of the component connection $\nabla^{(r)}(\epsilon)$ on the punctured base $C^{(r)}$.
The rank-one local system $\mathcal L_{\rm ab}^{(r)}(\epsilon)$
lives on the spectral cover away from its branch points and
punctures, with monodromy $-1$ around each branch point, and $\operatorname{Loc}_1^{-}$ denotes rank-one local systems with monodromy $-1$ around the branch points. No WKB asymptotics of its holonomies are assumed here.

At a node, we work in the common tangential block. After identifying the loop orientations, its normalized monodromy is $M_k=\exp(2\pi B^{(k)}/\epsilon)$.
Under our Hermitian regular-locus assumptions, its eigenvalues
$m_i=\exp(2\pi\mu_i^{(k)}/\epsilon)$ are distinct. The prescribed invariant boundary flag therefore has a unique monodromy-invariant splitting, given by
\begin{equation}
    \Pi_i=\prod_{j\neq i}\frac{M_k-m_jI}{m_i-m_j}.
\end{equation}
This identifies its graded Abelian channels with the eigenlines
$L_i^{(k)}$. Fix their scalar normalizations using the tangential
spectral-line frames above, and denote the resulting boundary
identifications by $J_-$ and $J_+$.
Since $\mathcal G_k$ matches the corresponding eigenlines,
\begin{equation}
    J_+^{-1}\mathcal G_kJ_-=\bigoplus_i g_{k,i},
\end{equation}
where $g_{k,i}$ are the spectral-line gluing maps.

For each component path $\eta$, let $\mathcal F_\eta^{\rm fr}$
be its GMN transport expressed in these boundary frames, including
all endpoint changes from the original GMN charts. Local exact
reconstruction then gives
\begin{equation}
    \mathcal T_\eta
    =J_{\rm end}\mathcal F_\eta^{\rm fr}J_{\rm start}^{-1}.
\end{equation}
At external endpoints, retain the prescribed canonical Stokes
frames. Substitution into (\ref{eq: cat-transport-limit}) cancels
the intermediate identifications and yields, in these external
frames,
\begin{equation}
    \mathop{\rm catlim}\mathcal T_\rho^{\rm reg} =
    \mathcal F_{\rho_m}^{\rm fr} \left(\bigoplus_i g_{e_m,i}\right)
    \mathcal F_{\rho_{m-1}}^{\rm fr}\cdots \left(\bigoplus_i g_{e_1,i}\right)
    \mathcal F_{\rho_0}^{\rm fr}. \label{eq: cat-abelian}
\end{equation}
Thus componentwise GMN transport and spectral-line matching
already recover the full regularized transport; no continuation
of wall germs through the nodes is required. All componentwise
lifts below are understood with these endpoint identifications.
This construction does not require a termwise degeneration of
the smooth GMN detours.

The subsequent path-lifting rule uses (\ref{eq: cat-transport-limit}), $(\ref{eq: cat-abelian})$: one first fixes the broken representative of the original path class, then performs the GMN detours within each component and inserts the matching operator at every node. 

\subsection{The path-lifting rule of the caterpillar degenerate Spectral networks}

In this section we study the path lifting rule in the caterpillar limit. Under the caterpillar limit the original spectral curve decomposes into a chain of genus $0$ components $\Gamma^{(j)}$, which are $j$-sheeted covers of the base curve $\mathbb{P}\backslash\{0,\infty  \}$. In the original spectral network, the entry $(S_{\pm})_{i,j}$ corresponds to the path-lifting contribution that starts from sheet $j$ and ends at sheet $i$. However, sheet $j$ does not exist on the components $\Gamma^{(k)}, k<j$, which means we need to modify the path-lifting rule in the Caterpillar limit, or equivalently we need to recognize how a path starting from $\infty^+_{k+1}$ and ending at $\infty^-_{k}$ (denoted by $\rho_{k,k+1}$) behaves in the caterpillar limit(we mention that $+$ and $-$ are inherited labels from the original Stokes arc, not local Stokes sectors on the limiting components). In the caterpillar limit, $\rho_{k,k+1}$ is the skeleton
\begin{equation}
    s^{(k+1)}_{k+1,+} \to s^{(k+1)}_{k} \xrightarrow{\rm neck} p^{(k)}_{k} \to s_{k,-}^{(k)}, \label{eq:brokenSkeleton}
\end{equation}
 here $(s^{(k+1)}_{k},p_{k}^{(k)})$ is a pair of shared punctures. Strictly speaking, it is not a smooth path, but a broken path plus neck gluing datum.

\begin{Lem} \label{lem: brokenpath}
Fix a generic phase $\vartheta$ and the marked Stokes arc $\rho_{t,\vartheta}$, truncated at $p_{\pm}=Re^{i\theta_{\pm}(\vartheta)}$. Let $\mathcal B_t$ be the branch locus. Suppose that the chosen broken representative admits a smoothing $\rho_{t,\vartheta}^{\rm std}$ and endpoint connectors $a_{t,\vartheta},b_{t,\vartheta}$ such that
\begin{equation}
    \widehat\rho_{t,\vartheta}
    =b_{t,\vartheta}\circ\rho_{t,\vartheta}^{\rm std}\circ a_{t,\vartheta}
\end{equation}
and $\rho_{t,\vartheta}$ lie in a common simply connected domain $U_t\subset\mathbb C^{\times}\setminus\mathcal B_t$ and agree near their endpoints. Then $[\widehat\rho_{t,\vartheta}]=[\rho_{t,\vartheta}]$ relative to endpoint neighbourhoods in $\mathbb C^{\times}\setminus\mathcal B_t$. The homotopy preserves the prescribed winding and tangential data and lifts to a homotopy on the spectral cover after fixing the initial sheet. The corresponding non-Abelian transports agree after retaining the endpoint frame identifications.
\end{Lem}

\begin{Prf}
Remove the common initial and terminal segments of the two paths. The remaining paths have the same endpoints in $U_t$. Since $U_t$ is simply connected, they are homotopic relative to these endpoints. Reattaching the common segments gives a homotopy $H$ from $\rho_{t,\vartheta}$ to $\widehat\rho_{t,\vartheta}$, fixed near $p_{\pm}$ and entirely contained in $U_t$. In particular, no intermediate path meets a branch point or the singularity $0$. A logarithm exists on $U_t$, so its endpoint values remain fixed throughout the homotopy. Thus the prescribed winding and tangential data are preserved.

The restriction $\pi^{-1}(U_t)\to U_t$ is an unramified covering. After fixing the initial sheet, the covering homotopy property gives a unique lift of $H$. Since the endpoints of $H$ are fixed and their fibers are discrete, the lifted endpoints remain fixed as well. Hence the lifted paths are relatively homotopic, and no additional nontrivial spectral excursion is introduced. This statement concerns the ordinary lifts before adding wall detours.

By flatness of the original non-Abelian connection,
\begin{equation}
    \mathcal T_{\rho_{t,\vartheta}}
    =\mathcal T_{b_{t,\vartheta}}
     \mathcal T_{\rho_{t,\vartheta}^{\rm std}}
     \mathcal T_{a_{t,\vartheta}}.
\end{equation}
Set $J_+=\mathcal T_{a_{t,\vartheta}}F_+(p_+)$ and $J_-=\mathcal T_{b_{t,\vartheta}}^{-1}F_-(p_-)$. Then
\begin{equation}
    F_-(p_-)^{-1}\mathcal T_{\rho_{t,\vartheta}}F_+(p_+)
    =J_-^{-1}\mathcal T_{\rho_{t,\vartheta}^{\rm std}}J_+.
\end{equation}
The endpoint frame matching and the regularized neck transport are then treated as in Propositions~\ref{prop: cat-connection-matching} and~\ref{prop: cat-regularized-transport}. The argument applies separately to every generic $\vartheta$, with its corresponding marked Stokes directions.
\end{Prf}

\begin{Prop}
Let $\rho$ be the large arc going from $\infty^+$ to $\infty^-$ in the lower half-plane, and assume that the chosen Stokes sector is the $S_+$-sector. In the caterpillar chamber, the dominance order is $1<2<\dots<n$. Then the canonical caterpillar lifts of $\rho$ contributing to the adjacent entry $(S_+)_{k,k+1}$ are strictly supported on the two adjacent components $\Gamma^{(k+1)}\cup \Gamma^{(k)}$. More precisely, there exists exactly one skeleton:
\begin{equation}
    \infty^{(k+1)}_{k+1,+} \to \infty^{(k+1)}_{k} \xrightarrow{\rm neck} 0^{(k)}_{k} \to \infty_{k,-}^{(k)}. \nonumber
\end{equation}
Therefore the path-lifting only supports on the two adjacent components and involves only one neck. \label{prop: sklton}
\end{Prop}

\begin{Prf}
The lifting paths of $\rho$ contributing to the Stokes entry $(S_+)_{k,k+1}$ are precisely those starting from $\infty^{+}_{k+1}$ and ending at $\infty^-_k$. By (\ref{eq: structure}) and the structure of $\Gamma^{(j)}$, there is actually no $k$-th sheet on component $\Gamma^{(j)}$, $j \leq k-1$. We need to decide the limiting points of $\infty^{+}_{k+1}$ and $\infty^-_k$ in the caterpillar limit. The components containing sheet $k+1$ are $\Gamma^{(k+1)}, \dots \Gamma^{(n)}$. For $j \geq k+2$, $\infty^{(j)}_{k+1}$ is actually a nodal branch and is glued to $0^{(j-1)}_{k+1}$. Therefore, $\infty^{(k+1)}_{k+1}$ is the only external infinity of sheet $k+1$. Therefore $\infty^{(k+1)}_{k+1}$ is the limiting point of $\infty_{k+1}$. Similarly, $\infty^{(k)}_k$ is the limiting point of $\infty_{k}$. The original Stokes arc $\rho$ determines a fixed relative homotopy class between its two marked endpoints. By lemma \ref{lem: brokenpath} (the existence is guaranteed, as $\rho$ is the large arc, while the branch cuts are finite), after the specialization of the endpoint lifts, the canonical stable representative is therefore the reduced broken path joining the corresponding points on $\Gamma^{(k+1)}$ and $\Gamma^{(k)}$.

Any excursion through an additional neck gives one of two possibilities. If the excursion is homotopically trivial, it can be contracted and does not change the reduced representative. If it is homotopically nontrivial, then it changes the relative homotopy class and therefore cannot arise from the degeneration of the original arc $\rho$. Hence the canonical stable lift contains no additional neck excursion.

The GMN lifting rule (see \cite{Gaiotto_2013} or Appendix C of \cite{alekseev2024wkbasymptoticsstokesmatrices}) modifies the fixed base path only by local detours at intersections with $S$-walls. In section~4.3, we apply the non-abelianization statement componentwise, therefore all detours remain inside a single component and cannot create an additional transition between different
components. 

It remains to determine the lift on $\Gamma^{(k+1)}\cup\Gamma^{(k)}$. In the $S_+$-sector, the dominance order $1<2<\cdots<n$ fixes the allowed orientation of the $S$-wall detours, so a lift cannot detour along a wall in the opposite direction. Among the primary walls on $\Gamma^{(k+1)}$, only the walls of type $(k+1,k)$ are compatible
with a lift beginning on sheet $k+1$ and reaching $\infty_k^{(k+1)}$, and then glued to $0^{(k)}_k$. 

Finally, the relevant primary walls have a common source or a common
target and are therefore non-composable under the GMN joint rule. Hence
no secondary wall is generated. The canonical contributing skeleton is thus unique and is
\begin{equation}
    \infty^{(k+1)}_{k+1,+} \longrightarrow \infty^{(k+1)}_{k} \xrightarrow{\rm neck}
    0^{(k)}_{k} \longrightarrow \infty^{(k)}_{k,-}.
    \nonumber
\end{equation}
\end{Prf}

We now examine what happens at the neck. In the strict caterpillar limit, the tube/neck is actually a node, or a pair of punctures identified in the nodal curve. In the framework of \cite{Gaiotto_2013}, the Stokes matrix can be written as
\begin{equation}
    (S_+)_{k,k+1}= \langle e_k ,\mathcal{T}_{\rho_{\rm cat}} e_{k+1}\rangle, 
\end{equation}
here $\mathcal{T}_{\rho_{\rm cat}}$ is the parallel transport (of the non-Abelian connection) along the path $\rho_{\rm cat}$, and $\rho_{\rm cat}$ is the caterpillar limiting path of $\rho$. Since $\Gamma^{(k+1)}$ and $\Gamma^{(k)}$ share the neck data, namely the Gelfand-Zeitlin coordinate and the eigenline space 
\begin{equation}
V_k=\bigoplus_{i=1}^k L_i^{(k)}, \,\, L_i^{(k)}=\mathbb C v_i^{(k)} ,
\end{equation}
here $v_{i}^{(k)}$ is the eigenvector corresponding to the eigenvalue $\mu^{(k)}_i$. The transition from $e_{k+1}$ to $e_k$ must be decomposed into three steps
\begin{equation}
    \mathbb{C} e_{k+1} \to V_k\to V_k \to \mathbb{C}e_k. 
\end{equation}
 Write
\begin{equation}
B^{(k+1)}= \begin{pmatrix} B^{(k)} & u_k\\
u_k^\ast & b
\end{pmatrix}.
\end{equation}
The residue coupling into the common block is
$\operatorname{pr}_{V_k}B^{(k+1)}e_{k+1}=u_k$.
In the endpoint-normalization convention used below, the
channel-dependent component connection factors are included
in $g_i^{(k)}$, while the remaining common factor is denoted
by $N_k^{\rm reg}$. Then the first arrow is given by this projection, and the third arrow is given by $v\mapsto \langle e_k,v \rangle$. 

We now explain why the neck contribution can be written in terms of a
gluing operator on the spectral eigenlines. The Stokes matrix is
defined as a matrix coefficient of the non-Abelian parallel transport, while in the stable caterpillar limit, the $\rho_{\rm cat}$ breaks into
componentwise paths together with one crossing of the normalized
neck. The part along the componentwise skeleton will be absorbed into the normalization factor $N_k^{\rm reg}$. The remaining parallel transport across the neck degenerates to the gluing isomorphism of the spectral line bundle between the two normalized branches of the node in the caterpillar limit. By Corollary~\ref{lem: nAconnection}, the regularized neck transport is the gluing operator $\mathcal G_k$ on the shared
eigenline space $V_k$.

Therefore in the strict caterpillar limit the canonical contribution takes the form 
\begin{equation}
    (S_+)_{k,k+1}= N_k^{\rm reg} \langle e_k, \mathcal{G}_k u_k \rangle,
\end{equation}
here $\mathcal{G}_k: V_k\to V_k$ should be viewed as a neck gluing operator, and $N_k$ denotes the regularized common, $i$-independent contribution of the geometric skeleton  and of the chosen endpoint framing. 

In the caterpillar limit, the gluing operator only matches the shared formal channels and does not mix the formal eigenspaces. Therefore $[\mathcal{G}_k,B^{(k)}]=0$ at the neck, and on the regular locus
\begin{equation}
    B^{(k)} =\sum_{i=1}^k  \mu_i^{(k)}P^{(k)}_i,\,\, P^{(k)}_i=v^{(k)}_i(v^{(k)}_i)^*.
\end{equation}
Since $\mathcal{G}^k$ commutes with $B^{(k)}$, 
\begin{equation}
    \mathcal{G}_k =\sum g^{(k)}_i P^{(k)}_i.
\end{equation}
Finally, 
\begin{equation}
    (S_+)_{k,k+1}= N_k^{\rm reg} \sum_{i=1}^k g^{(k)}_i\langle e_k, P^{(k)}_i u_k \rangle. \label{eq: summing}
\end{equation}
Note this summation is over the gluing channels, rather than over different path-liftings. 

\begin{Rmk}
    The scalar $g_i^{(k)}$ is the $i$-dependent endpoint-normalization
factor in the chosen abelianization/framing convention. Changing endpoint
framings multiplies $g_i^{(k)}$ by framing factors, while the common
$i$-independent contribution is absorbed into $N_k^{\rm reg}$. We also emphasize that,
in this convention, the gluing angle is encoded in the phase of the complex
overlap $\langle e_k,P_i^{(k)}u_k\rangle$. Indeed, see Example \ref{Ex: explicit}, equation (\ref{eq: rotation}). 
\end{Rmk}

\begin{Rmk}[Regularized nature of the strict caterpillar lift]
The spectral coordinates appearing in the strict caterpillar
path-lifting rule are regularized limiting coordinates.  Indeed, the
strict nodal curve no longer contains the nonzero plumbing parameters
or the finite lengths of the corresponding cylinders.  The transport
along a smooth neck is replaced by a framed identification of the
spectral-line fibers at the two normalized branches of the node.
Thus the gluing operator $ \mathcal G_k=\sum_i g_i^{(k)}P_i^{(k)} $
and its coefficients $g_i^{(k)}$ encode the finite regularized neck
data. The raw smooth coordinate is recovered only after restoring the
omitted plumbing powers. Hence what we obtain corresponds to the regularized Stokes matrices as defined in \cite{alekseev2024wkbasymptoticsstokesmatrices}. \label{Rmk: regularizednature}
\end{Rmk}

This motivates the following definition of caterpillar path lifting rule: 

\begin{Def} \label{def: CatPathLifting}
Let $\mathcal{W}_{\mathrm{cat}}(\vartheta)$ defined in (\ref{eq: structure}) be a stable caterpillar spectral network. Each pair of adjacent components is associated with a gluing operator $ G_r=\sum_{i=1}^{r}g_i^{(r)}P_i^{(r)}$. 

Let $\rho_{\rm cat}$ be a chosen broken representative of the original path class, with
component paths $\rho_0,\dots,\rho_m$ and node crossings $e_1,\dots,e_m$. On every component, let $\operatorname{Lift}_{\mathcal{W}^{(r)}}$ be the ordinary GMN path-lifting map for the local spectral network
$\mathcal{W}^{(r)}(\vartheta)$. The caterpillar lift of
$\rho_{\mathrm{cat}}$ is defined by applying the local GMN lifting rule
componentwise and inserting the corresponding neck gluing operator
whenever the path crosses a node:
\begin{equation}
    \operatorname{Lift}_{\mathrm{cat}}(\rho_{\mathrm{cat}})  =
    \operatorname{Lift}_{\mathcal{W}^{(r_m)}}(\rho_m) \circ G_{e_m} \circ\cdots\circ
    G_{e_1} \circ \operatorname{Lift}_{\mathcal{W}^{(r_0)}}(\rho_0). \label{eq: componentSeq}
\end{equation}

All compositions at nodes contract the shared eigenline indices. Thus an incoming channel $L_i^-$ is matched only with $L_i^+$, and the full composition sums over all compatible intermediate channels. Reverse crossings use $G_e^{-1}$. Wall detours are generated only within the components; a node contributes its framed gluing map and no additional GMN detour.
\end{Def}

\begin{Rmk}[Path-lifting rule for the (2,2) system] \label{Rmk: 22plr}
    Recall that the $(2,2)$ type spectral curve splits into two spectral curves with only one irregular singularity via sending $t$ to $0$. Moreover, the nature of componentwise nodal spectral networks yields the fact that we need not to consider the S-walls across the neck. Therefore, under the limit $t\to 0$, the path-lifting is taken componentwise with neck gluing operators inserted at the neck. Still, only terms with matching sheet labels on the two sides of every neck are retained. 

    The gluing across the node is an isomorphism between two rank-$n$ fibers, but it is required to preserve the spectral eigenline decomposition. Thus, after choosing right and left normalized eigenvectors near the two punctures, the neck gluing operator is $ \mathcal{G}_{R\leftarrow L}= \sum_{i=1}^{n} g_i\, v_i^{(n)}(v_i^{(n)})^*$. This is a torus gluing datum in the eigenline basis, not an additional factorization coordinate. 
\end{Rmk}

Finally, we mention that although we use the existence of the Abelianized connection in this section, the equation (\ref{eq: summing}) does not require the exact WKB conjecture. The reason is that the two-component statement of \ref{prop: sklton} can be proved alternatively using the non-Abelian connection.

\subsection{Computations of the endpoint normalization factor}

We begin with the following computations:

\begin{Lem} 
Define $c_{i}^{(k)}=\langle e_k, v_{i}^{(k)}\rangle $, here $v_i^{(k)}$ are the unit eigenvectors. For Hermitian/anti-Hermitian $B$, we claim that 
\begin{equation}
    \lvert  c^{(k)}_i \rvert^2= \frac{\prod_{l=1}^{k-1}(\xi_i^{(k)}-\xi^{(k-1)}_l) }{\prod_{j\neq i}(\xi^{(k)}_i-\xi^{(k)}_j)}, 
\end{equation}  
here $\xi^{(k)}_i$ denotes the $i$-th eigenvalue of the $k$-th principal minor of $B$.
\label{Lem: cconst}
\end{Lem}

\begin{Prf}
Consider the diagonal matrix element of the resolvent
\begin{equation}
    R_k(t):= \left\langle e_k,(tI_k-B^{(k)})^{-1}e_k\right\rangle .    
\end{equation}
We compute $R_k(t)$ in two different ways. First, by the cofactor formula for the inverse matrix, the $(k,k)$-entry of $(tI_k-B^{(k)})^{-1}$ is given by the ratio of the $(k,k)$-cofactor to the determinant:
\begin{equation}
R_k(t)=\frac{\det(tI_{k-1}-B^{(k-1)})}{\det(tI_k-B^{(k)})}. 
\end{equation}
Here $B^{(k-1)}$ is precisely the principal minor obtained by deleting the
last row and the last column of $B^{(k)}$. Second, since $B^{(k)}$ is normal (therefore the following derivations also apply to anti-Hermitian matrices, namely $iB$) with simple spectrum, it has an orthonormal eigenbasis $v_1^{(k)},\dots,v_k^{(k)}$. Hence
\begin{equation}
(tI_k-B^{(k)})^{-1}=\sum_{i=1}^k\frac{v_i^{(k)}(v_i^{(k)})^\ast}{t-\xi_i^{(k)}}.
\end{equation}
Taking the matrix element against $e_k$, we get
\begin{equation}
R_k(t)=\sum_{i=1}^k\frac{\left|\langle e_k,v_i^{(k)}\rangle\right|^2}{t-\xi_i^{(k)}}=\sum_{i=1}^k\frac{|c_i^{(k)}|^2}{t-\xi_i^{(k)}}.
\end{equation}
Comparing the residues of the two expressions for $R_k(t)$ at $t=\xi_i^{(k)}$, we obtain
\begin{equation}
|c_i^{(k)}|^2=\operatorname*{Res}_{t=\xi_i^{(k)}}\frac{\det(tI_{k-1}-B^{(k-1)})}{\det(tI_k-B^{(k)})}.
\end{equation}
A straightforward calculation gives
\begin{equation}
    |c_i^{(k)}|^2=\frac{\prod_{l=1}^{k-1}\left(\xi_i^{(k)}-\xi_l^{(k-1)}\right)}{\prod_{j\neq i}\left(\xi_i^{(k)}-\xi_j^{(k)}\right)}.
\end{equation}
This proves the claim.
\end{Prf}

The following Lemma \ref{lem: schendpoint} is explicitly a statement on the Hermitian/anti-Hermitian real slice equipped with its natural Hermitian metric. Upon complexification, one should replace the Hermitian norm-square by the product of the corresponding left and right endpoint evaluations. In the following local calculation, the Schlesinger shift $T_{k,i}$ is local in the sense that all other spectral parameters
held fixed; the external endpoint overlap $\langle e_k,P_i^{(k)}u_k\rangle$ is treated separately. This can be justified as follows: after the Schlesinger shift, the endpoint normalizations are generically changed. Namely, $G^{(k)}_i:=g^{(k)}_iP^{(k)}_i$ is mapped to $G^{(k),\prime}_i: L_{i,-}^{(k),\prime} \to L_{i,+}^{(k),\prime}$. Denote the linear isomorphism by $U_{i,\pm}: L_{i,\pm}^{(k)}  \to L_{i,\pm}^{(k),\prime}$. Therefore what we used to be compared with $G^{(k)}$ is $\widehat{G}^{(k),\prime}_i:U_{i,+}^{-1}G_i^{(k),\prime}U_{i,-}$. If we choose a frame $f_{i,\pm} \in L^{(k)}_{i,\pm}$, $f_{i,\pm}^{\prime}\in L^{(k),\prime}_{i,\pm}$. Then
\begin{equation}
    \widehat{G}^{(k),\prime}_i f_{i,-}= U_{i,+}^{-1}G^{(k),\prime}_i f_{i,-}^{\prime}= U_{i,+}^{-1} g^{(k),\prime}_i f_{i,+}^{\prime}=g^{(k),\prime}_i f_{i,+}. 
\end{equation}
Here $g^{(k),\prime}_i$ is exactly $T_{k,i}g^{(k)}_i$. Therefore, at a generic locus, 
\begin{equation}
    \frac{\langle e_k, \widehat{G}^{(k),\prime}_i  u_k \rangle}{\langle e_k, G^{(k)}_i u_k \rangle}= \frac{T_{k,i}g^{(k)}_i}{g^{(k)}_i}. 
\end{equation}

\begin{Lem} \label{lem: schendpoint}
Let $T_{k,i}: \xi_i^{(k)}/\epsilon \longmapsto \xi_i^{(k)}/\epsilon+1$ ($\epsilon \in \mathbb{R}^{\times}$)
be the elementary Schlesinger shift of the intermediate eigenline
$L_i^{(k)}$. Then under the Hermitian endpoint-normalization prescription introduced in the proof below, the
elementary Schlesinger shift satisfies
\begin{equation}
\frac{T_{k,i}g_i^{(k)}}{g_i^{(k)}} = c_{i,k} |\ell^-_{i,k}|^2 |\ell^+_{i,k}|^2,
\label{eq: endpoint-norm-cocycle}
\end{equation}
where $\ell^-_{i,k}= h(e_k,-)|_{L_i^{(k)}}, \, \ell^+_{i,k} =h(u_k,-)|_{L_i^{(k)}}$, 
and $c_{i,k}$ is $\epsilon$-periodic, determined by the
regularized endpoint trivializations.
\end{Lem}
\begin{Prf}
Let $p_-$ and $p_+$ denote the two normalized branches of the node met
by the canonical caterpillar lift. We treat the two endpoints separately.
For $\sigma\in\{-,+\}$, choose a local coordinate $z_-=z, z_+=v=1/s$, and denote the corresponding endpoint fiber of the intermediate eigenline by $L_{i,\sigma}^{(k)}:= L_i^{(k)}|_{p_\sigma}$. 

Along the $i$-th formal channel, the local formal solution contains the
factor $z_\sigma^{-\xi_i^{(k)}/\epsilon}$. Under the elementary shift $T_{k,i}$, one has
\begin{equation}
z_\sigma^{-\xi_i^{(k)}/\epsilon} \longmapsto z_\sigma^{-(\xi_i^{(k)}+\epsilon)/\epsilon}
= z_\sigma^{-1} z_\sigma^{-\xi_i^{(k)}/\epsilon}.
\end{equation}
Hence the induced change of the local lattice is a length-one elementary
modification, whose quotient is
\begin{equation}
Q_{i,-}^{(k)} := \left( z^{-1}\mathcal O_{p_-}/ \mathcal O_{p_-}
\right) \otimes L_{i,-}^{(k)} \simeq L_{i,-}^{(k)}, \,Q_{i,+}^{(k)} := (\mathcal{O}_{p_+}/s\mathcal{O}_{p_+})\otimes L_{i,+}^{(k)} \simeq L_{i,+}^{(k)}.
\end{equation}
Thus the quotient line associated with each endpoint occurs with
multiplicity one. More precisely, we have two short exact sequences
\begin{equation}
 \begin{aligned}
     &0 \to \mathcal{O}_{p_-} \otimes L_{i,-}^{(k)} \to  z^{-1}\mathcal{O}_{p_-}\otimes L_{i,-}^{(k)} \to Q_{i,-}^{(k)} \to 0, \\
     &0 \to s\mathcal{O}_{p_+} \otimes L^{(k)}_{i,+} \to \mathcal{O}_{p_+} \otimes L^{(k)}_{i,+} \to Q_{i,+}^{(k)} \to 0.
 \end{aligned}   
\end{equation}
Let $D_{\sigma}$ denote the determinant line of endpoint normalization, and $D_{\sigma}^{\prime}$ denote the same determinant line after modification. The above exact sequences then yield $D_-^{\prime} \otimes D^{\vee}_- \simeq Q_{i,-}^{(k)}$ and $D_+^{\prime} \otimes D_+^{\vee} \simeq Q_{i,+}^{(k), \vee}$. The relative determinant line of an oriented gluing from $-$ to $+$ is then $D_G= \Hom (D_-,D_+)=D_+ \otimes D_-^{\vee}$. Therefore
\begin{equation}
    D_G^{\prime} \otimes D_G^{\vee} \simeq (D_+^{\prime} \otimes D_+^{\vee}) \otimes(D_-^{\prime} \otimes D_-^{\vee})^{\vee} \simeq Q_{i,+}^{(k), \vee} \otimes Q_{i,-}^{(k), \vee}. \label{eq: factorMul}
\end{equation}

Let $h$ be the ambient Hermitian metric on $V_k$. Its restriction to
$L_i^{(k)}$ induces Hermitian metrics $h_{i,\sigma}^{(k)}:=h|_{L_{i,\sigma}^{(k)}}$
on the two endpoint fibers. These in turn induce dual Hermitian metrics
on $(L_{i,\sigma}^{(k)})^\vee$. Explicitly, if
$v_{i,\sigma}^{(k)}$ is any unit vector in $L_{i,\sigma}^{(k)}$, then
for $\ell\in(L_{i,\sigma}^{(k)})^\vee$,
\begin{equation}
|\ell|^2_{(h_{i,\sigma}^{(k)})^\vee} := |\ell(v_{i,\sigma}^{(k)})|^2.
\end{equation}
This definition is independent of the chosen unit vector, since another
unit vector differs from it by a phase.

The endpoint framings determine covectors $\ell^-_{i,k}\in (L_{i,-}^{(k)})^\vee,
\, \ell^+_{i,k} \in(L_{i,+}^{(k)})^\vee$, given, under the chosen identifications of the endpoint fibers with
$L_i^{(k)}$, by $ \ell^-_{i,k}(v) = \langle e_k,v\rangle,\, \ell^+_{i,k}(v) =
\langle u_k,v\rangle$. 
We use the convention that the Hermitian pairing $h(-,-)$ is
conjugate-linear in the first argument and linear in the second.

\textbullet\,\textbf{Hermitian endpoint-normalization prescription.}
We use the natural Hermitian structure inherited from the ambient
space $V_k$. The regularized endpoint trivializations are chosen to
have unit Hermitian norm, so their residual ambiguity is unitary.
In this convention, the relative gluing phase is carried by
$\langle e_k,P_i^{(k)}u_k\rangle$, whereas the scalar change of the
endpoint normalization is measured by the induced Hermitian norm on
the endpoint quotient line. This prescription is proved to be compatible with the analytic results of Stokes matrices in Appendix $B$. 

\, \\
Since the elementary modification contributes exactly one copy of
$Q_{i,\sigma}^{(k)}$, the corresponding endpoint evaluation is linear
in $\ell^\sigma_{i,k}$. Under the Hermitian endpoint-normalization
prescription, its scalar contribution is obtained by pairing this
linear factor with its Hermitian conjugate. Hence
\begin{equation}
    C_\sigma = c_{i,\sigma}\, h_{i,\sigma}^{(k),\vee} \bigl(\ell^\sigma_{i,k},\ell^\sigma_{i,k}\bigr) = c_{i,\sigma} \|\ell^\sigma_{i,k}\|^2.
\end{equation}
Because the quotient line occurs with multiplicity one, this
contribution has bidegree $(1,1)$; higher powers would correspond to
additional copies of the quotient line and therefore do not occur.

By the uniqueness of the canonical caterpillar skeleton, each endpoint is
met exactly once. All channel-independent skeleton and transport
factors are absorbed into $N_k^{\mathrm{reg}}$. Therefore no further
$i$-dependent endpoint factor occurs, and by (\ref{eq: factorMul})
\begin{equation}
    \frac{T_{k,i}g_i^{(k)}}{g_i^{(k)}} = c_{i,k} \|\ell^-_{i,k}\|^2\|\ell^+_{i,k}\|^2,
    \qquad c_{i,k} := c_{i,-}c_{i,+}.
\end{equation}

Finally, for a unit eigenvector $v_i^{(k)}$ of $L_i^{(k)}$,
\begin{equation}
\begin{aligned}
    \|\ell^-_{i,k}\|^2 \|\ell^+_{i,k}\|^2 &= |\langle e_k,v_i^{(k)}\rangle|^2
    |\langle v_i^{(k)},u_k\rangle|^2\\
    &= \left| \langle e_k,P_i^{(k)}u_k\rangle \right|^2.
\end{aligned}
\end{equation}
This proves the claim.
\end{Prf}

Also, $c_{i,k}$ may contain a $\epsilon$-periodic factor, so we first quotient such factors out. Then the constant $c_{i,k}$ reflects only the choice of regularized endpoint trivializations. We henceforth choose a shift-compatible framing convention in which $c_{i,k}=1$. For the adjacent entry $(S_+)_{k,k+1}$, the unique caterpillar lift contains a single normalized neck. Hence the residual constant phase of its endpoint cocycle is purely local and can be fixed to one by a shift-compatible choice of endpoint trivializations. For non-adjacent entries, such a simultaneous normalization additionally requires compatibility among the phases associated with different multi-neck lifts; equivalently, the resulting discrete framing cocycle must have trivial holonomy.

In our system (\ref{eq: symODE}) what really appears is $\frac{B}{\epsilon}$. Therefore we use the following data:
\begin{equation}
    B\to \frac{B}{\epsilon},\, \xi^{(k)}_i \to \frac{\xi^{(k)}_i}{\epsilon}, \, u_k\to \frac{u_k}{\epsilon},\, P^{(k)}_i\to P^{(k)}_i,\, e_k \to e_k. 
\end{equation}

\begin{Prop} \label{prop: variantEquation} Let $ T_{k,i}: \frac{\xi_i^{(k)}}{\epsilon}\longmapsto \frac{\xi_i^{(k)}}{\epsilon}+1 $ be the elementary Schlesinger shift ($\epsilon \in \mathbb{R}^{\times}$) of the formal exponent associated with the intermediate Gelfand--Zeitlin eigenline $v_i^{(k)}$. Then 
\begin{equation} 
\frac{T_{k,i}g_i^{(k)}}{g_i^{(k)}}= \pm\frac{ \prod_{l=1}^{k-1}(\frac{\xi_i^{(k)}-\xi_l^{(k-1)}}{\epsilon}) \prod_{l=1}^{k+1}(\frac{\xi_i^{(k)}-\xi_l^{(k+1)}}{\epsilon})}{ \left[\prod_{j\neq i}(\frac{\xi_i^{(k)}-\xi_j^{(k)}}{\epsilon}) \right]^2}, 
\label{eq: schsft} 
e\end{equation} 
for anti-Hermitian, Hermitian $B/\epsilon$ respectively. \label{Prop: schshift} \end{Prop}

\begin{Prf}
It remains to compute these two projector matrix elements. Since
$B^{(k)}$ is normal with simple spectrum, we may take $P_i^{(k)}=v_i^{(k)}(v_i^{(k)})^\ast$.
For the left endpoint, applying Lemma \ref{Lem: cconst} yields
\begin{equation}
    \langle e_k, v^{(k)}_i(v_i^{(k)})^* e_k  \rangle= |\langle e_k, v_i^{(k)} \rangle|^2=\frac{
\prod_{l=1}^{k-1}
\left(\xi_i^{(k)}/\epsilon-\xi_l^{(k-1)}/\epsilon\right)
}{
\prod_{j\neq i}
\left(\xi_i^{(k)}/\epsilon-\xi_j^{(k)}/\epsilon\right)
}.
\end{equation}

For the right endpoint, use the Schur complement formula:
\begin{equation}
\det(tI_{k+1}-B^{(k+1)}/\epsilon) = \det(tI_k-B^{(k)}/\epsilon) \left( t-\frac{b}{\epsilon}-\eta \frac{u_k^\ast}{\epsilon}(tI_k-B^{(k)}/\epsilon)^{-1}\frac{u_k}{\epsilon}
\right),
\end{equation}
here $\eta=1$ for Hermitian $B$, and $\eta=-1$ for anti-Hermitian $B$.
Expanding in the eigenbasis of $B^{(k)}$,
\begin{equation}
\ u_k^\ast(tI_k-B^{(k)}/\epsilon)^{-1}u_k = \sum_{a=1}^k\frac{ \langle u_k,P_a^{(k)}u_k\rangle}{
t-\xi_a^{(k)}/\epsilon }.
\end{equation}
Evaluating the Schur complement at $t=\xi_i^{(k)}/\epsilon$, one obtains
\begin{equation}
\langle u_k/\epsilon,P_i^{(k)}u_k/\epsilon\rangle=-\eta \frac{\det(\frac{\xi_i^{(k)}}{\epsilon}I_{k+1}-\frac{B^{(k+1)}}{\epsilon})
}{\left. \frac{d}{dt}\det(tI_k-\frac{B^{(k)}}{\epsilon})\right|_{t=\xi_i^{(k)}/\epsilon} }.
\end{equation}
Thus
\begin{equation}
\langle u_k/\epsilon,P_i^{(k)}u_k/\epsilon\rangle= -\eta \frac{ \prod_{l=1}^{k+1} \left( \frac{\xi_i^{(k)}-\xi_l^{(k+1)}}{\epsilon}\right) }{ \prod_{j\neq i} \left( \frac{\xi_i^{(k)}-\xi_j^{(k)}}{\epsilon}\right) }.
\end{equation}

Multiplying the two endpoint cocycles gives
\begin{equation}
\frac{T_{k,i}g_i^{(k)}}{g_i^{(k)}} =-\eta
\frac{ \prod_{l=1}^{k-1} \left(\xi_i^{(k)}/\epsilon-\xi_l^{(k-1)/\epsilon}\right) \prod_{l=1}^{k+1}
\left(\xi_i^{(k)}/\epsilon-\xi_l^{(k+1)}/\epsilon\right) }{ \left[\prod_{j\neq i}\left(\xi_i^{(k)}/\epsilon-\xi_j^{(k)}/\epsilon\right) \right]^2}.
\end{equation}
This concludes the proof. 
\end{Prf}

Now we solve the shift equation from (\ref{eq: schsft})
\begin{equation}
    \text{log} \frac{Z(\xi_i^{(k)}/\epsilon+1) }{Z(\xi_{i}^{(k)}/\epsilon)}= \log \frac{ \prod_{l=1}^{k-1} \left( \frac{\xi_i^{(k)}-\xi_l^{(k-1)}}{\epsilon}\right) \prod_{l=1}^{k+1}
\left( \frac{\xi_i^{(k)}-\xi_l^{(k+1)}}{\epsilon}\right) }{ \left[\prod_{j\neq i}\left( \frac{\xi_i^{(k)}-\xi_j^{(k)}}{\epsilon}\right) \right]^2}. \label{eq: exactShft}
\end{equation}
Note the Gamma function satisfies $\Gamma(1+x)=x\Gamma(x)$. We apply this identity to (\ref{eq: exactShft}), as we may write 
\begin{equation}
\begin{aligned}
    \text{log}\,Z\left( \frac{\xi_i^{(k)}}{\epsilon}\right) &=\sum_{l=1}^{k-1}\text{log}\,\Gamma \left( \xi_i^{(k)}/\epsilon-\xi^{(k-1)}_l/\epsilon \right)+\sum_{l=1}^{k+1}\text{log}\, \Gamma\left(\xi_i^{(k)}/\epsilon-\xi^{(k+1)}_l/\epsilon\right) \\&-2\sum_{j\neq i}\text{log} \Gamma\,\left(\xi^{(k)}_i/\epsilon-\xi^{(k)}_j/\epsilon\right) . 
    \end{aligned}
\end{equation}
Therefore
\begin{equation}
    Z^{\text{loc}}_i= \frac{\prod_{l=1}^{k-1}\Gamma(\xi_i^{(k)}/\epsilon-\xi^{(k-1)}_l/\epsilon)\prod_{l=1}^{k+1}\Gamma(\xi_i^{(k)}/\epsilon-\xi^{(k+1)}_l/\epsilon)}{\left[\prod_{j\neq i}\Gamma(\xi^{(k)}_i/\epsilon-\xi^{(k)}_j/\epsilon) \right] ^2}. \label{eq: rew}
\end{equation}

\begin{Thm}  \label{prop:entryFormula}
    The $(k,k+1)$ Stokes matrix entry of the ODE system (\ref{eq: symODE}) in the sector $S_+$ can be written as (up to an $\epsilon$-periodic factor)
    \begin{equation}
        (S_+)_{k,k+1}= N_k^{\rm reg} \sum_{i}^k  \langle e_k, P^{(k)}_i \frac{u_k}{\epsilon} \rangle   \frac{\prod_{l=1}^{k-1}\Gamma(i\mu_i^{(k)}/\epsilon-i\mu^{(k-1)}_l/\epsilon)\prod_{l=1}^{k+1}\Gamma(i\mu_i^{(k)}/\epsilon-i\mu^{(k+1)}_l/\epsilon)}{\left[\prod_{j\neq i}\Gamma(i\mu^{(k)}_i/\epsilon-i\mu^{(k)}_j/\epsilon) \right] ^2}. 
        \label{eq: StokesMatrix}
    \end{equation}
    Here $i\mu^{(k)}_i$ are the eigenvalues of principal minors of $iB$ appearing in \ref{eq: symODE}. 
\end{Thm}

See appendix \ref{app: checkXu} for a consistency check.

\subsection{Choosing the minors}
In this section we assume the exact WKB conjecture. In the GMN spectral networks defined on a smooth Riemann surface(with possible punctures), there is an equivalence of spectral networks. Specifically, the branch cuts can be moved through other branch cuts, namely if one moves a branch cut $B_{12}$ across a branch cut $B_{13}$, then this operation has the effect of changing $B_{13}$ from type $(13)$ to type $(23)$. Under a series of such operations, all the branch cuts of type $(i,k+1)$ can be changed to type $(k,k+1)$. Namely, there are $k$ branch cuts of type $(k,k+1)$ in total (these $k$ branch cuts are labeled by $B_{1,k+1}, B_{2,k+1}, \dots B_{k,k+1}$). In the strict caterpillar limit, the component $\Gamma^{(k+1)}$ contains exactly these $k$ branch cuts. As a result, the path $b_{i,k+1}$ should be thought of going from $\infty_{k+1}^+$ to $\infty_{k}^+$ through the branch cut $B_{ik}$ (with type $(i,k+1)$ and type $(k,k+1)$ after the operations) from component $\Gamma^{(k+1)}$ to $\Gamma^{(k)}$.    

We then prove the strict-caterpillar/nodal analogue of conjecture 4.24 of \cite{alekseev2024wkbasymptoticsstokesmatrices}. 

\begin{Prop}
Fix a generic non-critical phase $\vartheta$, and work in the ANXZ
factorization chart. Let $\alpha_{ij}$, $1\leq j\leq i\leq n-1$,
be the factorization coordinate corresponding to the $j$-th occurrence
of the simple-root factor $M_{i,i+1}$. Let $b_{ij}$ be the elementary
lifted path associated with the corresponding primary wall in the strict
caterpillar spectral network. Then
\begin{equation}
     \alpha_{ij}=X_{b_{ij}} .
\end{equation}
For a lateral critical phase, the full lift has the form
\begin{equation}
    \alpha_{ij}=X_{b_{ij}}+\cdots,
\end{equation}
where the extra terms are supported on local finite-web corrections and
satisfy $\operatorname{Re}Z(\mu)<\operatorname{Re}Z(b_{ij}) $ in the $S_+$-sector.
\end{Prop}

\begin{Prf}
In the ANXZ factorization chart, the branch points are arranged in rows.
The $i$-th row contains $i$ occurrences of the simple-root wall of
type $i<i+1$. The factorization coordinate $\alpha_{ij}$ is, by
definition, the parameter of the $j$-th occurrence of the elementary
factor $M_{i,i+1}$ in the chosen reduced-word factorization of $S_+$.

In the strict caterpillar leading network, the only wall contributing to
this elementary factor is the corresponding primary wall associated to the path $b_{ij}$. Therefore the path-lifting rule gives
\begin{equation}
    M_{i,i+1}(\alpha_{ij}) = M_{i,i+1}(X_{b_{ij}}). \nonumber
\end{equation}
By uniqueness of the reduced-word factorization, the parameters agree:
\begin{equation}
    \alpha_{ij}=X_{b_{ij}}. \nonumber    
\end{equation}

At a critical phase one has to choose one of the two lateral networks.
The additional lifts arise from finite-web corrections. By the WKB
ordering in the $S_+$-sector and proposition C.12 of \cite{alekseev2024wkbasymptoticsstokesmatrices}, each such term has strictly smaller real central charge than the primary detour $b_{ij}$. Hence
\begin{equation}
    \alpha_{ij}=X_{b_{ij}}+\cdots,
\end{equation}
where the omitted terms are exponentially smaller.
\end{Prf}

It is proved in \cite{BERENSTEIN199649} that once we fix a reduced word $\mathbf{i}$, then the factorization coordinates $\alpha_{i}$, $1\leq i \leq N$, where $N=\frac{n(n-1)}{2}$, are determined by the associated reduced word factorization. We mention that this is exactly the number of branch cuts and the maximal length of a Weyl group element and hence the length of a reduced word. From the localization of the branch cuts/the Gelfand Zeitlin pattern, we choose a reduced word
\begin{equation}
    \mathbf{i}^{\rm GZ}= (n-1,n-2, \dots 1 )(n-1, n-2,\dots,2) \dots(n-1,n-2)(n-1).
\end{equation}

Define $B_k=(n-1,\dots,k)$, then $\mathbf{i}^{GZ}=B_1B_2\dots B_{n-1}$. Then via looking at a letter in $r \in B_k$, $s_{n-1}\dots s_{r+1}(e_r-e_{r+1})=e_r-e_n$. Also, since
\begin{equation}
    B_1 \dots B_{k-1}(e_m)=e_{m-k+1}, 
\end{equation}
then letter $r$ in the $k$th block corresponds to the root $ B_1 \dots B_{k-1}(e_r-e_n)=e_{r-k+1}-e_{n-k+1}$. Hence, the $k$th block gives the roots 
\begin{equation}
    e_{n-k}-e_{n-k+1}, \,\, e_{n-k-1}-e_{n-k+1}, \,\,\dots,\,\,e_1-e_{n-k+1}.    
\end{equation}
Therefore for $k=1,2,\dots, n-1$ we obtain the Gelfand-Zeitlin pattern. 
\begin{Ex}
Consider the reduced word $(212)$ compatible with the
Gelfand--Zeitlin pattern. Let
\begin{equation}
    S_+=
    \begin{pmatrix}
        d_1 & s_{12} & s_{13} \\
        0   & d_2    & s_{23} \\
        0   & 0      & d_3
    \end{pmatrix}
    =DU,
    \,
    D=\operatorname{diag}(d_1,d_2,d_3),
    \, U= \begin{pmatrix}
        1 & \dfrac{s_{12}}{d_1} & \dfrac{s_{13}}{d_1} \\
        0 & 1                   & \dfrac{s_{23}}{d_2} \\
        0 & 0                   & 1
    \end{pmatrix}.
\end{equation}
Using the factorization associated with the reduced word $(212)$,
we have
\begin{equation}
\begin{aligned}
    U&=x_2(\alpha_1)x_1(\alpha_2)x_2(\alpha_3)=
    \begin{pmatrix}
        1 & \alpha_2 & \alpha_2\alpha_3\\
        0 & 1        & \alpha_1+\alpha_3\\
        0 & 0        & 1
    \end{pmatrix}.
\end{aligned}
\end{equation}
Therefore, on the factorization chart where $s_{12}\neq0$,
\begin{equation}
    \alpha_2=\frac{s_{12}}{d_1}, \qquad \alpha_3=\frac{s_{13}}{s_{12}},
    \qquad \alpha_1= \frac{s_{12}s_{23}-d_2s_{13}}{d_2s_{12}}.
\end{equation}
\end{Ex}

Since the upper-triangular elementary matrices satisfy $E_{ab}E_{cd}=\delta_{bc}E_{ad}$, the contribution of $(S_+)_{k,k+1}$ only comes from $x_{k,k+1}$. Hence
\begin{equation}
    \prod_i^k x_{k,k+1} (\alpha_{i,k+1}) = I+  \left( \sum_i^{k} \alpha_{i,k+1}   \right) E_{k,k+1}, 
\end{equation}
and in the caterpillar limit we have
\begin{equation}
    (S_+)_{k,k+1} =N_k \sum_i g_i^{(k)}  \langle e_k, P^{(k)}_i u_k \rangle  = d_k\sum^k_i \alpha_{i,k+1}
\end{equation}

\begin{Rmk}
The branch-cut moves considered above are equivalences of spectral networks which preserve the canonical framings at the irregular ends. Therefore the Stokes matrix remains unchanged. The corresponding changes in the open-path description can be understood through closed-cycle spectral coordinates. After identifying the sheet labels and tangential endpoints, let $b$ and $b'$ be two reference lifts, and set $\gamma=b'-b$. With the fixed holonomy conventions, we have $X_{b'}=X_{\gamma}X_b$. If an open framed transport is written as $cX_b$, where $c$ denotes its endpoint normalization, replacing the reference lift by $b'$ requires $c'=cX_{\gamma}^{-1}$. Hence $c'X_{b'}=cX_b$. Thus the closed-cycle factors arising from the change of reference lifts are compensated by the corresponding framing factors. This explains how the open-path and gluing descriptions remain compatible under the branch-cut move, while $(S_+)_{k,k+1}$ is preserved.
\end{Rmk}

Hence each $\alpha_{i,k+1}$ should contain the neck gluing factor (and hence the Gamma functions) associated to the eigenline $L^{(k)}_i$(up to a permutation). These data are used to construct the $1$-loop part of the Nekrasov partition function. 

For a $(2,2)$-type spectral curve, as we have explained in remark \ref{Rmk: 22plr}, the S-walls will not cross the neck. Therefore, we can choose the factorization coordinates piecewise. Under a double caterpillar limit, the factorization coordinates on both sides should admit good asymptotic behaviors.

\section{Physical implication}
We first review the physical interpretations of a $(1,2)$ spectral curve (\ref{eq:intro-two-components}). In the framework of class S theory \cite{Gaiotto_2012}, it corresponds to $4$d $\mathcal{N}=2$ gauge theory with quiver

\begin{tikzpicture}[
    node distance=1.8cm,
    gauge/.style={
        circle,
        draw,
        thick,
        minimum size=1.05cm,
        inner sep=1pt
    },
    flavor/.style={
        rectangle,
        draw,
        thick,
        minimum width=1.25cm,
        minimum height=0.9cm,
        inner sep=2pt
    },
    edge/.style={thick}
]
    \node[flavor] (Un) {$U(n)$};
    \node[gauge, right=of Un] (Unm1) {$U(n-1)$};
    \node[right=of Unm1] (dots) {$\cdots$};
    \node[gauge, right=of dots] (U2) {$U(2)$};
    \node[gauge, right=of U2] (U1) {$U(1)$};

    \draw[edge] (Un) -- (Unm1);
    \draw[edge] (Unm1) -- (dots);
    \draw[edge] (dots) -- (U2);
    \draw[edge] (U2) -- (U1);

    \node[below=0.15cm of Un] {\scriptsize flavor};
    \node[below=0.15cm of Unm1] {\scriptsize gauge};
    \node[below=0.15cm of U2] {\scriptsize gauge};
    \node[below=0.15cm of U1] {\scriptsize gauge};
\end{tikzpicture}

 In the language of class $S$, a long tube on the Riemann surface $C$ corresponds to a gauge group node. And since the spectral curve is a covering of $C$, the tube can be lifted to the spectral curves. In the process of caterpillar separation, there are $n-1$, $n-2$, $\dots$, $1$ tubes when we are doing the first, second, $\dots$ , $n-1$th separation respectively. So basically these tubes give gauge groups $U(n-1)$, $U(n-2)$, $\dots$, $U(1)$. This could also be viewed as a pants decomposition.    

Each edge gives a bifundamental hypermultiplet, and for each gauge node $U(k)$ we may introduce the Coulomb parameters(or the vacuum expectation values of the adjoint scalar)
\begin{equation}
    \vec{a}^{(k)}= (a_1^{(k)}, \dots, a_k^{(k)}),
\end{equation}
and the flavor node $U(n)$ gives mass parameters
\begin{equation}
    \vec{a}^{(n)}= (a_1^{(n)}, \dots, a_n^{(n)}).
\end{equation}
This comes from the fact that we have identified the vanishing cycles with the $A$-cycles of the spectral curve, then eigenvalues are naturally the periods (the central charges) of electric charges. In our case, the periods corresponding to the outer component is $a^{(k)}_i= \frac{1}{2\pi i} \oint \lambda_{\rm SW}=-\mu^{(k)}_i$. For the anti-Hermitian case $iB$, we still use $a^{(k)}_i=-\mu^{(k)}_i$, as we can define $i\hbar=\epsilon$ to compensate the additional imaginary factor. For a bifundamental hypermultiplet connecting $U(k+1)$ and $U(k)$, the central charge has the form
\begin{equation}
    Z_{\text{bifund}}= \lambda^{(k+1)}_{\alpha}-\lambda_{\beta}^{(k)}+m_k, 
\end{equation}
here $m_k$ is the mass parameter. Further if we let $m_k=0$ and using the fact that bifundamental hypermultiplets are BPS, these differences between eigenvalues actually give the masses of bifundamental hypermultiplets. In the rest part of this section, we set $m_k=0$. 

Then we need to clarify the gauge couplings. The first observation is that the $SU(k)$ part of the gauge symmetry is conformal, while the $U(1)$ part is not. For an $SU(k)$ node, the $4$d $\mathcal{N}=2$ one-loop beta function is $\beta=2h^{\vee} (SU(k))-N_f$, 
here $h^{\vee}$ is the dual Coxeter number and $N_f$ denotes the number of flavors. In our case since the node $SU(k)$ are connected to $SU(k-1)$ and $SU(k+1)$, from the point of view of $SU(k)$, $U(k+1)$ contributes $k+1$ fundamental hypermultiplets and $U(k-1)$ contributes $k-1$ fundamental hypermultiplets. Therefore $N_f=2k$, and for $A_l$ type the dual Coxeter number is $l+1$. Hence $\beta=0 \implies \text{Conformal}$.

This reminds us that the non-conformal $U(1)$ parts should be treated separately. In general physics literature $q_k=e^{2\pi i \tau_k}$. The caterpillar limit $z_2, \dots,z_{n-1} \to \infty$ is actually a weak coupling limit. Then we define 
\begin{equation}
    q_k^{SU(k)} = \frac{1}{z_k}. \label{eq: gaugeCoupling}
\end{equation}
Then in the caterpillar limit $z_k\to \infty$, $q_k^{SU(k)} \to 0$ as desired. The $U(1)$ part would contribute an additional factor in the Nekrasov partition function \cite{nekrasov2002seibergwittenprepotentialinstantoncounting} and see Appendix \ref{app: E}. 

As discussed in \cite{alekseev2024wkbasymptoticsstokesmatrices} and in section $2$, we have the decomposition of the form(similarly for the $(2,1)$ type spectral curve)
$\Gamma_{1,2}^{(k)} \to \Gamma^{(k)} \cup \Gamma^{(k-1)}_{1,2} $, with the $U(k- 1)$ node corresponding to the spectral curve $\Gamma^{(k-1)}_{1,2}$, or more precisely to the $(E^{(k-1)}, \Phi^{(k-1)}_{1,2})$ effective rank $k-1$ system. The eigenline channels then decompose with respect to the Higgs field $\Phi^{(k-1)}_{1,2}$. Then $E^{(k-1)}$ admits the decomposition
\begin{equation}
    E^{(k-1)}=\bigoplus_{i=1}^{k-1} L^{(k-1)}_i,
\end{equation}
and therefore the adjoint bundle $\text{End}(E^{(k-1)})$ also admits such decompositions
\begin{equation}
    \text{End}(E^{(k-1)}) = \bigoplus_{i,j} \text{Hom} (L^{(k-1)}_i,L_j^{(k-1)}).
\end{equation}
We also have the coupling channels between the adjacent components, namely
\begin{equation}
    \text{Hom}(E^{(k-1)},E^{(k)})= \bigoplus_{j}^{k-1} \bigoplus_i^k \text{Hom}(L_j^{(k-1)} , L_i^{(k)}).
\end{equation}

Now we take $F:= \text{Hom}(L_b,L_a)$, then the elementary modification at $p$ can be written as the short exact sequence
\begin{equation}
0 \to L_a \to L_a(p):=L_a\otimes\mathcal O(p)
\to i_{p*}\left(L_{a,p}\otimes K_p^{-1}\right)\to 0,
\end{equation}
where $i_p:{p}\hookrightarrow C$ denotes the inclusion.

Define $F^{\prime}= \text{Hom}(L_b, L_a(p))  $, then 
\begin{equation}
    0 \to F \to F^{\prime} \to Q_p \to  0, \label{eq:shortexact}
\end{equation}
here $Q_p= F \otimes(\mathcal{O}(p)/\mathcal{O})= F^{\prime}/F$.  

The $1$-loop part of the Nekrasov partition function, $Z_{\rm 1-loop}$ can be viewed as the Gaussian fluctuation determinant of the vector multiplets and Hypermultiplets in the Coulomb moduli space. Since the gauge group is $U(k), k=1,2 \dots n-1$, the Coulomb moduli are parameterized by $\mu^{(k)}_i$. The Schlesinger shift of $\mu^{(k)}_i$ by $\epsilon$ induces a Hecke modification, which is a special case of the elementary modification, and affect the following channels,
\begin{equation}
    \mathcal{G}^{\rm vir}_{k,i} = \left[ \bigoplus_{j}^{k-1}
 \text{Hom}(L^{(k-1)}_j, L^{(k)}_i)    \right] \oplus \left[\bigoplus_j^{k+1} \text{Hom}(L^{(k+1)}_j,L^{(k)}_i) \right] \ominus 2 \left[ \bigoplus_{j\neq i}^k \text{Hom}(L^{(k)}_j,L^{(k)}_i) \right]. \label{eq: virtual}
 \end{equation}
note that this is exactly the $q\to 1, \mu_e\to 1$ 4d limit of (3.26) in \cite{Kimura_2016}, as the Cartan coefficients in this limit become $(2,-1,-1)$(we take its negative). Denote $F^{(l)}_j:=\text{Hom}(L^{(l)}_j,L^{(k)}_i) $, then each corresponds to a short exact sequence of the form (\ref{eq:shortexact}). Now we calculate the determinant of each factor in (\ref{eq: virtual}). By the additivity of the determinant functor \cite{5f8535f3-14dc-32c9-b9a7-71c41c9f6c11}(realized as tensor product), the fluctuation factor is
\begin{equation}
    \frac{\text{det}\, R\Gamma(F^{\prime})}{\text{det}\, R\Gamma(F)}= \text{det}\, R\Gamma(Q_p) 
\end{equation}
here $\text{det} \,R\Gamma(F)$ is the determinant of cohomology of $F$, namely
\begin{equation}
    \text{det} \,R\Gamma(F) = \text{det} \,H^0(C,F) \otimes \text{det} \,H^1(C,F)^{-1}
\end{equation}
And since $Q_p$ supports only on a point, we have $\text{det} \,R\Gamma(Q_p)=\text{det}\,Q_p$. , we then obtain
\begin{equation}
    \frac{\text{det}\, R\Gamma(F^{\prime})}{\text{det}\, R\Gamma(F)}= \text{det}\, Q_p=Q_p. \label{eq: dlineshift}
\end{equation}
Therefore, $e_T(Q_{j}^{(l)}) = \mu^{(k)}_i-\mu^{(l)}_j+\epsilon$. Combining the above expressions we obtain equivariant Euler class
\begin{equation}
    e_T(\mathcal{G}^{\rm vir}_{k,i} ) = \frac{ \prod_{l=1}^{k-1} \left(\mu_i^{(k)}-\mu_l^{(k-1)}+\epsilon\right) \prod_{l=1}^{k+1}
\left(\mu_i^{(k)}-\mu_l^{(k+1)}+\epsilon\right) }{ \left[\prod_{j\neq i}\left(\mu_i^{(k)}-\mu_j^{(k)}+\epsilon\right) \right]^2},
\end{equation}
which almost coincides with the right hand side of (\ref{eq: exactShft}) up to a shift of $\epsilon$. This is not a problem, as we can re-index the same quotient at $T^{-1}_{k,i} \mu$, so the overall $\epsilon$ shift disappears. From now on, $\mathcal{G}_{k,i}^{\rm vir}$ denotes the re-indexed complex $\mathcal{G}_{k,i}^{\rm vir}(T^{-1}_{k,i} \mu)$.

Now, in the spirit of
\cite{nekrasov2002seibergwittenprepotentialinstantoncounting,nekrasov2003seibergwittentheoryrandompartitions},
the perturbative contribution is associated with the formal
$T^2$-equivariant completion
\begin{equation}
\mathcal{G}^{\rm vir}_{k,i}
\otimes
\mathbb{C}[z_1,z_2].
\end{equation}
Let $x$ denote a bare equivariant weight. Before imposing the self-dual specialization, the centered weight $x$ generates the formal modes
\begin{equation}
x+m\epsilon_1+n\epsilon_2,
\qquad m,n\geq 0.
\end{equation}
Their regularized determinant is encoded by
$\gamma_{\epsilon_1,\epsilon_2}(x;\Lambda)$, which may equivalently
be expressed in terms of the Barnes double-gamma function, up to
regularization-dependent polynomial terms.

On the self-dual locus $\epsilon_1=\epsilon, \, \epsilon_2=-\epsilon$, the formal modes become
\begin{equation}
x+(m-n)\epsilon.
\end{equation}
For fixed $m-n$, the corresponding one-parameter weight has infinite
multiplicity. Consequently, the restriction to the self-dual one-parameter torus
cannot be interpreted as an ordinary weight decomposition with
finite-dimensional weight spaces. The appropriate prescription is to
define the regularized function for generic
$(\epsilon_1,\epsilon_2)$ and then analytically continue it to the
self-dual locus, following the regularized perturbative construction
of
\cite{
nekrasov2003seibergwittentheoryrandompartitions}.

The resulting self-dual regularized function
$\gamma_{\epsilon}(y;\Lambda)$ satisfies
\begin{equation}
\gamma_{\epsilon}(x+\epsilon;\Lambda) + \gamma_{\epsilon}(x-\epsilon;\Lambda)
- 2\gamma_{\epsilon}(x;\Lambda) = \log\left(\frac{x}{\Lambda}\right).
\label{eq:selfdual-pert-difference}
\end{equation}
Equivalently, defining
\begin{equation}
\mathcal{Z}_{\epsilon}(x) := \exp\left( \gamma_{\epsilon}(x;\Lambda) \right),
\end{equation}
we obtain
\begin{equation}
\frac{\mathcal{Z}_{\epsilon}(x+\epsilon)\mathcal{Z}_{\epsilon}(x-\epsilon)}{
\mathcal{Z}_{\epsilon}(x)^2} = \frac{x}{\Lambda}. \label{eq: normalizationfc}
\end{equation}
Thus the equivariant weight $x$ produced by the
elementary modification is the centered second discrete difference
of the logarithm of the regularized perturbative determinant.

More generally, let us choose an orientation of the linear quiver and denote by $\mathcal W_{\rm quiver}$ the corresponding global set of perturbative weights. Each ordered vector root and each bifundamental weight associated with an oriented edge is included exactly once. We define the global perturbative determinant by
\begin{equation}
\log Z_{\rm 1-loop}:= \sum_{w\in\mathcal W_{\rm quiver}} \sigma_w \gamma_{\epsilon}(x_w;\Lambda), \qquad \sigma_w\in\mathbb Z.
\end{equation}
The local virtual character $\mathcal G^{\rm vir}_{k,i}$ should not be regarded as an independent summand of the global fluctuation complex. Rather, it records the weights of the global quiver character which are incident on the variable $\mu_i^{(k)}$. In particular, although the same bifundamental weight appears in the local characters associated with the two endpoints of an edge, the corresponding perturbative factor is included only once in $Z_{\rm 1-loop}$.

Recall that $T_{k,i}$ denotes the Schlesinger shift $ T_{k,i}: \, \mu_i^{(k)}
\longmapsto \mu_i^{(k)}+\epsilon$. For every global weight $x_w$ depending on $\mu_i^{(k)}$, we have $ T_{k,i}x_w = x_w+s_w\epsilon, \, s_w\in\{1,-1\}$. Since the finite-difference equation is centered, the sign $s_w$ does not affect the result. Indeed,
\begin{equation}
\begin{aligned}
\gamma_{\epsilon} \left( x_w+s_w\epsilon;\Lambda \right) + \gamma_{\epsilon} \left(
x_w-s_w\epsilon;\Lambda \right) - 2\gamma_{\epsilon} \left( x_w;\Lambda \right)
\, = \log \left( \frac{x_w}{\Lambda} \right).
\end{aligned}
\end{equation}
It follows that the equivariant Euler class of the local virtual character is obtained as the centered second discrete difference of the global perturbative determinant:
\begin{equation}
\frac{(T_{k,i}Z_{\mathrm{1-loop}}) (T_{k,i}^{-1}Z_{\mathrm{1-loop}})} {Z_{\mathrm{1-loop}}^2} = \Lambda^{ -\operatorname{rk}_{\rm vir} \mathcal G^{\rm vir}_{k,i}}
e_T \left( \mathcal G^{\rm vir}_{k,i} \right).
\end{equation}
Here $\mathcal G^{\rm vir}_{k,i}$ determines the local response of the global perturbative determinant under the shift of $\mu_i^{(k)}$, rather than an independent perturbative factor to be multiplied over all $(k,i)$.

For the virtual character considered above, we have
\begin{equation}
\begin{aligned}
\operatorname{rk}_{\rm vir} \mathcal G^{\rm vir}_{k,i} &= (k-1)+(k+1)-2(k-1) =2.
\end{aligned}
\end{equation}
Therefore the resulting factor $\Lambda^{-2}$ is independent of the Gelfand--Zeitlin variables and may be absorbed into the overall normalization. Hence the equivariant Euler class supplied by the elementary modification determines the centered second discrete derivative of the global perturbative partition function.

\subsection{Nekrasov partition function for $(2,2)$ type theory}

For (2,2) type spectral curve, it corresponds to a quiver gauge theory of the form:
\begin{equation}
    U(1)_{\text{gauge}} -\dots -U(n-1)_{\text{gauge}}-U(n)_{\text{gauge}}-U(n-1)_{\text{gauge}} -\dots - U(1)_{\text{gauge}}. 
\end{equation}
The parameter $t$ should be regarded as gauge-coupling parameter as well. This can be seen from the viewpoint of Seiberg-Witten period. Through calculating the $B$-period directly one can show $a_D \sim \frac{1}{2\pi i} a \text{log}\,t+\text{regular terms}$. Therefore the period matrix $\tau_{\text{gauge}}=\frac{\partial a_D}{\partial a}\sim \frac{1}{2\pi i}\text{log} \,t$, which implies
\begin{equation}
    t\sim \text{exp} \, 2\pi i \tau_{\text{gauge}}. 
\end{equation}
Again, as $t\to0$ (the weak-coupling limit), the $n$ tubes are stretched to a length $-\text{log} \,t$. These $n$ tubes are the lift of a single tube on the base surface $C$, and correspond to a $U(n)$ gauge group. And when $t\to 0$, the tubes eventually pinch off, leaving two punctures on the two components after separation. Both punctures have $n$ preimages on the spectral curve, so there is a $U(n)$ flavor symmetry group arising on each side. Physically, the $U(n)$ gauge group is ungauged. 

The picture should look like:
\begin{figure}[htbp]
\centering
\resizebox{0.9\linewidth}{!}{
\begin{tikzpicture}[
    gauge/.style={
        circle,
        draw,
        thick,
        minimum size=8mm,
        inner sep=1pt,
        font=\scriptsize
    },
    flavor/.style={
        rectangle,
        draw,
        thick,
        minimum width=9mm,
        minimum height=8mm,
        inner sep=2pt,
        font=\scriptsize
    },
    dots/.style={
        font=\scriptsize
    },
    every node/.style={align=center}
]

\node[gauge] (A1) at (-4.8,1.4) {\(U(1)\)};
\node[dots]  (Ad1) at (-3.6,1.4) {\(\cdots\)};
\node[gauge] (A2) at (-2.2,1.4) {\(U(n\!-\!1)\)};
\node[gauge] (A3) at (0,1.4) {\(U(n)\)};
\node[gauge] (A4) at (2.2,1.4) {\(U(n\!-\!1)\)};
\node[dots]  (Ad2) at (3.6,1.4) {\(\cdots\)};
\node[gauge] (A5) at (4.8,1.4) {\(U(1)\)};

\draw[thick] (A1) -- (Ad1) -- (A2) -- (A3) -- (A4) -- (Ad2) -- (A5);

\draw[thick,->] (0,0.75) -- (0,0.05);

\node[gauge]  (B1) at (-4.8,-1.0) {\(U(1)\)};
\node[dots]   (Bd1) at (-3.6,-1.0) {\(\cdots\)};
\node[gauge]  (B2) at (-2.2,-1.0) {\(U(n\!-\!1)\)};
\node[flavor] (BF1) at (-0.55,-1.0) {\(U(n)\)};

\draw[thick] (B1) -- (Bd1) -- (B2) -- (BF1);

\node[dots] at (0.55,-1.0) {\(\sqcup\)};

\node[flavor] (BF2) at (1.65,-1.0) {\(U(n)\)};
\node[gauge]  (B3) at (3.0,-1.0) {\(U(n\!-\!1)\)};
\node[dots]   (Bd2) at (4.15,-1.0) {\(\cdots\)};
\node[gauge]  (B4) at (5.35,-1.0) {\(U(1)\)};

\draw[thick] (BF2) -- (B3) -- (Bd2) -- (B4);

\end{tikzpicture}
}
\end{figure}

Similar to the argument we made before, the central $U(n)$ gauge node is not conformal. To be more precise, the $SU(n)$ part of the middle node is asymptotically free, while the $U(1)$ part is neither conformal nor asymptotically free. This can be seen from the fact that 
\begin{equation}
    \beta= 2n-N_{f}=2n-2(n-1)=2>0. 
\end{equation}
The $U(1)$-factor is an IR free node. Accordingly, we should introduce a cutoff $\Lambda$ for the $U(n)$ node. 

We mention that the gluing process has a natural explanation on the double Gelfand-Zeitlin side, see proposition \ref{Prop: WeylMeasureDH} and remark \ref{Rmk: DHmeasure}. Now we explain this in more detail. 

\textbullet ,\textbf{Quiver interpretation of the gluing measure}.
Recovering the original $(2,2)$ theory amounts to gauging the diagonal
subgroup of these two flavor symmetries. Equivalently, one identifies
their Cartan parameters with the Coulomb parameter $a$ of the middle
$U(n)$ vector multiplet and integrates over $a$ modulo the Weyl group.

Schematically, the gluing takes the form
\begin{equation}
Z_{(2,2)}(t) = \int_{\mathfrak t/W} d\mu_{\rm glue}(a)\, Z_L(a)\, Z_{\rm mid}(a;t)\,
Z_R(a^*), \label{eq: schematicForm}
\end{equation}
where $Z_L$ and $Z_R$ are the contributions of the two tails and
$Z_{\rm mid}(a;t)$ is the contribution of the middle vector multiplet.
The measure required by diagonal gauging is the Weyl measure
\begin{equation}
d\mu_{\rm glue}(a) \propto \Delta(a)^2\,da.
\end{equation}
Proposition \ref{Prop: WeylMeasureDH} shows that precisely this measure
is obtained by pushing forward the Liouville measure of the double
Gelfand--Zeitlin system. Thus the double Gelfand--Zeitlin
Duistermaat--Heckman measure has the direct physical interpretation as
the classical gluing measure for the middle $U(n)$ gauge node.

The plumbing parameter $t$ does not belong to this measure. It controls
the propagation through the neck, and hence enters the classical factors contained in $Z_{\rm mid}(a;t)$. In particular, $t\to0$ suppresses the gauged interaction and restores the two separate $U(n)$ flavor symmetries.

\begin{Rmk}
The same gluing interpretation persists after quantization. The
double Gelfand--Zeitlin polarization gives
\begin{equation}
Q\left(T^*U(n)\right) = \bigoplus_{\lambda} V_\lambda\otimes V_\lambda^*.
\end{equation}
Thus the neck carries the same intermediate $U(n)$ representation
$\lambda$ on the two sides, with the opposite orientation producing
the dual representation $V_\lambda^*$. Gauging the diagonal symmetry
amounts to pairing the left and right factors and projecting onto
diagonal $U(n)$ invariants. The classical density
$\Delta(a)^2\,da$ is the Duistermaat--Heckman, or semiclassical, limit
of this representation-theoretic gluing rule.
\end{Rmk}

\subsection{Relation to the isomonodromy $\tau$ function}

The Nekrasov partition function appearing in the gauge-theoretic construction above is already established and will not be treated conjecturally here. The only additional assumption concerns its identification with the coefficient normalization in a Kyiv-type Fourier expansion of the isomonodromic tau function. Such Fourier expansions are well established in a number of Painlevé and isomonodromic settings \cite{Gamayun_2012,Iorgov_2014,Gamayun_2013,Its_2014,Gavrylenko_2018}, but the absolute normalization relevant to the present higher-rank irregular problem is not fixed by these results.

Let $C_{\rm pert}(a)$ denote the perturbative coefficient in such a Fourier expansion, and let $Z_{\rm pert}(a)$ be the perturbative part of the Nekrasov partition function obtained above. For $\nu_{ij}=h_i-h_j$, we conjecture that the canonical normalization of the tau function can be chosen compatibly with the factorization coordinates so that
\begin{equation}
\frac{C_{\rm pert}(a+\epsilon\nu_{ij})}{C_{\rm pert}(a)} =
\widehat\alpha_{ij}(a) = \frac{Z_{\rm pert}(a+\epsilon\nu_{ij})}{Z_{\rm pert}(a)}.
\label{eq: conjectural-tau-factorization}
\end{equation}
Here $\widehat\alpha_{ij}$ denotes the factorization coordinate after removal of the prescribed classical, endpoint and Fourier-phase factors. The factorization coordinates themselves are determined by the Stokes and spectral-network analysis discussed above; see in particular \cite{alekseev2024wkbasymptoticsstokesmatrices}. The second equality in (\ref{eq: conjectural-tau-factorization}) follows from the gauge-theoretic construction developed above. The conjectural content is only the first equality, namely the identification of the same discrete ratio with the perturbative Fourier coefficient of the isomonodromic tau function. This distinction is relevant because the monodromy-dependent normalization of the tau function is an additional datum \cite{Bertola_2009,bertola2020corrigendumdependencemonodromydata,Its_2018}.

Assuming (\ref{eq: conjectural-tau-factorization}), the perturbative tau coefficient is determined by our factorization formulas through finite-difference equations on the root lattice. For the simple roots $\alpha_k=h_k-h_{k+1}$, set $R_k(a)=\widehat\alpha_{k,k+1}(a)$. Then
\begin{equation}
C_{\rm pert}(a+\epsilon\alpha_k) = R_k(a)C_{\rm pert}(a),
\,\, k=1,\ldots,n-1.
\label{eq: perturbative-difference}
\end{equation}
Equivalently, $\Delta_k\log C_{\rm pert}=\log R_k$, where $\Delta_kF(a)=F(a+\epsilon\alpha_k)-F(a)$. The compatibility condition is
\begin{equation}
R_k(a+\epsilon\alpha_\ell)R_\ell(a) = R_\ell(a+\epsilon\alpha_k)R_k(a).
\label{eq: perturbative-flatness}
\end{equation}
The explicit gauge-theoretic solution implies the compatibility condition identically. Consequently, any other nonzero solution differs from it by a function periodic under all shifts in the generated lattice.

Thus, conditional only on the tau-function identification in (\ref{eq: conjectural-tau-factorization}), the factorization coordinates computed in this paper give the discrete logarithmic derivatives of the perturbative Fourier coefficient. In particular, the explicit Nekrasov expression obtained from the gauge-theoretic construction provides a solution of these finite-difference equations; the remaining conjecture identifies this solution with the perturbative Fourier coefficient of the isomonodromic tau function, up to a factor periodic under the specified sector shifts. Fixing this residual ambiguity requires an additional normalization convention.

In rank two, the resulting expression agrees with the Painlevé III Kyiv coefficient under the parameter identification given in Appendix~\ref{app:PIII_Kiev_check}. This agreement provides an explicit consistency check of the proposed identification. The general-rank identification remains conjectural.

\subsection{The isomonodromy $\tau$ functions of (1,2) system}

If we are considering a rank-2 $(1,2)$ system, say
\begin{equation}
    \epsilon \frac{dF}{dz}=  \left( iu- i\frac{B}{z} \right)F, \label{eq: 541}
\end{equation}
the genus is then $\frac{(2-2)(2-1)}{2}=0$. Since we are considering the system (\ref{eq: 541}), $\frac{u_2-u_1}{\epsilon}$ is a natural choice of the time $Q$. In this case there is no compact action-angle pair and hence no nontrivial Coulomb-shift lattice. Therefore, we do not expect that the isomonodromy tau function admits a summation like the Kiev type formula \cite{Gamayun_2012}, \cite{Iorgov_2014}. Instead, it should be a single term
\begin{equation}
    \tau=C_{\text{mono}} \times f(Q).
\end{equation}
Here $C_{\text{mono}}$ is a constant depending on the factorization coordinates/Stokes entries and $f(Q)$ is a function of $Q$. Then $C_{\text{mono}}$ should satisfy the shift equation of the Gamma function part, which cannot be determined from solving JMU tau alone. Up to a normalization, the Stokes matrix entry (which equals the factorization coordinates for $n=2$) should be written in (see appendix \ref{app: checkXu} and (\ref{eq: StokesMatrix}))
\begin{equation}
    (S_+)_{12} = \frac{(\frac{u_2-u_1}{\epsilon})^{-\frac{t_2-t_1}{ i \epsilon}} }{\Gamma\left(1-\frac{\mu_1-t_1}{ i \epsilon}\right)\Gamma\left(1-\frac{\mu_2-t_1}{ i \epsilon}\right)},
\end{equation}
where we just defined $t_i=B_{ii}$, so $t_1+t_2=\mu_1+\mu_2$. The correspondence is as follows
\begin{equation}
    \sigma=\frac{t_1-t_2}{2}, m=\frac{\mu_1-\mu_2}{2},  \, M_i=\mu_i-t_1, \,\hbar=-i \epsilon. \label{eq: parameters}
\end{equation}
Then
\begin{equation}
    (S_+)_{12}=Q^{-\frac{2\sigma}{\hbar}} \frac{1}{\Gamma\left(1+\frac{M_1}{\hbar}\right)\Gamma\left(1+\frac{M_2}{\hbar}\right)},
\end{equation}

Then we formulate the shift equation. The shift direction is $t_1 \to t_1-\hbar$, so we write
\begin{equation}
    \frac{C_{\rm mono}(t_1)}{C_{\rm mono}(t_1-\hbar)}= \frac{1 }{\Gamma\left(1+\frac{\mu_1-t_1}{\hbar}\right)\Gamma\left(1+\frac{\mu_2-t_1}{ \hbar}\right)}. \label{eq: shift1}
\end{equation}
Let $x_i=\frac{\mu_i-t_1}{\hbar}$, then as $t_1\to t_1-\hbar$, $x_i \to x_i+1$. The solution to (\ref{eq: shift1}) is thus
\begin{equation}
    C_{\rm mono}(t_1)= G(1+x_1)G(1+x_2).
\end{equation}
This can be checked directly
\begin{equation}
    \frac{C_{\rm mono}(t_1)}{C_{\rm mono}(t_1-\hbar)}= \prod_{i=1}^2 \frac{G(1+x_i)}{G(2+x_i)} = \prod_{i=1}^2 \frac{1}{\Gamma(1+x_i)}
\end{equation}
as desired. 

We can compute directly the Jimbo-Miwa-Ueno tau function(JMU tau) in Appendix \ref{app: D}:
\begin{equation}
    \tau(Q)= C Q^{-\frac{M_1M_2}{\hbar^2}},
\end{equation}
where $C$ is a constant determined from the Stokes data. To be more precise, as we just assumed in section~5.2 it should satisfy the Stokes shift equation, we identify $C$ with $C_{\text{mono}}$ up to a normalization factor. 

We also mention that the second term matches with $U(1)$ partition function under an identification $1-q=cQ$ as we compute in Appendix \ref{app: E}. As we just mentioned before, the rank $2$, $(1,2)$ system has genus $0$. Therefore the Coulomb moduli space has dimension $0$. In fact, there are no non-abelian instantons in our case, the U(1) partition function actually gives the whole partition function of the Abelian sector. Therefore it naturally equals the $Q$-dependent part of the isomonodromy $\tau$ function.

\subsection{The rank $2$, (2,2) isomonodromy tau-function}

The rank $2$ spectral curve $\Gamma_{2,2}^{(2)}$ has genus $1$. Therefore it has one Coulomb parameter and one angle coordinate $\theta$. We should expect that the rank $2$, (2,2) system should produce a Kiev type formula:
\begin{equation}
    \tau_{(2,2)}= \sum_{l}Z_{U(1)}e^{2\pi i \theta l} Z_{\text{cl}} (a_{l})C_{\text{mono}}^{\text{in}}(\sigma_{\rm in}(a_l))C_{\text{mono}}^{\text{out}}(\sigma_{\rm out}(a_l)) Z^{\text{vec}}_{\text{FP}} \Phi_{\text{inst}}(a_l), \label{eq: 22isotau}
\end{equation}
here $a_l=a+\hbar l\nu$ here $\nu$ denotes a shift direction of the Fourier lattice, say the root $e_i-e_j$. The term 
\begin{equation}
    Z_{U(1),L}:= Q^{-\frac{M_1M_2}{\hbar^2}}
\end{equation}
should be regarded as the contribution of the $U(1)$ normalization. In the weak-coupling limit of $(2,2)$ system, there are two $U(1)$ nodes, therefore two copies of the $U(1)$ normalizations arise. Therefore we write $Z_{U(1)}=Z_{U(1),L}Z_{U(1),R}$. 

Now, $t$ is the genuine weak-coupling parameter. Therefore we should expect that the shift equation of $t$-related parts in the Stokes factors should provide a classical-like part. We solve the half shift equation(or the symmetry difference equation)
\begin{equation}
    \frac{Z_{\text{cl}}(a+\frac{1}{2}\hbar\nu,t)}{Z_{\text{cl}}(a-\frac{1}{2}\hbar\nu;t)}=X_{\nu}^{\text{neck}} \label{eq:symdiff}.
\end{equation}
Since the rank is $2$ there is only one direction $e_1-e_2$. The neck part is given by
\begin{equation}
\begin{aligned}
    X_{ij}^{\text{neck}} &\sim \exp \left( \frac{1}{i\hbar} \int_{\gamma_{ij}^{\text{neck}}} (\lambda_i-\lambda_j) -\int_{\gamma_{ij}^{\text{neck}}} \alpha\right)     =\exp \left( -\frac{1}{\hbar} (\mu^{(2)}_i-\mu^{(2)}_j)\log t -\int_{\gamma_{ij}^{\text{neck}}} \alpha  \right) \\
    &= t^{-\frac{\mu^{(2)}_i-\mu^{(2)}_j}{\hbar}} \exp \left(- \int_{\gamma_{ij}^{\text{neck}}} \alpha \right) 
    = t^{\frac{(a^{(2)},\nu)}{\hbar}}\exp \left(- \int_{\gamma_{ij}^{\text{neck}}} \alpha \right), 
\end{aligned} \label{eq: neckcontribution}
\end{equation}
here $\nu=e_1-e_2$, and $\exp \left( -\int_{\gamma_{ij}^{\text{neck}}}\alpha \right)$ only gives constant term (containing no $t$-dependence) related to the frame $G$ of $A$.  
If we want to use the standard Nekrasov normalization (see Appendix B.2 of \cite{Alday_2010})
\begin{equation}
    Z_{\rm cl}^{\rm std}(a;t)=t^{\frac{(a,a)}{2\hbar^2}}, \label{eq:45}
\end{equation}
then we must use the symmetry difference equation (\ref{eq:symdiff}), which provides exactly (\ref{eq:45}). 

The vector factor has a direct interpretation in the gluing process or, as we have discussed in section $5.1$ and section $3.2$, a interpretation in the DGZ system.
After cutting the full system along the internal neck, the two
components define wavefunctions
$\Psi_{\rm in}(a)$ and $\Psi_{\rm out}(a)$ in the Hilbert space
associated with the common $U(2)$ boundary. Reconstructing the full
system amounts to gauging this common boundary symmetry and hence to
contracting the two wavefunctions with the inverse quantum metric of
the intermediate vector channel (\ref{eq: schematicForm}):
\begin{equation}
Z_{\rm glued} = \int_{\mathfrak t/W} d\mu_{\rm glue}(a)\, \Psi_{\rm out}(a) \Psi_{\rm in}(a).
\end{equation}

At a fixed Cartan value $a$, the off-diagonal vector and ghost modes
decompose into oriented root channels. For a root $\nu$, let $x_\nu=\frac{(a,\nu)}{\hbar}$. The local contribution of the corresponding normal fluctuation
complex to the inverse gluing metric is the inverse Euler factor
$x_\nu^{-1}$. The first discrete primitive,
\begin{equation}
\frac{K_\nu(x_\nu+1)}{K_\nu(x_\nu)}
=
\frac{1}{x_\nu},
\end{equation}
incorporates the tower of modes propagating along the neck and gives $K_\nu(x_\nu)=\Gamma(x_\nu)^{-1}$. This Gamma function is the transition kernel between two adjacent root-charge sectors. The quantity entering the sewing formula is,
however, the normalization of the intermediate sector itself.
It is therefore obtained by a second discrete primitive,
\begin{equation}
\frac{Z_\nu(x_\nu)}{Z_\nu(x_\nu-1)}
=
K_\nu(x_\nu),
\end{equation}
whose standard solution is $ Z_\nu(x_\nu) =G(1+x_\nu)^{-1}$. Consequently, the quantum inverse metric of the vector channel is
\begin{equation}
Z_{\rm vec}^{\rm neck}(a) = \prod_{\nu\in\Delta}
G\left(1+\frac{(a,\nu)}{\hbar}\right)^{-1}. \label{eq: middleGlue}
\end{equation}
The two discrete primitives successively lift the local inverse
Euler factor of the single neck vector/ghost complex first to its
one-step propagation kernel and then to the absolute sewing
normalization of the intermediate state.

Therefore, the isomonodromic tau function can be written as
\begin{equation}
    \tau_{(2,2)}= \sum_{l}Z_{U(1)}e^{2\pi i \theta l} Z_{\text{cl}} (a_{l})Z_{\rm pert}(a_l) \Phi_{\text{inst}}(a_l), \label{eq: isotaualt}
\end{equation}
by defining $Z_{\rm pert}(a)=C_{\text{mono}}^{\text{in}}(a_l^{\rm in})C_{\text{mono}}^{\text{out}}(a_l^{\rm out}) Z^{\text{neck}}_{\text{vec}}(a_l) $. For a consistency check, see Appendix \ref{app:PIII_Kiev_check}.

From the physical point of view, the instanton expansion does not originate from the classical spectral network or from the spectral line bundle data alone. These data determine the semiclassical gluing parameters, the Coulomb modulus $(a)$, the angle coordinate, and the perturbative monodromy factors. The non-perturbative instanton sector appears only after the middle $U(2)$ symmetry is gauged. In the $\Omega$-background, the path integral of this gauged $U(2)$ node localizes to fixed points labelled by pairs of Young diagrams $Y=(Y_1,Y_2)$. Therefore the Young-diagram sum should be understood as the equivariant instanton sum of the internal gauge node, while the two genus-zero components contribute only the left and right wavefunctions coupled to this internal channel.

\begin{equation}
     \Phi_{\rm inst}(a)=\mathcal{Z}_{(2,2)}^{\rm glued, rank2}= \sum_{Y} \Lambda^{2 \lvert Y \rvert} \mathcal{Z}_{\rm vec} (a,Y) \Psi_{\rm out}(a,Y) \Psi_{\rm in}(a,Y).
\end{equation}

\begin{Rmk}
    The parameter $t$ should be interpreted as the weak-coupling, or plumbing, parameter of the internal gauge channel. From the four-dimensional gauge-theoretic point of view, this parameter is identified with the instanton counting parameter of the asymptotically free $SU(2)$ node. More precisely, for an asymptotically free $SU(2)$ theory the natural instanton counting parameter is $\Lambda^{\beta}$, where $\beta$ is the one-loop beta-function coefficient. Therefore the plumbing parameter $t$ should be identified, up to a normalization constant, with $\Lambda^2$ (as $\beta=2$ in our case): $t=c\Lambda^2$. The constant $c$ depends on the normalization of the irregular types, the $U(1)$ factor, and the choice of normalization of the left and right wavefunctions, and can be absorbed into these conventions. With this identification, the classical neck contribution $Z_{\rm cl}(a;t)\sim t^{\frac{a^2}{\hbar^2}}$ takes the standard asymptotically free form
\begin{equation}
Z_{\rm cl}(a;\Lambda)\sim \left(\Lambda^2\right)^{\frac{a^2}{\hbar^2}},
\end{equation}
and the instanton expansion $\sum_{\vec Y}\Lambda^{2|\vec Y|}(\cdots)$
can equivalently be written as an expansion in powers of the plumbing parameter $t$.
\end{Rmk}

\begin{Rmk}
    For rank $2$, the $(2,2)$ system has two irregular singularities of Poincaré rank one, located at $0$ and $\infty$. Hence, after passing to the associated $SL_2$-oper, it belongs to the Painlevé III type class. The corresponding isomonodromy tau function admits the standard $c=1$ irregular conformal block expansion \cite{Jimbo:1981tov}, \cite{gaiotto2009asymptoticallyfreen2theories}, \cite{Its_2014}, \cite{Bershtein_2017}. 

The full tau function is obtained by the Fourier completion over the shifted internal momentum,
\begin{equation}
\tau_{(2,2)}(t)= \sum_{l\in\mathbb Z} e^{2\pi i l\theta} Z_{U(1)} Z_{\rm cl}(a_l;t)
Z_{\rm pert}(a_l) \Phi_{\rm inst}(a_l;t), \qquad a_l=a+\hbar l\nu .
\end{equation}
Here $Z_{\rm pert}(a) = C_{\rm mono}^{\rm in}(a) C_{\rm mono}^{\rm out}(a) Z_{\rm vec}^{\rm neck}(a)$. This is the standard Painlevé III/$c=1$ irregular conformal block representation of the isomonodromy tau function, written in the normalization adapted to the double-caterpillar decomposition. We shall not rederive the general CFT proof here; the essential point for our purposes is the identification of the Fourier parameter $a$, its dual angle $\theta$, and the perturbative factors with the spectral-network and Stokes data constructed above.
\end{Rmk}

\subsection{General n, (2,1)-system}
As discussed in Section~5.2, we assume that the normalized factorization coordinates determine the local finite-difference responses of the perturbative determinant. The local factors $\mathcal R_{k,i}$ should not be regarded as independent cocycles whose primitives are subsequently multiplied over all $(k,i)$. Indeed, the bifundamental contribution associated with an adjacent pair of gauge nodes appears in the local shift equations at both endpoints of the corresponding edge, whereas the global quiver partition function contains only one bifundamental determinant for each oriented edge.

We therefore choose an orientation of the linear quiver and first define the global set of perturbative weights. Each ordered root of a gauge node and each bifundamental weight associated with an oriented edge is included exactly once. Let $ \mathcal M_{\rm quiver} = \mathcal M_{\rm vec} \sqcup \mathcal M_{\rm bif}$, and assign the multiplicities $\sigma_w =\pm 1$ for $w \in \mathcal{M}_{\rm bif}, \mathcal{M}_{\rm vec}$ respectively. For a dimensional equivariant weight $w$, define the corresponding dimensionless variable by $x_w=\frac{w}{\hbar}$. Then  the Barnes factor associated with the centered second primitive of the equivariant weight $w$ is 
\begin{equation}
G\left( 1+\frac{w}{\hbar} \right).
\end{equation}
The appearance of the argument $1+w/\hbar$ follows directly from the two consecutive first differences entering the centered second difference, rather than from a choice of the origin of the Schlesinger lattice.

For the linear quiver associated with the $(2,1)$ system, the vector weights are
$ w_{\alpha\beta}^{(k)} = \mu_\alpha^{(k)}-\mu_\beta^{(k)}, \, \alpha\neq\beta$, 
while, after choosing the orientation $(k-1)\to k$, the bifundamental weights are
$w_{\alpha\beta}^{(k,k-1)} = \mu_\alpha^{(k)} - \mu_\beta^{(k-1)}$. Hence, up to the elementary $\Lambda$-dependent normalization arising from (\ref{eq: normalizationfc}), the $1$-loop partition function may be written directly in terms of the Barnes $G$-function as
\begin{equation}
\begin{aligned}
Z_{\rm 1-loop}^{(2,1),n} ={}& \prod_{k=2}^{n-1} \prod_{\substack{\alpha,\beta=1, \alpha\neq\beta}}^{k} G \left( 1+ \frac{\mu_\alpha^{(k)} - \mu_\beta^{(k)}}{
\hbar} \right)^{-1}&\times \prod_{k=2}^{n} \prod_{\alpha=1}^{k}
\prod_{\beta=1}^{k-1} G \left( 1+ \frac{ \mu_\alpha^{(k)} - \mu_\beta^{(k-1)}}{
\hbar} \right).
\end{aligned}
\end{equation}

We emphasize that the shift $T_{k,i}$ acts on all global weights containing $\mu_i^{(k)}$. Thus, the local virtual character $\mathcal G^{\rm vir}_{k,i}$ describes the centered second-difference response of the global perturbative determinant in the $(k,i)$ direction. It should not be regarded as an independent fluctuation complex whose primitive is subsequently multiplied over all $(k,i)$. In particular, although a bifundamental weight contributes to the local shift equations at both endpoints of the corresponding quiver edge, the associated Barnes factor is included only once in $Z_{\rm 1-loop}$.

There is a subtlety for the classical part. As we discussed in remark \ref{Rmk: regularizednature}, the regularized Stokes entries along with the regularized factorization coordinates forget the plumbing parameters which contribute to the classical partition function. In order to recover these data we can either read it from the Thimm torus action or keep this information in the near-caterpillar regime(this is what we did in (\ref{eq: neckcontribution})). In this section we will use the first approach. Alekseev et al. \cite{alekseev2024wkbasymptoticsstokesmatrices} define the regularized Stokes matrices by
\begin{equation}
    S^{\rm reg} = VSV^{-1}, 
\end{equation}
where $V:=\prod_{k=2, \dots, n-1}^{\xrightarrow {}} \rho_k^{\frac{1}{2\pi i}\frac{\text{log}\, \delta_k(S_-)\delta_k(S_+)}{\epsilon}}$, and $\rho_k = \frac{u_{k}-u_{k-1}}{u_{k+1}-u_k}$. For convenience we define $H_k:=\delta_k(S_-)\delta_k(S_+)$, then 
$H_k|_{\mathbb{C}^k} =S^{(k)}$. Now since
\begin{equation}
    \log \,H_k =  \sum_{i=1}^k \log \, \lambda^{(k)}_i(S) P^{(k)}_i +\dots, 
\end{equation}
we obtain that
\begin{equation}
    \rho_k^{\frac{1}{2\pi i}\frac{\text{log}\, \delta_k(S_-)\delta_k(S_+)}{\epsilon}}|_{L^{(k)}_i}= \rho_k^{ \frac{1}{2\pi i}\frac{\log \, \lambda^{(k)}_i(S)}{\epsilon} }.
\end{equation}
On the other hand, $(2.4.7)$ of \cite{alekseev2024wkbasymptoticsstokesmatrices} states that $\log\, \lambda^{(k)}_i (S)= -\frac{2\pi \mu^{(k)}_i}{\epsilon}$, which means that the Thimm character is
\begin{equation}
    \omega^{(k)}_i= \rho_k^{ i\frac{\mu^{(k)}_i}{\epsilon^2}}.
\end{equation}

Now, the Thimm torus action has the effect: $\mu^{(k)}_i \mapsto \mu^{(k)}_i,\, \theta^{(k)}_i \mapsto \theta^{(k)}_i+\varphi^{(k)}_i$, where $\varphi^{(k)}_i:= \frac{\mu^{(k)}_i}{\epsilon^2} \log \rho_k$. Since the Thimm torus action fixes the action coordinates, it must possess a local generating function, say $\mathcal{F}_k$, which satisfies
\begin{equation}
    \frac{\mu^{(k)}_i}{\epsilon^2} \log \rho_k= \varphi^{(k)}_i = \frac{\partial \mathcal{F}_k(\mu^{(k)})}{\partial\mu^{(k)}_i}.
\end{equation}
Solving this equation, we obtain $\mathcal{F}_k(\mu^{(k)})=\frac{\log \rho_k}{2\epsilon^2} \sum_i (\mu^{(k)}_i)^2+\text{const}$. Setting $\hbar= -i\epsilon $ and taking the exponential on both sides yields
\begin{equation}
    e^{\mathcal{F}_k(\mu^{(k)})} =\text{const.} \times \rho_k^\frac{\sum_i (\mu^{(k)}_i)^2}{2\epsilon^2}  = \text{const.} \times \rho_k^{-\frac{1}{2\hbar^2}\sum_i (\mu^{(k)}_i)^2 }.
\end{equation}
 On the other hand, we may deduce the local cocycle, namely it is the integration
 \begin{equation}
     \log \mathcal{X}^{(k)}_i(\mu) = \int^{\mu^{(k)}_i+\frac{\hbar}{2}}_{\mu^{(k)}_i-\frac{\hbar}{2}}ds \,  \varphi^{(k)}_i (s) = -\int^{\mu^{(k)}_i+\frac{\hbar}{2}}_{\mu^{(k)}_i-\frac{\hbar}{2}}ds \, \frac{s}{\hbar^2} \log \rho_k= -\frac{\mu^{(k)}_i \log \rho_k }{\hbar}.
 \end{equation}
Equivalently, $\mathcal{X}^{(k)}_i =\rho_k ^{-\frac{\mu^{(k)}_i}{\hbar}}$. Now for $(k,i)\neq (l,j)$, we have $T_{l,j} \mathcal{X}^{(k)}_i=X^{(k)}_i$. Therefore 
$T_{l,j} \mathcal{X}^{(k)}_i \mathcal{X}^{(l)}_j =T_{k,i} \mathcal{X}^{(l)}_j\mathcal{X}^{(k)}_i$. This guarantees that there exists a global discrete potential. Define
\begin{equation}
    Z_{\rm cl,k } =  \rho_k^{\frac{1}{2\hbar^2}   \sum^k_{i=1} (\mu^{(k)}_i)^2 },
\end{equation}
which is $e^{-\mathcal{F}_k}$ up to a constant, and it satisfies the difference equation (we use $-\mu^{(k)}_i$ because $a^{(k)}_i=-\mu^{(k)}_i$)
\begin{equation}
    \frac{Z_{\rm cl,k}(-\mu^{(k)}_i+\frac{\hbar}{2}) }{Z_{\rm cl,k}(-\mu^{(k)}_i-\frac{\hbar}{2})} = \rho_k^{-\frac{\mu^{(k)}_i}{\hbar}}. 
\end{equation}
Note that the negative sign of $e^{-\mathcal{F}_k}$ comes from the inverse shift direction. Restricting to the $SU(k)$ subgroup, then the root cocycles become
\begin{equation}
    \mathcal{X}_{k,ij}= \frac{\mathcal{X}_{k,i}}{\mathcal{X}_{k,j}} =\rho_k^{-\frac{(\mu^{(k)}, \nu_{ij})}{\hbar}},
\end{equation}
where $\nu_{ij}=e_i-e_j$ is a root. Then we may write $\mathcal{X}_{k,\nu}=\rho_k^{-\frac{(\mu^{(k)},\nu)}{\hbar}} $. Then
\begin{equation}
    Z^{SU(k)}_{\text{cl},k}= \rho_k^{\frac{1}{2\hbar^2}    \left[ \sum_{i=1}^k (\mu^{(k)}_i)^2 -\frac{1}{k}(\sum_{i=1}^k \mu^{(k)}_i)^2\right]}. \label{eq: classical}
\end{equation}
\begin{Rmk}
    Note that in the weak-coupling limit, $\rho_k$ can be identified with $q_k^{SU(k)}$ defined in (\ref{eq: gaugeCoupling}). 
\end{Rmk}
\subsection{The general rank-$n$ $(2,2)$ system}

For the rank-$n$ spectral curve of type $(2,2)$, the
factorization data consist of the left and right Gelfand--Zeitlin
tails together with the gluing data of the middle neck. Restricting to $SU(n)$:
\begin{equation}
    \bigl\{\alpha_{ij}^{L}\bigr\}
    \ \cup\
    \frac{(\mathbb C^\times)^n}{\mathbb C^\times}
    \ \cup\
    \bigl\{\alpha_{ij}^{R}\bigr\}.
\end{equation}
We represent the middle gluing angle by
$(\theta_1,\ldots,\theta_n)$, modulo their common rescaling.  Equivalently,
one may impose $\prod_{i=1}^{n}e^{i\theta_i}=1$. Thus the middle neck carries $n-1$ independent angle variables, as expected for the root lattice $Q(A_{n-1})$.

Let 
\begin{equation}
    \Lambda_{\rm cat}= \bigoplus_{k=2}^{n-1}Q(A_{k-1})_L \oplus Q(A_{n-1})_M \oplus
    \bigoplus_{k=2}^{n-1}Q(A_{k-1})_R. \label{eq: lattice}
\end{equation}
We collect the traceless action variables into $ \boldsymbol{\mathfrak a} = \left( \{\boldsymbol a_L^{(k)}\}_{k=2}^{n-1}, \boldsymbol a_M, \{\boldsymbol a_R^{(k)}\}_{k=2}^{n-1} \right)$, and define $ \boldsymbol{\mathfrak a}_{\boldsymbol\ell} = \boldsymbol{\mathfrak a} +\hbar\boldsymbol\ell, \, \boldsymbol\ell\in\Lambda_{\rm cat}$. The conjectural Fourier expansion of the isomonodromic tau function
then takes the form
\begin{equation}
\begin{aligned}
\label{eq: FourierCompletion}
    \tau_{(2,2)}^{(n)} ={}& Z_{U(1)} \sum_{\boldsymbol\ell\in\Lambda_{\rm cat}}
    \exp\left( 2\pi i \left\langle \boldsymbol\theta, \boldsymbol\ell \right\rangle
    \right)
    \\
    &\times Z_{\rm cl} \left( \boldsymbol{\mathfrak a}_{\boldsymbol\ell}; \boldsymbol q_L,t,\boldsymbol q_R \right) Z_{\rm pert} \left( \boldsymbol{\mathfrak a}_{\boldsymbol\ell} \right) Z_{\rm inst} \left( \boldsymbol{\mathfrak a}_{\boldsymbol\ell}; \boldsymbol q_L,t,\boldsymbol q_R \right).
\end{aligned}
\end{equation}
Here \(Z_{U(1)}\) denotes the root-lattice-invariant Abelian
normalization.  Any factor which depends nontrivially on
\(\boldsymbol{\mathfrak a}_{\boldsymbol\ell}\) should instead be
included in the coefficient inside the Fourier sum.

The classical factor decomposes node by node: $Z_{\rm cl}= Z_{\rm cl}^{L} Z_{\rm cl}^{M}
Z_{\rm cl}^{R}$, where
\begin{equation}
    Z_{\rm cl}^{M} = t^{\frac{ (\boldsymbol a_M,\boldsymbol a_M) }{2\hbar^2}},
\end{equation}
and $Z_{\rm cl}^{L,R}$ take the form (\ref{eq: classical}). 

The signs of the exponents may be reversed by changing the
orientation of the corresponding plumbing coordinate.  Linear
powers of the plumbing parameters can likewise be redistributed
between the Fourier phase and the classical factor by a
time-dependent affine normalization of the angle coordinates.

The perturbative Nekrasov factor has the gluing decomposition
\begin{equation}
    Z_{\rm pert} = Z_{\rm 1-loop}^{L}\, Z_{\rm vec}^{\rm neck}\, Z_{\rm 1-loop}^{R}, \label{eq: pertDecomp}
\end{equation}
with the middle term defined in (\ref{eq: middleGlue}) up to an $a$-independent normalization and the separately treated central $U(1)$ factor. Under the assumption made in section~5.2, this factor supplies the perturbative coefficients of $\tau$. The two tail factors are the global Barnes primitives reconstructed from the left and right factorization-coordinate cocycles. The middle factor must be included exactly once. Indeed, by the discussion in section $3.2$ there is only one summation over the shared $a^{(n)}$, since there is only one representation. Its precise Gamma- or Barnes-\(G\) normalization is fixed by the finite-difference equation of the middle gluing coordinates.

\section*{Acknowledgements}

The author would like to thank Xiaomeng Xu and Zikang Wang for many helpful discussions and comments, and Xu in particular for pointing out the connection to the double Gelfand-Zeitlin system. The author would also like to thank Andrew Neitzke and Yan Zhou for taking the time to listen to a presentation of this work and for offering valuable comments and suggestions.

\begin{appendices}

\appendix

\section{Explicit calculations of n=2}
In this section we are considering the system 
\begin{equation}
    \epsilon\frac{dF}{dz} = \left( iu-(i\frac{B}{z}+i\frac{A}{z^2})   \right)F
\end{equation}
or the spectral curve
\begin{equation}
    \det \left( xI_2 -\left( iu- i\frac{B}{z}-i\frac{A}{z^2}\right)  \right)=0,
\end{equation}
so that there is an involution symmetry $(x,z) \to (-\bar{x},\bar{z})$ of the spectral curve. 

If $n=2$, then $\Gamma(u,A,B)$ has 4 branch points. Also we have $8$ marked points in total. Now consider the (long) exact sequence
\begin{equation}
   0 \to H_1\bigl(\hat{\Gamma}(u,A,B)\bigr)
\to H_1\bigl(\hat{\Gamma}(u,A,B); L(v)\bigr)
\to H_0\bigl(L(v)\bigr)
\to H_0\bigl(\hat{\Gamma}(u,A,B)\bigr)
\to 0.
\end{equation}
We have $\text{dim} \,H_0(\hat{\Gamma}(u,A,B))=1, \text{dim}\,H_0(L(v))=8$, and here $\hat{\Gamma}(u,A,B)$ is a genus one curve with four boundary circles, therefore 
\begin{equation}
    \text{dim}\,\, H_1(\hat{\Gamma}(u,A,B)) = 2g+ \text{\# of boundary circles}-1=2+4-1=5.
\end{equation}
Therefore dim $H_1(\hat{\Gamma}(u,A,B),L(v))=8+5-1=12$. Also dim $H_1(\Gamma(u,A,B))= 2g + p-1=2+4-1=5$, here $p$ is the number of punctures. 

We first consider the degenerate case $v=0$ and choose the branch cuts connecting vertical branch points. A basis of $H_1(\hat{\Gamma}(u,A,B),L(v))$ can be described as follows. First there are $5$ closed cycles. 

\textbullet $\alpha$ corresponds to the first dimension of a torus. Namely, it encircles the left branch cut on sheet $1$. 

\textbullet $\beta$ corresponds to the second dimension of a torus. Namely, it starts from the middle region of two branch cuts on sheet $2$, passing through the right branch cut and entering sheet $1$. Then it encircles the lower $2$ branch points and goes back to the middle region from the left branch cut. 

\textbullet The three boundary circles. We can choose them to be the two boundary circles at $0$ lying over sheet $1,2$, and one boundary circle at $\infty$ lying over sheet $1$. We label them by $\partial 0_1, \partial 0_2, \partial \infty_1$. Note that $\partial \infty_2$ are related to the other $5$ closed cycles. 

\textbullet Seven open cycles remain and they are labeled on the graph. Namely $\sigma_i$ goes from $\infty^+_i$ to $\infty^-_i$ without passing the branch cuts, $\gamma_i$ goes from $0^+_i$ to $0^-_i$ without passing the branch cuts, $\tau_i$ goes from $0^-_i$ to $\infty^-_i$ without passing the branch cuts, and $\delta$ goes from $\infty_2^+$ to $\infty_1^-$, passing from the right branch cut. 

\begin{figure}[htbp]
    \centering
    \begin{subfigure}{0.45\textwidth}
        \centering
        \includegraphics[width=\linewidth]{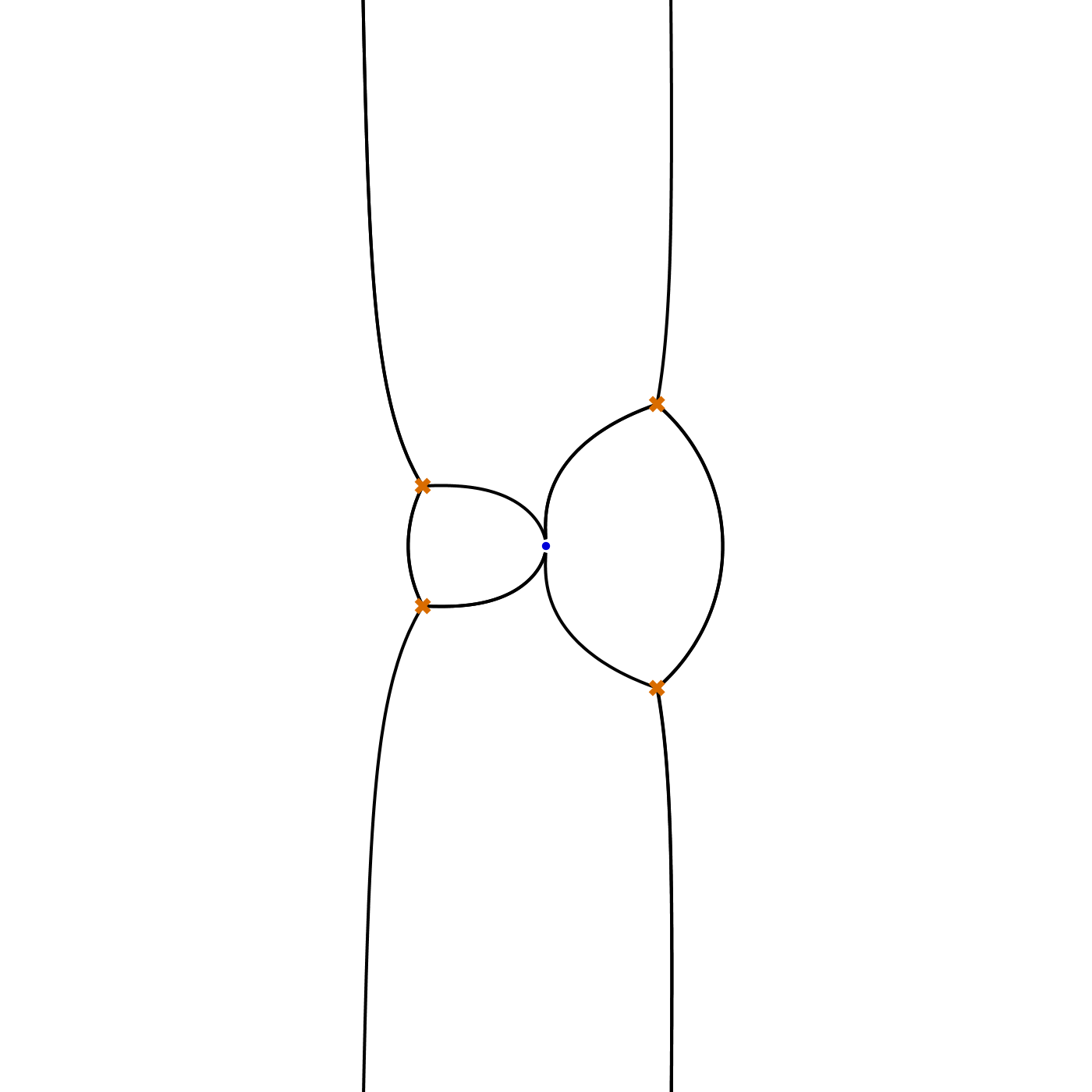}
        \caption{The $n=2,v=0$ spectral network}
        \label{fig:left}
    \end{subfigure}
    \hfill  
    \begin{subfigure}{0.45\textwidth}
        \centering
        \includegraphics[width=\linewidth]{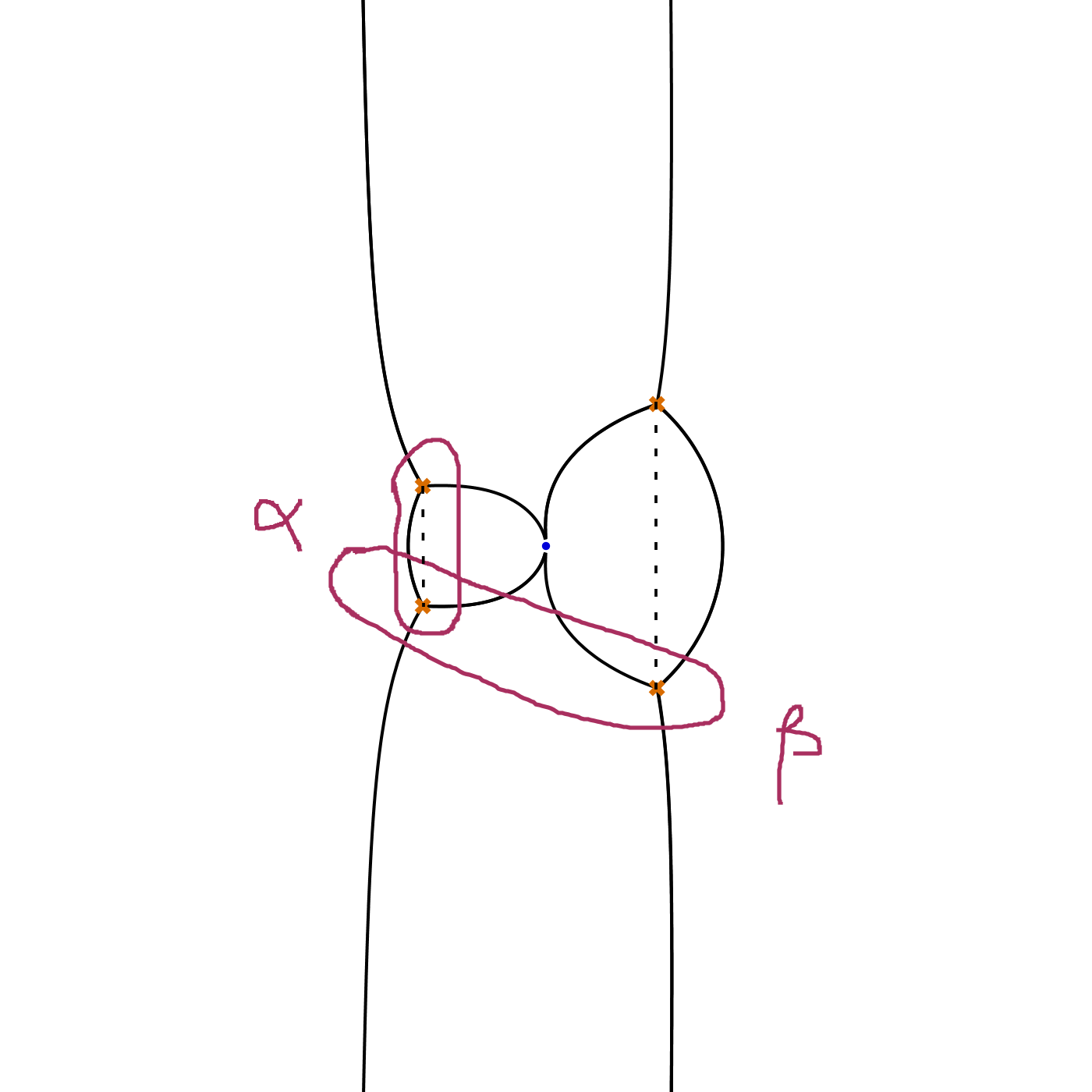}
        \caption{Closed cycles $\alpha$ and $\beta$}
        \label{fig:right}
    \end{subfigure}
    \hfill
    \begin{subfigure}{0.45\textwidth}
        \centering
        \includegraphics[width=\linewidth]{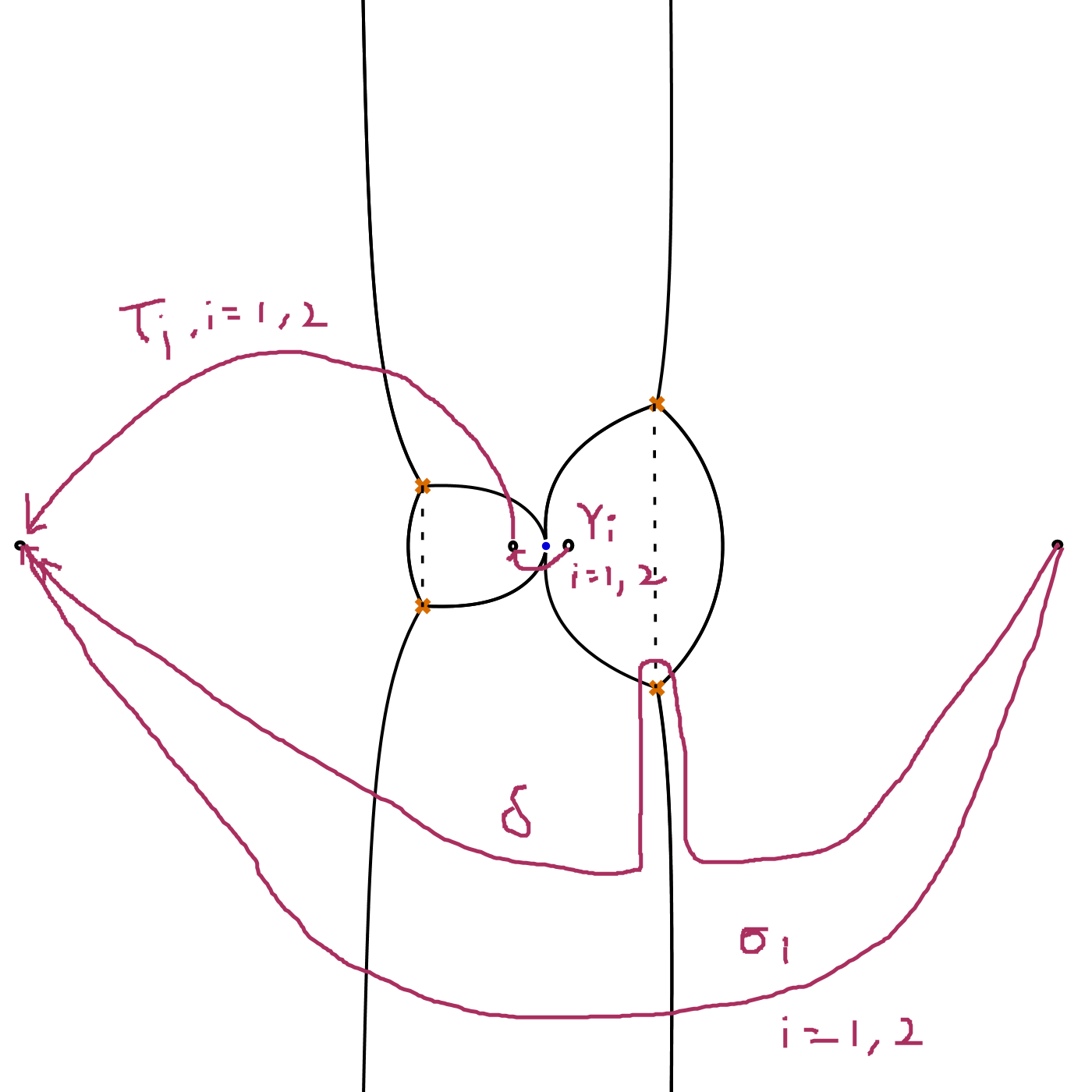}
        \caption{Open cycles $\sigma_i, \gamma_i, \tau_i, \delta, i=1,2$} 
        
    \end{subfigure}
    \hfill
    \begin{subfigure}{0.45\textwidth}
        \centering
        \includegraphics[width=\linewidth]{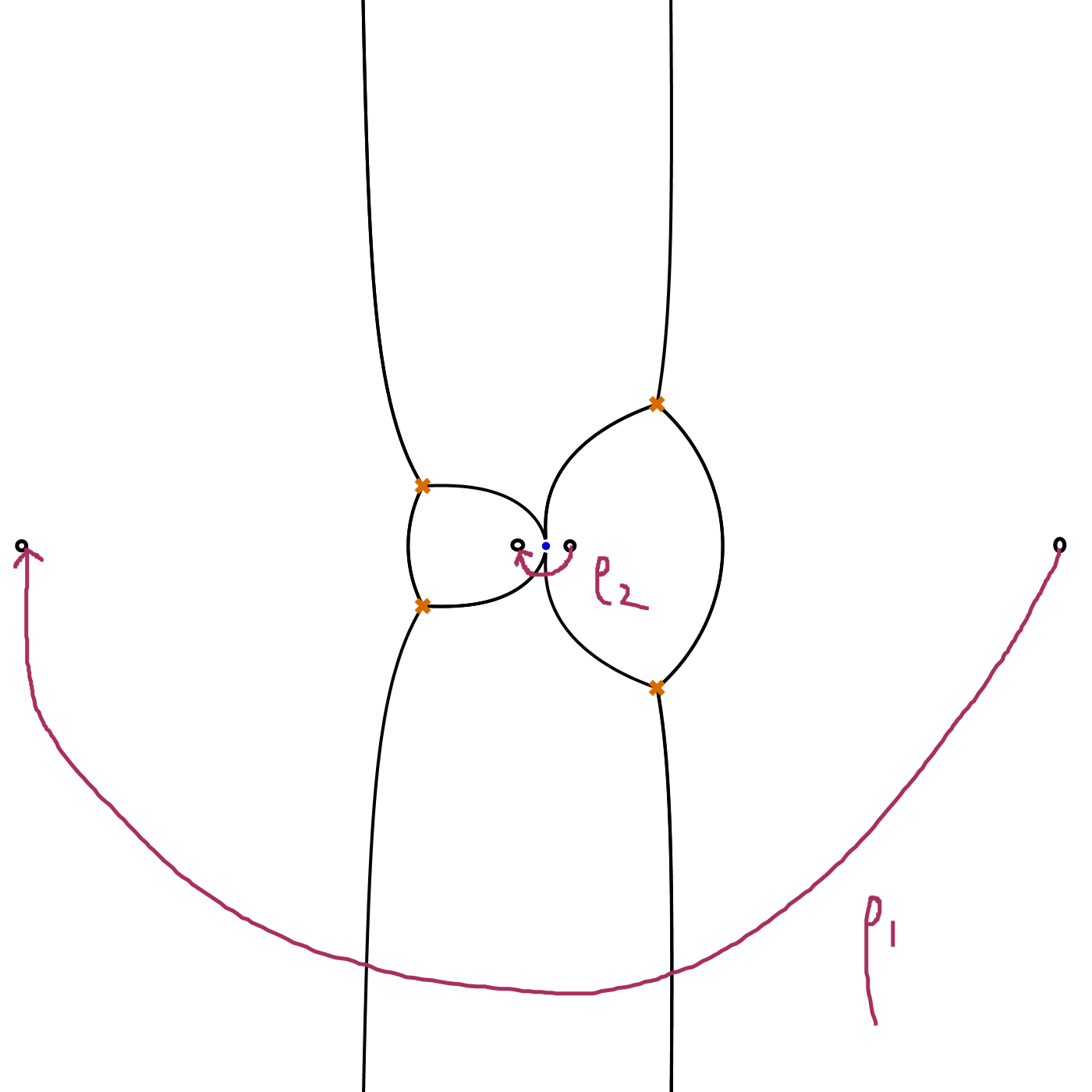}
        \caption{The two paths corresponding to the Stokes matrices} 
    \end{subfigure}
    \label{fig:both}
\end{figure}

\begin{Prop}
Let $q_v=-e^{-2iv}\frac{P_4(z)}{z^4}\,dz^2$ be the rank-two quadratic differential associated with a generic anti-Hermitian $(2,2)$ system. Assume that $P_4$ has simple zeroes and that the Hermitian pencil $H(x)=ux^2-Bx-A$ has simple spectrum for every $x\in\mathbb R$. Let $N_0$, $N_\infty$ and $S$ denote respectively the numbers of critical trajectories ending at $0$, ending at $\infty$, and saddle connections. Whenever the critical graph has no recurrent or finite-cylinder component, one has
\begin{equation}
    N_0=N_\infty=6-S .
\end{equation}
\end{Prop}

\begin{Prf}
Since $H(x)$ is Hermitian for $x\in\mathbb R$, its two eigenvalues are
real. The absence of a real eigenvalue collision gives
\[
P_4(x) = \bigl(\lambda_1(H(x))-\lambda_2(H(x))\bigr)^2 >0, \qquad x\in\mathbb R .
\]
Hence $\sqrt{P_4(x)}$ admits a distinguished positive branch along the
real axis.

Consider the real oriented blow-up of $\mathbb P^1$ at $0$ and
$\infty$. It is an annulus, and the horizontal foliation of $q_v$
defines a line field on this annulus. Near the two boundary components,
\[
q_v \sim -e^{-2iv}c_0\,\frac{dz^2}{z^4}, \qquad q_v \sim -e^{-2iv}c_\infty\,\frac{d\zeta^2}{\zeta^4}, \quad \zeta=z^{-1},
\]
where
\[
c_0=P_4(0)>0, \qquad c_\infty=[z^4]P_4(z)>0.
\]
The positive branch of $\sqrt{P_4(x)}$ along $\mathbb R_{>0}$ identifies
the two asymptotic square-root framings. Consequently, after taking
account of the opposite orientations of the two boundary circles, the
boundary indices of the horizontal line field at $0$ and $\infty$ are
equal for every $v$.

We now apply the turning-number formula to a regular neighbourhood of
the compactified critical graph. The two sides of every finite edge
occur with opposite orientations and cancel, while each critical
trajectory ending at a boundary component contributes one boundary
corner. Therefore
\[
\frac{N_0-N_\infty}{2}= \operatorname{ind}_{\partial_0}(q_v)-
\operatorname{ind}_{\partial_\infty}(q_v) =0.
\]
Thus $N_0=N_\infty$.

Finally, the four simple zeroes of $q_v$ produce twelve critical
half-trajectories. A saddle connection consumes two such
half-trajectories, and hence $N_0+N_\infty+2S=12$. Together with $N_0=N_\infty$, this gives $N_0=N_\infty=6-S$.
\end{Prf}

Due to the involution symmetry, $N_0$ must be even, which means S is even as well. This leads immediately to the following proposition: 
\begin{Prop}
    The spectral network with an involution symmetry ($v$=0) contains an even number of saddle connections.
\end{Prop}

 Since \(v=0\) is critical, Proposition A.2 implies \(S\ge2\). The two symmetry-forced saddle connections join the two conjugate pairs of branch points. Any additional saddle connection joins different conjugate pairs and is accompanied by its complex conjugate. Its existence imposes the additional real period condition \(\operatorname{Im}Z_\delta=0\), which is not forced by the involution. Hence \(S=4\) and \(S=6\) occur only on higher-codimension loci, or in excluded finite-cylinder configurations. Therefore \(S=2\) for generic involution-symmetric parameters.

Now we take $(a)$ as an example. Then $\alpha$ is the cycle encircling the saddle connection. Using the path-lifting rules, 
\begin{equation}
    \text{Lift} (\rho_1)= \sigma_1+\sigma_2+\delta+ (\delta-\beta)+ \dots
\end{equation}
Then $(S_{\infty,+})_{ii}=X^0_{\sigma_i}$ and $(S_{\infty,+})_{1,2}=X^0_{\delta-\beta}+X^0_{\delta}+\dots$. Applying the main conjecture between spectral networks and WKB asymptotics, 
\begin{equation}
    (S_{\infty,+})_{12} \sim C_1\text{exp} \left( \frac{Z(\delta)}{\epsilon} \right)+C_2\text{exp}\left( \frac{Z(\delta-\beta)}{\epsilon} \right)+\text{exponentially small terms.}
\end{equation}

Because of the involution symmetry, 
\begin{equation}
    \text{Re} \,Z(\gamma)= \frac{1}{2}Z(\gamma-\iota(\gamma)). 
\end{equation}
Then $\text{Re} (Z(\delta))=\frac{1}{2}Z(\delta-\iota(\delta))= \frac{1}{2}(Z(\alpha)+Z(\partial0_1))$ . Similarly, $\text{Re}(Z(\delta-\beta))=\frac{1}{2}(Z(\eta)+Z(\partial0_2))$. Here $\eta$ is the clockwise cycle encircling the right branch cut on sheet 2.

However, since
\begin{equation}
\exp\left(\frac{Z(\delta)}{\epsilon}\right) = \exp\left(\frac{Z(\delta-\beta)}{\epsilon}\right) \exp\left(\frac{Z(\beta)}{\epsilon}\right),
\end{equation}
the relative exponential ordering is determined by
$\operatorname{Re}(Z(\beta)/\epsilon)$. Since $\vartheta=0$ is a critical phase, this ordering should be understood laterally. Writing
$\epsilon=|\epsilon|e^{iv}$, the neck contribution in the caterpillar limit is
\begin{equation}
Z(\beta)= i\bigl(\mu_2^{(2)}-\mu_1^{(2)}\bigr) \log\frac{1}{|t|} +O(1).
\end{equation}
Hence
\begin{equation}
\operatorname{Re}\bigl(e^{-iv}Z(\beta)\bigr) = \bigl(\mu_2^{(2)}-\mu_1^{(2)}\bigr)
\sin v\, \log\frac{1}{|t|} +O(1).
\end{equation}
Thus, for either lateral phase $\vartheta=0^\pm$ and sufficiently small $t$, one of the two spectral channels is exponentially dominant; the dominant channel is exchanged when the critical phase is crossed. We choose the lateral chamber in which
$C_1\exp(Z(\delta)/\epsilon)$ is dominant. At the critical phase itself the two channels may have the same exponential order, and the lateral prescription is understood. For convenience, we choose $v=0^{+}$, then $\text{Re} (e^{-iv}Z(\beta))>0$.  

\begin{Rmk}
    There is an ambiguity here. We assume the fact the right pair of branch points stay in the macroscopic scale under the caterpillar limit. If, instead, the right pair of branch points are sent to the inner spectral network, then the leading term is exchanged. Therefore the relative ordering does not change. \label{B.3}
\end{Rmk}

For $\text{Re} (Z(\delta))=\frac{1}{2}Z(\delta-\iota(\delta))= \frac{1}{2}(Z(\alpha)+Z(\partial0_1))$, the second part gives the residues of $0$ on sheet $1$, and the first part gives the period of the left branch cut. 

We calculate
\begin{equation}
    x_{\pm}= \frac{i}{2} \text{Tr} \, \Phi(z) \pm  \frac{i}{2z^2} \sqrt{P(z)},
\end{equation}
here 
\begin{equation}
    P(z):= (\Delta u z^2-\Delta Bz-\Delta A)^2+4(b_{12}z+a_{12})(\bar{b}_{12}z+\bar{a}_{12}).
\end{equation}
Then the branch points are the zeroes of $P(z)$. In order to calculate $Z(\alpha)$ we choose the simplest representative, namely the line, then we obtain the elliptic integral:
\begin{equation}
    Z(\alpha)= \int \lambda =\int(x_+-x_-)\,dz =i \int \frac{\sqrt{P(z)}}{z^2} \,dz.
\end{equation}
\begin{Rmk}
In physics, the cycle encircling the saddle connection is a $4$d BPS charge. $Z(\alpha)$ is thus the central charge/mass of this BPS charge. It's very common that in $4$d supersymmetric theories the BPS masses are represented by elliptic integrals. There is one more BPS charge: $\partial\infty_1-\alpha-\partial0_1$.     
\end{Rmk}

By Proposition \ref{prop: 4.2}, under the caterpillar degeneration,
$\partial_{0,1}$ becomes the boundary circle surrounding the first
puncture over $v=0$ in the inner spectral curve, while $\alpha$
degenerates to its unique branch-cut cycle. With the orientations fixed
above, their contributions combine as
\begin{equation}
\begin{aligned}
    2\operatorname{Re} Z(\delta) = Z(\alpha)+Z(\partial_{0,1})  = \bigl(
    Z(\partial_{\infty,1})-Z(\partial_{0,1}) \bigr) +Z(\partial_{0,1}) = \mu_1^{(2)}.
\end{aligned}
\end{equation}
Here the boundary periods are understood with the residue normalization
used in Proposition \ref{prop: 4.2}. If $\delta-\beta$ gives the leading
contribution instead, Remark \ref{B.3} similarly gives
\begin{equation}
    2\operatorname{Re}Z(\delta-\beta)=\mu_2^{(2)}.
\end{equation}

We next consider the imaginary part. For definiteness, suppose that
$Z(\delta)$ gives the leading contribution. Adapting the rank-two
calculation of
\cite{alekseev2024wkbasymptoticsstokesmatrices}, one obtains
\begin{equation}
\begin{aligned}
    \operatorname{Im}Z_{\infty}^{\rm reg}(\delta)= (B_{22}-B_{11})
    \log\frac{e(u_2-u_1)}{|B_{12}|} +\frac{\mu_2^{(2)}-\mu_1^{(2)}}{2}
    \log \frac{B_{11}-\mu_1^{(2)}}{\mu_2^{(2)}-B_{11}} .
\end{aligned}
\label{eq: imaginary-regularized-period}
\end{equation}
We briefly explain the dependence on $u_2-u_1$. Set $t:=u_2-u_1>0$. Near $z=\infty$, the difference of the two Seiberg--Witten differentials has the expansion
\begin{equation}
    \lambda_{12} = i(B_{11}-B_{22})\frac{dz}{z} -it\,dz +O(z^{-2})\,dz.
\end{equation}
A corresponding singular primitive is therefore
\begin{equation}
    P_{12} = i(B_{11}-B_{22})\log z -it z +O(z^{-1}).
\end{equation}
Using the normalized coordinate $ \xi=t z$, we have $\log z=\log\xi-\log t$, and hence
\begin{equation}
\begin{aligned}
    P_{12}= i(B_{11}-B_{22})\log\xi-i\xi \, +i(B_{22}-B_{11})\log t +O(\xi^{-1}).
\end{aligned}
\end{equation}
The $\xi$-dependent singular terms are removed by the endpoint
regularization. The remaining scale-dependent contribution is therefore
$i(B_{22}-B_{11})\log t$, whose imaginary part is $(B_{22}-B_{11})\log t$. Since the corresponding spectral coordinate has the WKB behaviour
\begin{equation}
    X_\delta \sim \exp\left(\frac{Z_{\infty}^{\rm reg}(\delta)}{\epsilon}\right),
\end{equation}
this part of the regularized period contributes $t^{\,\frac{i(B_{22}-B_{11})}{\epsilon}}
$ to the Stokes entry.

To compare with the exact Stokes formula, one must in addition include
the $\epsilon$-dependent normalization of the formal endpoint basis.
In the conventions of Xu, the residue matrices are related by
$A_{\rm Xu}=-2\pi B$, and therefore
\begin{equation}
\begin{aligned}
    \left(\frac{u_2-u_1}{\epsilon}\right)^{
    \frac{(A_{\rm Xu})_{22}-(A_{\rm Xu})_{11}}
         {2\pi i\epsilon}}
    &=
    \left(\frac{u_2-u_1}{\epsilon}\right)^{
    \frac{i(B_{22}-B_{11})}{\epsilon}} .
\end{aligned}
\label{eq: Xu-scale-factor}
\end{equation}
Thus the regularized period determines the $u_2-u_1$ dependence,
whereas the additional factor of $\epsilon$ in the denominator is fixed
by the canonical formal normalization. The associated
$\epsilon^{-1}\log\epsilon$ terms cancel against the corresponding terms
in the Stirling asymptotics of the Gamma functions.

Finally, the remaining logarithmic terms in
\eqref{eq: imaginary-regularized-period} are reproduced collectively by
the Stirling asymptotics of the exact Gamma-function coefficient. In
making this comparison one also uses the rank-two identity
\begin{equation}
    (B_{11}-\mu_1^{(2)})
    (\mu_2^{(2)}-B_{11})
    =
    |B_{12}|^2.
\end{equation}
Hence the entire regularized period, rather than only the factor
$|B_{12}|$, is the leading WKB asymptotic of the exact Gamma-function
expression.

Similarly we may compute the Stokes matrix at $0$. Now we use the path lifting of $\rho_2$. Note that the wall flowing to $0$ in the lower half plane has dominance $2<1$. This can be justified as follows: first by looking at the behavior near $z=0$, $\Delta \lambda\sim -i\frac{\lambda_2-\lambda_1}{z^2}dz$, here $\lambda_1,\lambda_2$ are the eigenvalues of $A$. Then $\int^z \Delta \lambda \sim \frac{i(\lambda_2-\lambda_1)}{z}$. Write $z=y_0+iy$, then $\text{Re} \frac{i}{z}= \frac{y}{|z|^2}$. 
Also, the real part along a wall is monotonic and nonzero away from the branch points, therefore the dominance near $0$ is exactly the dominance near the branch point. We conclude that if the wall enters $0$ from the lower half plane ($y<0$), then $2<1$. We also note that if a wall starts from a branch point with imaginary part smaller than $0$ and enters $0$, it must enter $0$ from the lower half plane because of the involution symmetry and the fact that for a rank-2 spectral network the walls cannot intersect.

Then the real part is given by $\frac{1}{2}(Z(\eta)+Z(\partial 0_2)   )$ or $\frac{1}{2}(Z(\alpha)+Z(\partial 0_1))$. The point is that these periods are exactly the same as the real periods of the Stokes matrices at infinity. 

\begin{Lem}
    For a rank $2$, (2,2) type system with involution symmetry, if a branch point ejects a wall flowing to $0$ and a wall flowing to $\infty$, then the two walls have opposite dominance. Their limiting leading real parts coincide after identifying the corresponding residue cycles on the two normalized components.
\end{Lem}

Since $\rho_2$ is a microscopic arc connecting two anti-Stokes
directions at the real oriented blow-up of $z=0$, under the caterpillar
degeneration it becomes an open path entirely contained in the inner
spectral curve. Its regularized period is defined analogously by
\begin{equation}
\begin{aligned}
    Z_0^{\rm reg}(\gamma)
    :=
    \lim_{\substack{v_i\to 0_i^{\vartheta}\\
                    v_{i'}'\to 0_{i'}^{\vartheta'}}}
    \left[
        -P_{i'}^{(0)}(v_{i'}')
        +P_i^{(0)}(v_i)
        +\int_{\gamma(v_i,v_{i'}')}
        \lambda_{\rm in}
    \right].
\end{aligned}
\label{eq: inner-regularized-period}
\end{equation}
Here the superscripts $\vartheta,\vartheta'$ specify the chosen
anti-Stokes directions.

Let $G^{-1}A_0G= \Lambda = \operatorname{diag}(\lambda_1,\lambda_2), \, \beta:=G^{-1}BG$. In this eigenframe, the Seiberg--Witten differential on the $i$-th
inner sheet has the local expansion
\begin{equation}
    \lambda_{{\rm in},i} = \left( -\frac{i\lambda_i}{v^2} -\frac{i\beta_{ii}}{v}
    +O(1) \right)dv.
\end{equation}
A corresponding singular primitive is
\begin{equation}
    P_i^{(0)}(v)
    =
    \frac{i\lambda_i}{v}
    -i\beta_{ii}\log v
    +O(v).
\end{equation}

The rank-two calculation then gives
\begin{equation}
\begin{aligned}
    \operatorname{Im}Z_0^{\rm reg}(\gamma) = -(\beta_{22}-\beta_{11}) \log
    \frac{e(\lambda_2-\lambda_1)} {|\beta_{12}|} -\frac{\mu_2-\mu_1}{2}
    \log \frac{\beta_{11}-\mu_1}{\mu_2-\beta_{11}},
\end{aligned}
\label{eq: inner-imaginary-period}
\end{equation}
where $\mu_1<\mu_2$ are the eigenvalues of $\beta$.

To isolate the dependence on the irregular eigenvalue difference, set $s:=\lambda_2-\lambda_1>0, \,\xi:=\frac{s}{v}$. Then
\begin{equation}
    \log v=\log s-\log\xi.
\end{equation}
Consequently, the logarithmic part of the difference of the two
singular primitives contains
\begin{equation}
\begin{aligned}
    -i(\beta_{22}-\beta_{11})\log v =
    -i(\beta_{22}-\beta_{11})\log s  +i(\beta_{22}-\beta_{11})\log\xi.
\end{aligned}
\end{equation}
The $\xi$-dependent term is removed by the endpoint regularization,
leaving $-i(\beta_{22}-\beta_{11})\log s$. 
Since the corresponding spectral coordinate behaves as
\begin{equation}
    X_\gamma \sim \exp\left(\frac{Z_0^{\rm reg}(\gamma)}{\epsilon}\right),
\end{equation}
the scale-dependent part contributes $s^{-\frac{i(\beta_{22}-\beta_{11})}{\epsilon}}$. 

After including the $\epsilon$-dependent normalization of the formal
endpoint basis, the exact Stokes coefficient contains
\begin{equation}
    \boxed{ \left( \frac{\lambda_2-\lambda_1}{\epsilon}
    \right)^{ -\frac{i(\beta_{22}-\beta_{11})}{\epsilon} }}.
\label{eq: inner-scale-factor}
\end{equation}
The remaining logarithmic terms in
\eqref{eq: inner-imaginary-period} are reproduced by the Stirling
asymptotics of the Gamma-function factors in the exact Stokes
coefficient. Equivalently, in the spectral-network description they
are encoded by the regularized endpoint and neck normalizations.
Using
\begin{equation}
    (\beta_{11}-\mu_1) (\mu_2-\beta_{11}) = |\beta_{12}|^2,
\end{equation}
one obtains agreement with the rank-two specialization of the
Wang--Xu formula. Therefore we conclude that 
\begin{Prop}
    For a spectral curve/ODE of type $(2,2)$, in the double-caterpillar limit, 
the leading exponential asymptotics of the Stokes matrix at $\infty$ can be 
read off from the outer spectral network, while the leading exponential 
asymptotics of the Stokes matrix at $0$ can be read off from the inner 
spectral network, up to the standard choices of sector, normalization, and 
abelianization.
\end{Prop}

\section{Consistency check with Stokes matrices from isomonodromy deformation}
\label{app: checkXu}
In this appendix we compare our results with the Stokes matrix calculated from isomonodromy deformation \cite{xu2024regularizedlimitsstokesmatrices}. We consider the following term 
\begin{equation}
    \langle e_k, P^{(k)}_i u_k \rangle= \langle e_k, v_i^{(k)} \rangle \langle v_i^{(k)}, u_k \rangle. 
\end{equation}
This term can be derived from following expression
\begin{equation}
    e_k^* \text{adj} \left(\frac{\xi_i^{(k)}I_k -B^{(k)}}{\epsilon} \right) \frac{u_k}{\epsilon}=e_k^*\left( \prod_{j\neq i} (\xi^{(k)}_i/\epsilon -\xi^{(k)}_j/\epsilon)P^{(k)}_i \right) \frac{u_k}{\epsilon} = \frac{1}{\epsilon}\prod_{j\neq i} (\frac{\xi^{(k)}_i -\xi^{(k)}_j}{\epsilon}) \langle e_k, v_i^{(k)} \rangle \langle v_i^{(k)}, u_k \rangle. 
\end{equation}
Note this element is exactly the minor $(-1)^{k-1}\Delta^{1,\dots,k-1,k}_{1,\dots,k-1,k+1}(\frac{B-\xi^{(k)}_i \rm Id}{\epsilon})$ in \cite{xu2024regularizedlimitsstokesmatrices}. Also note that $|\langle e_k, v_i^{(k)} \rangle \langle v_i^{(k)}, u_k \rangle|^2=\langle e_k, P^{(k)}_i e_k \rangle \langle u_k, P^{(k)}_i u_k \rangle$. 

We may rewrite (\ref{eq: rew}) using the reflection formula of the Gamma function:
\begin{equation}
\begin{aligned}
    &Z^{\rm loc}_i \\
    = &\pi^2 \frac{\prod_{j\neq i}\text{sin}^2\left( \frac{\xi^{(k)}_i-\xi^{(k)}_j}{\epsilon} \pi \right) \Gamma^2(1+\xi^{(k)}_j/\epsilon-\xi^{(k)}_i/\epsilon)  }{\prod_{l=1}^{k-1} \sin \left( \frac{\xi^{(k)}_i- \xi^{(k-1)}_l}{\epsilon} \pi \right) \Gamma(1+\xi^{(k-1)}_l/\epsilon-\xi^{(k)}_i/\epsilon)\prod_{l=1}^{k+1} \sin \left( \frac{\xi^{(k)}_i-\xi^{(k+1)}_l} {\epsilon} \pi \right) \Gamma(1+\xi^{(k+1)}_l/\epsilon -\xi^{(k)}_i/\epsilon ) }. \label{eq: reflection}
\end{aligned}
\end{equation}
We write
\begin{equation}
    \mathcal{P}_i(\xi) = \frac{\prod_{j\neq i}\text{sin}^2\left( \frac{\xi^{(k)}_i-\xi^{(k)}_j}{\epsilon} \pi \right)}{\prod_{l=1}^{k-1} \sin \left( \frac{\xi^{(k)}_i- \xi^{(k-1)}_l}{\epsilon} \pi \right) \prod_{l=1}^{k+1} \sin \left( \frac{\xi^{(k)}_i-\xi^{(k+1)}_l} {\epsilon} \pi \right)}.
\end{equation}
Under a local Schlesinger shift, each $\sin$ factor changes by a sign, while the total number of $\sin$ factors is even, which means $T_{k,i}\mathcal{P}_i(\xi)=\mathcal{P}_i(\xi)$. Therefore the trigonometric terms $\mathcal{P}$ can be viewed as a $\epsilon$-periodic factor and can be eliminated under a proper choice of gauge. The remaining expression is
\begin{equation}
    \widetilde{Z}_i^{\rm loc} = \pi^2 \frac{\prod_{j\neq i}\Gamma^2(1+\xi^{(k)}_j/\epsilon-\xi^{(k)}_i/\epsilon)  }{\prod_{l=1}^{k-1}  \Gamma(1+\xi^{(k-1)}_l/\epsilon-\xi^{(k)}_i/\epsilon)\prod_{l=1}^{k+1} \Gamma(1+\xi^{(k+1)}_l/\epsilon -\xi^{(k)}_i/\epsilon ) }. \label{eq: normalizedreflection}
\end{equation}
Now we plug (\ref{eq: normalizedreflection}) into (\ref{eq: summing}), the contribution of the $i$-th channel is $\widetilde{Z}^{\rm loc}_i\langle e_k, P^{(k)}_i u_k \rangle $. We can further absorb $\pi^2$ to the shared factor $N_k$ and define
\begin{equation}
    g^{(k)}_i= \frac{\widetilde{Z}^{\rm loc}_i}{\pi^2}. 
\end{equation}
Using the observation we mentioned before, 
\begin{equation}
\begin{aligned}
    g^{(k)}_i\langle e_k, P^{(k)}_i u_k \rangle&= \frac{\prod_{j\neq i}\Gamma^2(1+\xi^{(k)}_j/\epsilon-\xi^{(k)}_i/\epsilon)  }{\prod_{l=1}^{k-1}  \Gamma(1+\xi^{(k-1)}_l/\epsilon-\xi^{(k)}_i/\epsilon)\prod_{l=1}^{k+1} \Gamma(1+\xi^{(k+1)}_l/\epsilon -\xi^{(k)}_i/\epsilon ) }(-1)^{k-1}\epsilon^k \\
    & \times\frac{1}{\prod_{j\neq i} (\xi^{(k)}_i-\xi^{(k)}_j)} \Delta^{1,\dots,k-1,k}_{1,\dots,k-1,k+1}(\frac{B-\xi^{(k)}_i \rm Id}{\epsilon})\\
    &= \frac{\prod_{j\neq i}\Gamma(1+\xi^{(k)}_j/\epsilon-\xi^{(k)}_i/\epsilon)\prod_{j\neq i}\Gamma(\xi^{(k)}_j/\epsilon-\xi^{(k)}_i/\epsilon)  }{\prod_{l=1}^{k-1}  \Gamma(1+\xi^{(k-1)}_l/\epsilon-\xi^{(k)}_i/\epsilon)\prod_{l=1}^{k+1} \Gamma(1+\xi^{(k+1)}_l/\epsilon -\xi^{(k)}_i/\epsilon ) } \prod_{j\neq i } \frac{\xi^{(k)}_j-\xi^{(k)}_i}{\epsilon} \\
    &\times \frac{1}{\prod_{j\neq i} (\xi^{(k)}_i-\xi^{(k)}_j)} \Delta^{1,\dots,k-1,k}_{1,\dots,k-1,k+1}(\frac{B-\xi^{(k)}_i \rm Id}{\epsilon}) (-1)^{k-1}\epsilon^k\\
    &= \epsilon \frac{\prod_{j\neq i}\Gamma(1+\frac{\xi^{(k)}_j-\xi^{(k)}_i}{\epsilon})\prod_{j\neq i}\Gamma(\frac{\xi^{(k)}_j-\xi^{(k)}_i}{\epsilon})  }{\prod_{l=1}^{k-1}  \Gamma(1+\frac{\xi^{(k-1)}_l-\xi^{(k)}_i}{\epsilon})\prod_{l=1}^{k+1} \Gamma(1+\frac{\xi^{(k+1)}_l -\xi^{(k)}_i}{\epsilon} ) }\Delta^{1,\dots,k-1,k}_{1,\dots,k-1,k+1}(\frac{B-\xi^{(k)}_i \rm Id}{\epsilon}), 
\end{aligned}
\end{equation}
or
\begin{equation}
    g^{(k)}_i \langle e_k, P^{(k)}_i \frac{u_k}{\epsilon} \rangle=\frac{\prod_{j\neq i}\Gamma(1+\frac{\xi^{(k)}_j-\xi^{(k)}_i}{\epsilon})\prod_{j\neq i}\Gamma(\frac{\xi^{(k)}_j-\xi^{(k)}_i}{\epsilon})  }{\prod_{l=1}^{k-1}  \Gamma(1+\frac{\xi^{(k-1)}_l-\xi^{(k)}_i}{\epsilon})\prod_{l=1}^{k+1} \Gamma(1+\frac{\xi^{(k+1)}_l -\xi^{(k)}_i}{\epsilon} ) }\Delta^{1,\dots,k-1,k}_{1,\dots,k-1,k+1}(\frac{B-\xi^{(k)}_i \rm Id}{\epsilon}).
\end{equation}
This result is comparable with the Stokes matrix given in \cite{xu2024regularizedlimitsstokesmatrices} and \cite{alekseev2024wkbasymptoticsstokesmatrices}. In fact, it reproduces exactly the summing channels of the regularized Stokes matrix entry given in Theorem ($2.11$) in \cite{alekseev2024wkbasymptoticsstokesmatrices}.

\section{Direct calculations of the isomonodromy tau functions of $(1,2)$ system}
\label{app: D}
We consider the semi-classical ODE system
\begin{equation}
    \epsilon \frac{dF}{dz}=  \left( iu- i\frac{B}{z} \right)F,
\end{equation}
which is equivalent to $\frac{dF}{dz}= \left( U+ \frac{V}{z} \right)F$, where we rename $U:=\frac{iu}{\epsilon}$ and $V:=-\frac{iB}{\epsilon}$. Now the only regular pole is $0$. Therefore the time is $Q=U_{2}-U_{1}$. For convenience, let 
\begin{equation}
    U= \begin{pmatrix}
-\frac{Q}{2} & 0 \\
0 & \frac{Q}{2}  
\end{pmatrix}.\nonumber
\end{equation}
In order to obtain the isomonodromy $\tau$, let $V=V(Q)$. Consider the deformation equation 
\begin{equation}
    \frac{\partial F}{\partial Q} = (zH+M)F, \qquad H=\begin{pmatrix}
-\frac{1}{2} & 0 \\
0 & \frac{1}{2}  
\end{pmatrix}. 
\end{equation}
The zero curvature condition is then 
\begin{equation}
    \partial_Q \left( U+\frac{V}{z} \right) -\partial_z (zH+M) + \left[U+\frac{V}{z},zH+M \right] =0. 
\end{equation}
Direct computations yield $ M_{12} =\frac{V_{12}}{Q}, \,M_{21}=\frac{V_{21}}{Q}, \, M_{11}=M_{22}=0$. Further
\begin{equation}
    Q\frac{dV_{11}}{dQ}=Q\frac{dV_{22}}{dQ}=0, \,\,Q\frac{dV_{12}}{dQ}= (V_{22}-V_{11})V_{12},\,\,     ,Q\frac{dV_{21}}{dQ}= (V_{11}-V_{22})V_{21}. 
\end{equation}
Therefore 
\begin{equation}
    Q\frac{d(V_{12}V_{21})}{dQ}= V_{21}Q\frac{dV_{12}}{dQ}+V_{12}Q \frac{dV_{21}}{dQ}=0. 
\end{equation}
Hence $V_{12}V_{21}$ is a constant. The JMU one-form is then given by
\begin{equation}
    d\log \tau = V_{12}V_{21} d\log Q \implies Q\frac{d}{dQ} \log \tau =V_{12}V_{21}. 
\end{equation}
Solving the differential equation gives
\begin{equation}
    \tau(Q) =C_{\text{mono}} Q^{V_{12}V_{21}},
\end{equation}
here is $C_{\text{mono}}$ is a constant depending on the Stokes data but not $Q$. Since
$V_{12}V_{21}=-\frac{B_{12}B_{21}}{\epsilon^2}$. Using (\ref{eq: parameters}) we rewrite
\begin{equation}
 B=   \begin{pmatrix}
t_1 & B_{12} \\
B_{21} & t_2  
\end{pmatrix}.
\end{equation}
Since $\mu_1, \mu_2$ are the eigenvalues,
\begin{equation}
    \det B=\mu_1\mu_2= t_1(\mu_1+\mu_2-t_1)-B_{12}B_{21} \implies B_{12}B_{21}=-(t_1-\mu_1)(t_1-\mu_2)=-M_1M_2 .\nonumber
\end{equation}
Then 
\begin{equation}
    V_{12}V_{21}=\frac{M_1M_2}{\epsilon^2}=-\frac{M_1M_2}{\hbar^2},\,\, \hbar=-i\epsilon. 
\end{equation}
Finally we conclude
\begin{equation}
    \boxed{\tau(Q) = C_{\text{mono}} Q^{-\frac{M_1M_2}{\hbar^2}} .}
\end{equation}

\section{U(1) partition function}
\label{app: E}

We first review the Nekrasov partition function\cite{nekrasov2002seibergwittenprepotentialinstantoncounting,nekrasov2003seibergwittentheoryrandompartitions}. Let
\begin{equation}
    Y^{(k)}=(Y^{(k)}_{1}, \dots Y^{(k)}_k)
\end{equation}
be a set of k-tuple Young diagrams associated to the gauge node $U(k)$. Also for the flavor node we let $Y^{(n)}=(\emptyset, \dots, \emptyset)$. 

We then define the Nekrasov factor \cite{nekrasov2002seibergwittenprepotentialinstantoncounting}
\begin{equation}
    N_{Y,W}(a)= \prod_{s\in Y} \left( a-\epsilon_1 L_W(s)+\epsilon_2(A_Y(s)+1)\right)\prod_{t\in W} \left(a-\epsilon_2A_W(t)+\epsilon_1(L_Y(t)+1)\right).
\end{equation}
Here $\epsilon_{i}$ are the parameters of the $\Omega$ background \cite{nekrasov2002seibergwittenprepotentialinstantoncounting}, and $A_Y(s),L_Y(s)$ denote the arm length and leg length of the Young diagram $s$, respectively. 

The $U(1)$-instanton partition function can be computed:
\begin{equation}
    \mathcal{Z}_{\text{inst}} (\sigma;m,-m;q)= \sum_{Y} q^{\lvert Y \rvert}\frac{N_{\emptyset,Y}(m-\sigma)N_{Y,{\emptyset}}(-m-\sigma)}{N_{Y,Y}(0)}=\sum_{Y} q^{\lvert Y \rvert}\frac{N_{\emptyset,Y}(M_1)N_{Y,{\emptyset}}(M_2)}{N_{Y,Y}(0)}.
\end{equation}
We use $(i,j)$ to denote the Young diagram. 
\begin{equation}
\begin{aligned}
    N_{\emptyset,Y}(\alpha)=\prod_{s=(i,j)\in Y} (\alpha-\epsilon_2A_{Y}(s)+\epsilon_1(L_{\emptyset}(s)+1))
    = \prod_{i\geq1} \prod_{j=1}^{Y_i} \left(  \alpha-\epsilon_2(Y_i-j) +\epsilon_1(1-i)  .       \right)
    \end{aligned}
\end{equation}
Here we use $L_{\emptyset}((i,j))=-i$ and $A_{Y}(i,j)=Y_i-j $. Let $\beta:=-\frac{\epsilon_2}{\epsilon_1}$, $\alpha=\epsilon_1u$ and $v=Y_i-j$ and we can rewrite
\begin{equation}
\begin{aligned}
    N_{\emptyset,Y}(\alpha)= \prod_{i\geq1} \prod_{v=0}^{Y_i-1}(\epsilon_1u-\epsilon_2v+\epsilon_1(1-i) )
    = \epsilon_1^{|Y|}\prod_{i\geq1} \prod_{v=0}^{Y_i-1} (u+\beta v+1-i). 
\end{aligned}
\end{equation}
We also rewrite the denominator. By definition
\begin{equation}
    \begin{aligned}
        N_{Y,Y}(0)&=\prod_{s\in Y}(-\epsilon_1L_W(s)+\epsilon_2(A_Y(s)+1))(-\epsilon_2A_W(s)+\epsilon_1(L_Y(s)+1))  \\
        &=(-1)^{\lvert Y\rvert} \epsilon_1^{2\lvert Y \rvert} \prod_{s\in Y} (L_Y(s)+\beta(A_Y(s)+1)) (L_Y(s)+1+\beta A_Y(s)).
    \end{aligned}
\end{equation}
Define 
\begin{equation}
    f_Y(\beta): = \prod_{s\in Y} (L_Y(s)+\beta(A_Y(s)+1)),\,\, g_Y(\beta)=\prod_{s\in Y}(L_Y(s)+1+\beta A_Y(s)),
\end{equation}
Then we rewrite 
\begin{equation}
    N_{Y,Y}(0)=(-1)^{\lvert Y\rvert} \epsilon_1^{2\lvert Y \rvert} f_Y(\beta)g_Y(\beta). 
\end{equation}
We also define 
\begin{equation}
    J_Y(u,\beta):=\prod_{i\geq1} \prod_{v=0}^{Y_i-1} (u+\beta v+1-i). 
\end{equation}
Therefore, 
\begin{equation}
\begin{aligned}
N_{Y,\varnothing}(M_2)
&=(-\epsilon_1)^{|Y|}J_Y(\beta-1-u_2,\beta),
\\
\frac{N_{\varnothing,Y}(M_1)N_{Y,\varnothing}(M_2)}
{N_{Y,Y}(0)}
&=\frac{J_Y(u_1,\beta)J_Y(\beta-1-u_2,\beta)}
{f_Y(\beta)g_Y(\beta)},
\\
\sum_Y q^{|Y|}
\frac{N_{\varnothing,Y}(M_1)N_{Y,\varnothing}(M_2)}
{N_{Y,Y}(0)}
&=
\sum_Y q^{|Y|}
\frac{J_Y(u_1,\beta)J_Y(\beta-1-u_2,\beta)}
{f_Y(\beta)g_Y(\beta)}.
\end{aligned}
\end{equation}
here $M_1=\epsilon_1u_1$, $M_2=\epsilon_1u_2$. The key fact is that $f_Y(\beta)g_Y(\beta)$ is precisely the Jack inner product, therefore we may apply the Cauchy identity for Jack functions,\
\begin{equation}
    Z_{\mathrm{inst}}=\sum_Y q^{|Y|}\frac{J_Y(u_1,\beta)J_Y(\beta-1-u_2,\beta)}{f_Y(\beta)g_Y(\beta)}=(1-q)^{-\frac{u_1(\beta-1-u_2)}{\beta}}=(1-q)^{-\frac{M_1(M_2+\epsilon_1+\epsilon_2)}{\epsilon_1\epsilon_2}}.
\end{equation}

We note that the correlator of $2$-point free-boson vertex operators is given by
\begin{equation}
    \langle V_{a_1}(z_1) V_{a_2}(z_2)   \rangle = (z_1-z_2)^{2a_1a_2}.
\end{equation}
Through appropriately choosing parameters the instanton part can be identified with a free-boson conformal block. 

\begin{Rmk}
The spectral-network data determine the semiclassical/perturbative factors.
The instanton or regular conformal-block factor must be supplied by the gauged channel / JMU regular part. Since the corresponding quiver is
\begin{equation}
U(2)_{\mathrm{flavor}}-U(1)_{\mathrm{gauge}}, \nonumber
\end{equation}
the parameter $q$ is not a genuine gauge coupling of a conformal four-dimensional gauge theory. Therefore this factor should not be interpreted as a physical Nekrasov partition function, but rather as a formal Nekrasov-type partition function or a conformal-block normalization factor.
\end{Rmk}

\section{Consistency check with the Painlev'e III Kiev coefficient}
\label{app:PIII_Kiev_check}

In this appendix, we compare the perturbative coefficient obtained from the double-degeneration and gluing construction with the structure coefficient in the Kiev-type expansion of the generic Painlev'e III tau function. This comparison is an independent consistency check: the endpoint monodromy factors and the vector factor are derived from the local Stokes data and the gluing process, rather than inferred from the Painlev'e III formula. We take the normalization $G(z+1)=\Gamma(z)G(z)$. 

\subsection{Perturbative coefficient from gluing}

Let $a$ be the traceless internal action variable and define the Coulomb-shift lattice by

\begin{equation}
a_l=a+l\hbar, \qquad \Sigma_l=\frac{a_l}{\hbar}=\Sigma+l, \qquad \Sigma=\frac{a}{\hbar}.
\label{eq:app_shifted_action}
\end{equation}

Let $\sigma_{\rm out}$ and $\sigma_{\rm in}$ be the traceless formal-monodromy parameters of the two genus-zero components. The endpoint monodromy factors are

\begin{equation}
C_{\rm mono}^{\alpha}(a_l;\sigma_\alpha) =
G\left( 1+\frac{a_l-\sigma_\alpha}{\hbar} \right) G\left( 1+\frac{-a_l-\sigma_\alpha}{\hbar} \right), \qquad \alpha\in{{\rm out},{\rm in}}.
\label{eq:app_endpoint_factor}
\end{equation}

Gluing the two components gauges their common $U(2)$ boundary symmetry. The central $U(1)$ part is treated separately, while the non-Abelian contribution is determined by the two oriented roots of $SU(2)$. On the traceless slice $(a_1,a_2)=(a_l,-a_l)$, the vector gluing factor is

\begin{equation}
Z_{\rm vec}^{\rm neck}(a_l) = \prod_{\nu\in\Delta} G\left( 1+\frac{(a_l,\nu)}{\hbar}
\right)^{-1} = \frac{1}{ G\left(1+\frac{2a_l}{\hbar}\right) G\left(1-\frac{2a_l}{\hbar}\right) }.
\label{eq:app_vector_factor}
\end{equation}

The complete perturbative coefficient is therefore

\begin{equation}
Z_{\rm pert}(a_l) = C_{\rm mono}^{\rm out}(a_l;\sigma_{\rm out})
Z_{\rm vec}^{\rm neck}(a_l) C_{\rm mono}^{\rm in}(a_l;\sigma_{\rm in}).
\label{eq:app_Zpert_definition}
\end{equation}

Explicitly,

\begin{equation}
Z_{\rm pert}(a_l) = \frac{ \displaystyle \prod_{\alpha={\rm out,in}} G\left( 1+\frac{a_l-\sigma_\alpha}{\hbar} \right) G\left( 1+\frac{-a_l-\sigma_\alpha}{\hbar}
\right)}{\displaystyle G\left(1+\frac{2a_l}{\hbar}\right) G\left(1-\frac{2a_l}{\hbar}\right)
}.
\label{eq:app_our_Zpert}
\end{equation}

\subsection{Comparison with the Kiev coefficient}

For the generic $P_{\mathrm{III}}(D_6)$ equation, denoted by $P_{\mathrm{III}'_1}$ in the Kiev convention, the structure coefficient is \cite{Gamayun_2013,Iorgov_2014}

\begin{equation}
C_{\mathrm{III}'_1} \left( \theta_*, \theta_\star, \Sigma \right)=\prod_{\varepsilon=\pm} \frac{ G(1+\theta_*+\varepsilon\Sigma)
G(1+\theta_\star+\varepsilon\Sigma) }{ G(1+2\varepsilon\Sigma)}.
\label{eq:app_Kiev_coefficient}
\end{equation}

The parameter identification is

\begin{equation}
\theta_* = -\frac{\sigma_{\rm out}}{\hbar},
\qquad \theta_\star = -\frac{\sigma_{\rm in}}{\hbar},
\qquad \Sigma_l = \frac{a_l}{\hbar}.
\label{eq:app_parameter_identification}
\end{equation}

Using this dictionary, the two endpoint factors become

\begin{equation}
C_{\rm mono}^{\rm out}(a_l;\sigma_{\rm out}) = \prod_{\varepsilon=\pm}
G\left( 1+\theta_*+\varepsilon\Sigma_l \right), \, C_{\rm mono}^{\rm in}(a_l;\sigma_{\rm in}) = \prod_{\varepsilon=\pm} G\left( 1+\theta_\star+\varepsilon\Sigma_l \right),
\end{equation}

while the vector factor becomes

\begin{equation}
Z_{\rm vec}^{\rm neck}(a_l) = \prod_{\varepsilon=\pm} G\left( 1+2\varepsilon\Sigma_l
\right)^{-1}.
\label{eq:app_vector_match}
\end{equation}

Consequently,

\begin{equation}
Z_{\rm pert}(a_l) = C_{\mathrm{III}'_1} \left( \theta_*, \theta_\star, \Sigma+l \right).
\label{eq:app_exact_match}
\end{equation}

Thus the perturbative coefficient obtained from the gluing construction agrees with the Painlev'e III Kiev coefficient sector by sector. The four Barnes $G$-functions in the numerator are supplied by the two endpoint systems, while the two Barnes $G$-functions in the denominator are supplied by the single gauged $SU(2)$ neck.

\begin{table}[h]
    \centering
    \renewcommand{\arraystretch}{1.3}
    \begin{tabular}{c|c|p{6.2cm}}
        \hline
        Our notation & Kiev notation & Interpretation \\
        \hline
        $a$ & $\hbar\Sigma$ &
        Internal traceless action or Coulomb parameter \\
        
        $a_l=a+l\hbar$ & $\hbar(\Sigma+l)$ &
        Shifted internal channel \\
        
        $l$ & $n$ &
        Fourier or Coulomb-shift lattice index \\
        
        $\sigma_{\rm out}$ & $-\hbar\theta_*$ &
        Outer formal-monodromy parameter \\
        
        $\sigma_{\rm in}$ & $-\hbar\theta_\star$ &
        Inner formal-monodromy parameter \\
        
        $e^{2\pi i\theta}$ & $s$ &
        Exponentiated angle coordinate \\
        
        $C_{\rm mono}^{\rm out}$ &
        $\prod_{\varepsilon=\pm}G(1+\theta_*+\varepsilon\Sigma_l)$ &
        Outer endpoint connection coefficient \\
        
        $C_{\rm mono}^{\rm in}$ &
        $\prod_{\varepsilon=\pm}G(1+\theta_\star+\varepsilon\Sigma_l)$ &
        Inner endpoint connection coefficient \\
        
        $Z_{\rm vec}^{\rm neck}$ &
        $\prod_{\varepsilon=\pm}G(1+2\varepsilon\Sigma_l)^{-1}$ &
        Vector gluing factor \\
        
        $Z_{\rm pert}$ &
        $C_{\mathrm{III}'_1}(\theta_*,\theta_\star,\Sigma_l)$ &
        Complete perturbative coefficient \\
        \hline
    \end{tabular}
    \caption{Dictionary between the gluing construction and the Kiev-type expansion.}
    \label{tab:app_PIII_dictionary}
\end{table}

\end{appendices}

% =================================================
% References
% =================================================
\bibliographystyle{unsrtnat}
\bibliography{references}

@article{alekseev2024wkbasymptoticsstokesmatrices,
author = {Alekseev, Anton and Neitzke, Andrew and Xu, Xiaomeng and Zhou, Yan},
year = {2024},
month = {10},
pages = {},
title = {WKB Asymptotics of Stokes Matrices, Spectral Curves and Rhombus Inequalities},
volume = {405},
journal = {Communications in Mathematical Physics},
doi = {10.1007/s00220-024-05133-0},
url={https://arxiv.org/abs/2403.17906}
}

@misc{xu2024regularizedlimitsstokesmatrices,
      title={Regularized limits of Stokes matrices, isomonodromy deformation and crystal basis}, 
      author={Xiaomeng Xu},
      year={2024},
      eprint={1912.07196},
      archivePrefix={arXiv},
      primaryClass={math.RT},
      url={https://arxiv.org/abs/1912.07196}, 
}

@article{crooks2023doublegelfandcetlinsysteminvariance,
  title={The double Gelfand–Cetlin system, invariance of polarization, and the Peter–Weyl theorem},
  author={Peter Crooks and Jonathan Weitsman},
  journal={Journal of Geometry and Physics},
  year={2023},
  url={https://api.semanticscholar.org/CorpusID:257900783}
}

@article{Gavrylenko_2018,
   title={Fredholm Determinant and Nekrasov Sum Representations of Isomonodromic Tau Functions},
   volume={363},
   ISSN={1432-0916},
   url={http://dx.doi.org/10.1007/s00220-018-3224-7},
   DOI={10.1007/s00220-018-3224-7},
   number={1},
   journal={Communications in Mathematical Physics},
   publisher={Springer Science and Business Media LLC},
   author={Gavrylenko, P. and Lisovyy, O.},
   year={2018},
   month=Aug, pages={1–58} }

@article{wang2025asymptoticmonodromyproblemshigherorder,
      title={Asymptotic and monodromy problems for higher-order Painlev\'e III equations}, 
      author={Zikang Wang and Xiaomeng Xu},
      year={2025},
      journal={Communications in Mathematical Physics, to appear},
      eprint={2512.19381},
      archivePrefix={arXiv},
      primaryClass={math.CA},
      url={https://arxiv.org/abs/2512.19381}, 
}

@article{Gaiotto_2012,
   title={N=2 dualities},
   volume={2012},
   ISSN={1029-8479},
   url={http://dx.doi.org/10.1007/JHEP08(2012)034},
   DOI={10.1007/jhep08(2012)034},
   number={8},
   journal={Journal of High Energy Physics},
   publisher={Springer Science and Business Media LLC},
   author={Gaiotto, Davide},
   year={2012},
   month=Aug }

@article{Gaiotto_2013,
   title={Spectral Networks},
   volume={14},
   ISSN={1424-0661},
   url={http://dx.doi.org/10.1007/s00023-013-0239-7},
   DOI={10.1007/s00023-013-0239-7},
   number={7},
   journal={Annales Henri Poincaré},
   publisher={Springer Science and Business Media LLC},
   author={Gaiotto, Davide and Moore, Gregory W. and Neitzke, Andrew},
   year={2013},
   month=Mar, pages={1643–1731} }

@article{nekrasov2002seibergwittenprepotentialinstantoncounting,
    author = "Nekrasov, Nikita A.",
    title = "{Seiberg-Witten prepotential from instanton counting}",
    eprint = "hep-th/0206161",
    archivePrefix = "arXiv",
    reportNumber = "ITEP-TH-22-02, IHES-P-04-22",
    doi = "10.4310/ATMP.2003.v7.n5.a4",
    journal = "Adv. Theor. Math. Phys.",
    volume = "7",
    number = "5",
    pages = "831--864",
    year = "2003"
}

@article{Hollands_2020,
   title={Exact WKB and Abelianization for the $T_3$ Equation},
   volume={380},
   ISSN={1432-0916},
   url={http://dx.doi.org/10.1007/s00220-020-03875-1},
   DOI={10.1007/s00220-020-03875-1},
   number={1},
   journal={Communications in Mathematical Physics},
   publisher={Springer Science and Business Media LLC},
   author={Hollands, Lotte and Neitzke, Andrew},
   year={2020},
   month=Oct, pages={131–186} }

@article{Iorgov_2014,
   title={Isomonodromic Tau-Functions from Liouville Conformal Blocks},
   volume={336},
   ISSN={1432-0916},
   url={http://dx.doi.org/10.1007/s00220-014-2245-0},
   DOI={10.1007/s00220-014-2245-0},
   number={2},
   journal={Communications in Mathematical Physics},
   publisher={Springer Science and Business Media LLC},
   author={Iorgov, N. and Lisovyy, O. and Teschner, J.},
   year={2014},
   month=Dec, pages={671–694} }

@article{Gamayun_2012,
   title={Conformal field theory of Painlevé VI},
   volume={2012},
   ISSN={1029-8479},
   url={http://dx.doi.org/10.1007/JHEP10(2012)038},
   DOI={10.1007/jhep10(2012)038},
   number={10},
   journal={Journal of High Energy Physics},
   publisher={Springer Science and Business Media LLC},
   author={Gamayun, O. and Iorgov, N. and Lisovyy, O.},
   year={2012},
   month=Oct }

@article{nekrasov2003seibergwittentheoryrandompartitions,
    author = "Nekrasov, Nikita and Okounkov, Andrei",
    title = "{Seiberg-Witten theory and random partitions}",
    eprint = "hep-th/0306238",
    archivePrefix = "arXiv",
    reportNumber = "ITEP-TH-36-03, PUDM-2003, IHES-P-03-43",
    doi = "10.1007/0-8176-4467-9_15",
    journal = "Prog. Math.",
    volume = "244",
    pages = "525--596",
    year = "2006"
}

@article{Alday_2010,
   title={Liouville Correlation Functions from Four-Dimensional Gauge Theories},
   volume={91},
   ISSN={1573-0530},
   url={http://dx.doi.org/10.1007/s11005-010-0369-5},
   DOI={10.1007/s11005-010-0369-5},
   number={2},
   journal={Letters in Mathematical Physics},
   publisher={Springer Science and Business Media LLC},
   author={Alday, Luis F. and Gaiotto, Davide and Tachikawa, Yuji},
   year={2010},
   month=Jan, pages={167–197} }

@article{Its_2014,
   title={Connection Problem for the Sine-Gordon/Painlevé III Tau Function and Irregular Conformal Blocks: Fig. 1.},
   volume={2015},
   ISSN={1687-0247},
   url={http://dx.doi.org/10.1093/imrn/rnu209},
   DOI={10.1093/imrn/rnu209},
   number={18},
   journal={International Mathematics Research Notices},
   publisher={Oxford University Press (OUP)},
   author={Its, Alexander and Lisovyy, Oleg and Tykhyy, Yuriy},
   year={2014},
   month=Nov, pages={8903–8924} }

@article{BERENSTEIN199649,
title = {Parametrizations of Canonical Bases and Totally Positive Matrices},
journal = {Advances in Mathematics},
volume = {122},
number = {1},
pages = {49-149},
year = {1996},
issn = {0001-8708},
doi = {https://doi.org/10.1006/aima.1996.0057},
url = {https://www.sciencedirect.com/science/article/pii/S0001870896900572},
author = {Arkady Berenstein and Sergey Fomin and Andrei Zelevinsky}
}

@article{Its_2018,
   title={Monodromy dependence and connection formulae for isomonodromic tau functions},
   volume={167},
   ISSN={0012-7094},
   url={http://dx.doi.org/10.1215/00127094-2017-0055},
   DOI={10.1215/00127094-2017-0055},
   number={7},
   journal={Duke Mathematical Journal},
   publisher={Duke University Press},
   author={Its, A. R. and Lisovyy, O. and Prokhorov, A.},
   year={2018},
   month=May }

@misc{nanopoulos2010hitchinequationirregularsingularity,
      title={Hitchin Equation, Irregular Singularity, and $N=2$ Asymptotical Free Theories}, 
      author={Dimitri Nanopoulos and Dan Xie},
      year={2010},
      eprint={1005.1350},
      archivePrefix={arXiv},
      primaryClass={hep-th},
      url={https://arxiv.org/abs/1005.1350}, 
}

@article{Witten_1997,
   title={Solutions of four-dimensional field theories via M-theory},
   volume={500},
   ISSN={0550-3213},
   url={http://dx.doi.org/10.1016/S0550-3213(97)00416-1},
   DOI={10.1016/s0550-3213(97)00416-1},
   number={1-3},
   journal={Nuclear Physics B},
   publisher={Elsevier BV},
   author={Witten, Edward},
   year={1997},
   month=Sept, pages={3–42} }

@article{Seiberg_1994,
   title={Electric-magnetic duality, monopole condensation, and confinement in N=2 supersymmetric Yang-Mills theory},
   volume={426},
   ISSN={0550-3213},
   url={http://dx.doi.org/10.1016/0550-3213(94)90124-4},
   DOI={10.1016/0550-3213(94)90124-4},
   number={1},
   journal={Nuclear Physics B},
   publisher={Elsevier BV},
   author={Seiberg, N. and Witten, E.},
   year={1994},
   month=Sept, pages={19–52} }

@article{Bertola_2009,
   title={The Dependence on the Monodromy Data of the Isomonodromic Tau Function},
   volume={294},
   ISSN={1432-0916},
   url={http://dx.doi.org/10.1007/s00220-009-0961-7},
   DOI={10.1007/s00220-009-0961-7},
   number={2},
   journal={Communications in Mathematical Physics},
   publisher={Springer Science and Business Media LLC},
   author={Bertola, M.},
   year={2009},
   month=Dec, pages={539–579} }

@article{bertola2020corrigendumdependencemonodromydata,
  title={Correction to: The Dependence on the Monodromy Data of the Isomonodromic Tau Function},
  author={Marco Bertola},
  journal={Communications in Mathematical Physics},
  year={2021},
  volume={381},
  pages={1445-1461},
  url={https://api.semanticscholar.org/CorpusID:231971662}
}

@article{Jimbo:1981tov,
    author = "Jimbo, Michio and Miwa, Tetsuji and Ueno, Kimio",
    title = "{Monodromy preserving deformation of linear ordinary differential equations with rational coefficients}: {I. General theory and {\ensuremath{\tau}}-function}",
    doi = "10.1016/0167-2789(81)90013-0",
    journal = "Physica D",
    volume = "2",
    number = "2",
    pages = "306--352",
    year = "1981"
}

@article{gaiotto2009asymptoticallyfreen2theories,
    author = "Gaiotto, Davide",
    editor = "Das, Sumit R. and Shapere, Alfred D.",
    title = "{Asymptotically free $\mathcal{N} = 2$ theories and irregular conformal blocks}",
    eprint = "0908.0307",
    archivePrefix = "arXiv",
    primaryClass = "hep-th",
    doi = "10.1088/1742-6596/462/1/012014",
    journal = "J. Phys. Conf. Ser.",
    volume = "462",
    number = "1",
    pages = "012014",
    year = "2013"
}

@article{5f8535f3-14dc-32c9-b9a7-71c41c9f6c11,
 ISSN = {00255521, 19031807},
 URL = {http://www.jstor.org/stable/24491170},
 author = {FINN KNUDSEN and DAVID MUMFORD},
 journal = {Mathematica Scandinavica},
 number = {1},
 pages = {19--55},
 publisher = {Mathematica Scandinavica},
 title = {THE PROJECTIVITY OF THE MODULI SPACE OF STABLE CURVES I: PRELIMINARIES ON "det" AND "Div"},
 urldate = {2026-07-08},
 volume = {39},
 year = {1976}
}

@article{Kimura_2016,
   title={Quiver W-algebras},
   volume={108},
   ISSN={1573-0530},
   url={http://dx.doi.org/10.1007/s11005-018-1072-1},
   DOI={10.1007/s11005-018-1072-1},
   number={6},
   journal={Letters in Mathematical Physics},
   publisher={Springer Science and Business Media LLC},
   author={Kimura, Taro and Pestun, Vasily},
   year={2018},
   month=Mar, pages={1351–1381} }

@article{Hitchin:1987mz,
    author = "Hitchin, Nigel J.",
    title = "{Stable bundles and integrable systems}",
    doi = "10.1215/S0012-7094-87-05408-1",
    journal = "Duke Math. J.",
    volume = "54",
    pages = "91--114",
    year = "1987"
}

@article{Ramanan1989,
author = {Ramanan, S. and Narasimhan, M.S. and Beauville, A.},
journal = {Journal für die reine und angewandte Mathematik},
pages = {169-179},
title = {Spectral curves and the generalised theta divisor.},
url = {http://eudml.org/doc/153148},
volume = {398},
year = {1989},
}

@article{biquard2002wildnonabelianhodgetheory, title={Wild non-abelian Hodge theory on curves}, volume={140}, DOI={10.1112/S0010437X03000010}, number={1}, journal={Compositio Mathematica}, author={Biquard, Olivier and Boalch, Philip}, year={2004}, pages={179–204}}

@article{GUILLEMIN1983106,
title = {The Gelfand-Cetlin system and quantization of the complex flag manifolds},
journal = {Journal of Functional Analysis},
volume = {52},
number = {1},
pages = {106-128},
year = {1983},
issn = {0022-1236},
doi = {https://doi.org/10.1016/0022-1236(83)90092-7},
url = {https://www.sciencedirect.com/science/article/pii/0022123683900927},
author = {V Guillemin and S Sternberg}
}

@article{gaiotto2011wallcrossinghitchinsystemswkb,
  title={Wall-crossing, Hitchin Systems, and the WKB Approximation},
  author={Davide Gaiotto and Gregory W. Moore and Andrew Neitzke},
  journal={Advances in Mathematics},
  year={2009},
  volume={234},
  pages={239-403},
  url={https://api.semanticscholar.org/CorpusID:115176676}
}

@article{Bershtein_2017,
   title={Bäcklund transformation of Painlevé III(D8) tau function},
   volume={50},
   ISSN={1751-8121},
   url={http://dx.doi.org/10.1088/1751-8121/aa59c9},
   DOI={10.1088/1751-8121/aa59c9},
   number={11},
   journal={Journal of Physics A: Mathematical and Theoretical},
   publisher={IOP Publishing},
   author={Bershtein, M A and Shchechkin, A I},
   year={2017},
   month=Feb, pages={115205} }

@book{KatoPerturbation,
  author    = {Tosio Kato},
  title     = {Perturbation Theory for Linear Operators},
  publisher = {Springer},
  address   = {Berlin, Heidelberg},
  series    = {Grundlehren der mathematischen Wissenschaften},
  volume    = {132},
  year      = {1966},
  doi       = {10.1007/978-3-662-12678-3}
}

@article{Gamayun_2013,
   title={How instanton combinatorics solves Painlevé VI, V and IIIs},
   volume={46},
   ISSN={1751-8121},
   url={http://dx.doi.org/10.1088/1751-8113/46/33/335203},
   DOI={10.1088/1751-8113/46/33/335203},
   number={33},
   journal={Journal of Physics A: Mathematical and Theoretical},
   publisher={IOP Publishing},
   author={Gamayun, O and Iorgov, N and Lisovyy, O},
   year={2013},
   month=July, pages={335203} }

@article{Sun_2016,
   title={A new integral formula for Heckman–Opdam hypergeometric functions},
   volume={289},
   ISSN={0001-8708},
   url={http://dx.doi.org/10.1016/j.aim.2015.09.037},
   DOI={10.1016/j.aim.2015.09.037},
   journal={Advances in Mathematics},
   publisher={Elsevier BV},
   author={Sun, Yi},
   year={2016},
   month=Feb, pages={1157–1204} }

@article{olshanski2013projectionsorbitalmeasuresgelfandtsetlin,
     author = {Grigori Olshanski},
     title = {Projections of {Orbital} {Measures,} {Gelfand-Tsetlin} {Polytopes,} and {Splines}},
     journal = {Journal of Lie Theory},
     pages = {1011--1022},
     year = {2013},
     volume = {23},
     number = {4},
     doi = {10.5802/jolt.762},
     zbl = {1281.22003},
     url = {https://jolt.centre-mersenne.org/articles/10.5802/jolt.762/}
}

@article{Okamoto1986StudiesOT,
  title={Studies on the Painlev{\'e} equations},
  author={Kazuo Okamoto},
  journal={Annali di Matematica Pura ed Applicata},
  year={1986},
  volume={146},
  pages={337-381},
  url={https://api.semanticscholar.org/CorpusID:125511090}
}

@article{Tang:2024qyr,
    author = "Tang, Qian and Xu, Xiaomeng",
    title = "{The Boundary Condition for Some Isomonodromy Equations}",
    eprint = "2402.07269",
    archivePrefix = "arXiv",
    primaryClass = "math.CA",
    month = "3",
    year = "2024"
}

@article{Boalch_2001,
   title={Stokes matrices, Poisson Lie groups and Frobenius manifolds},
   volume={146},
   ISSN={1432-1297},
   url={http://dx.doi.org/10.1007/s002220100170},
   DOI={10.1007/s002220100170},
   number={3},
   journal={Inventiones Mathematicae},
   publisher={Springer Science and Business Media LLC},
   author={Boalch, P.P.},
   year={2001},
   month=Dec, pages={479–506} }

@article{TOLEDANOLAREDO2023109189,
title = {Stokes phenomena, Poisson–Lie groups and quantum groups},
journal = {Advances in Mathematics},
volume = {429},
pages = {109189},
year = {2023},
issn = {0001-8708},
doi = {https://doi.org/10.1016/j.aim.2023.109189},
url = {https://www.sciencedirect.com/science/article/pii/S0001870823003328},
author = {Valerio {Toledano Laredo} and Xiaomeng Xu},
}

@article{Boalch_2007,
   title={Quasi-Hamiltonian geometry of meromorphic connections},
   volume={139},
   ISSN={0012-7094},
   url={http://dx.doi.org/10.1215/S0012-7094-07-13924-3},
   DOI={10.1215/s0012-7094-07-13924-3},
   number={2},
   journal={Duke Mathematical Journal},
   publisher={Duke University Press},
   author={Boalch, Philip},
   year={2007},
   month=Aug }

@inproceedings{Xu2024TheAD,
  title={The Alekseev-Meinrenken diffeomorphism arising from the Stokes phenomenon},
  author={Xiaomeng Xu},
  year={2024},
  url={https://api.semanticscholar.org/CorpusID:267094713}
}

@article{Iwaki:2014vad,
    author = "Iwaki, Kohei and Nakanishi, Tomoki",
    title = "{Exact WKB analysis and cluster algebras}",
    eprint = "1401.7094",
    archivePrefix = "arXiv",
    primaryClass = "math.CA",
    doi = "10.1088/1751-8113/47/47/474009",
    journal = "J. Phys. A",
    volume = "47",
    number = "47",
    pages = "474009",
    year = "2014"
}

@inproceedings{Xu2020RepresentationsOQ,
  title={Representations of quantum groups arising from the Stokes phenomenon},
  author={Xiaomeng Xu},
  year={2020},
  url={https://api.semanticscholar.org/CorpusID:269187555}
}

% =================================================
% Author information
% This appears strictly after the references.
% =================================================

\end{document}